\documentclass[preprint,12pt]{aastex}
 \usepackage{epstopdf}
 \usepackage{epsfig}
 \usepackage{longtable,lscape}
\shorttitle{Parallaxes for 70 Ultracool Dwarfs}
\shortauthors{Faherty et al.}
\begin{document} 

\title{The Brown Dwarf Kinematics Project (BDKP). III. Parallaxes for 70 Ultracool Dwarfs }

\author{Jacqueline K. Faherty\altaffilmark{1,2,4,5}, Adam J.\ Burgasser\altaffilmark{3,11}, Frederick M. Walter\altaffilmark{4}, Nicole Van der Bliek\altaffilmark{6}, Michael M. Shara\altaffilmark{1}, Kelle L. Cruz\altaffilmark{1,7}, Andrew A.\  West\altaffilmark{8}, Frederick J. Vrba\altaffilmark{9}, Guillem Anglada-EscudŽ\altaffilmark{10}}

\altaffiltext{1}{Department of Astrophysics, American Museum of Natural History, Central Park West at 79th Street, New York, NY 10034; jfaherty@amnh.org }
\altaffiltext{2}{Departmento de Astronomia, Universidad de Chile, Casilla 36-D, Santiago, Chile}

\altaffiltext{3}{Center of Astrophysics and Space Sciences, Department of Physics, University of California, San Diego, CA 92093, USA}
\altaffiltext{4}{Department of Physics and Astronomy, Stony Brook University Stony Brook, NY 11794-3800}
\altaffiltext{5}{Visiting astronomer, Cerro Tololo Inter-American Observatory, National Optical Astronomy Observatory, which are operated by the Association of Universities for Research in Astronomy, under contract with the National Science Foundation.}
\altaffiltext{6}{CTIO/National Optical Astronomy Observatory (Chile)}
\altaffiltext{7}{Department of Physics \& Astronomy, Hunter College, City University of New York, 695 Park Avenue, New York, NY 10065, USA}
\altaffiltext{8}{Department of Astronomy, Boston University, 725 Commonwealth Ave Boston, MA 02215}
\altaffiltext{9}{US Naval Observatory, Flagstaff Station, P.O. Box 1149, Flagstaff, AZ 86002}
\altaffiltext{10}{Department of Terrestrial Magnetism, Carnegie Institution of  Washington 5241 Broad Branch Road, NW, Washington, DC 20015 USA}
\altaffiltext{11}{Hellman Fellow}

\begin{abstract}
We report  parallax measurements for 70 ultracool dwarfs (UCDs) including 11 late-M, 32 L , and 27 T dwarfs.  Among this sample, 14 M and L dwarfs exhibit  low-surface gravity features, six are close binary systems, and two are metal-poor subdwarfs.  We combined our new measurements with 114 previously published UCD parallaxes and optical - mid-IR photometry to examine trends in spectral-type/absolute magnitude, and color-color diagrams.  We report new polynomial relations between spectral type and M$_{JHK}$.  Including resolved L/T transition binaries in the relations, we find no reason to differentiate between a ``bright" (unresolved binary) and ``faint" (single source) sample across the L/T boundary.  Isolating early T dwarfs, we find that the brightening of T0-T4 sources is prominent in M$_{J}$ where there is a [1.2 - 1.4] magnitude difference.  A similar yet dampened brightening of [0.3 - 0.5] magnitude happens at M$_{H}$ and a plateau or dimming of [-0.2 - -0.3] magnitude is seen in M$_{K}$.  Comparison with evolutionary models that vary gravity, metallicity, and cloud thickness verifies that for L into T dwarfs, decreasing cloud thickness reproduces brown dwarf near-IR color-magnitude diagrams. However we find that a near constant temperature of 1200 $\pm$100 K along a narrow spectral subtype of T0-T4 is required to account for the brightening and color magnitude diagram of the L-dwarf/T-dwarf transition. There is a significant population of both L and T dwarfs which are red or potentially ``ultra-cloudy" compared to the models, many of which are known to be young indicating a correlation between enhanced photospheric dust and youth. For the low surface-gravity or young companion L dwarfs we find that 8 out of 10 are at least [0.2-1.0] magnitude underluminous in  M$_{JH}$ and/or M$_{K}$ compared to equivalent spectral type objects.  We speculate that this is a consequence of increased dust opacity and conclude that low-surface gravity L dwarfs require a completely new spectral-type/absolute magnitude polynomial for analysis.  
\end{abstract}

\section{INTRODUCTION}
For any new class of astronomical objects, distances are crucial for investigating basic physical properties.   Brown dwarfs, low mass objects that lack sustained stable hydrogen burning in their cores, are a recent addition to the plethora of objects studied in astronomy.  They were first predicted by \citet{1962AJ.....67S.579K} and \citet{1963PThPh..30..460H} but not observationally confirmed until the late 1990s (\citealt{1995Natur.378..463N}; \citealt{1995Natur.377..129R}).   They have masses between $\sim$ 0.072M$_{\sun}$ and 0.012 M$_{\sun}$, straddling the boundary between the lowest mass stars and the highest mass exoplanets (\citealt{Saumon1996}; \citealt{Chabrier1997}).  In early 2000, the standard stellar spectral classification scheme was extended beyond M dwarfs to include ``L'' dwarfs, objects with temperatures ranging between 1300 and 2000 K and ``T-Y'' dwarfs, objects cooler than 1300K (see \citealt{2005ARA&A..43..195K} and references therein; \citealt{2011ApJ...743...50C}).   The L-spectral class encompasses both low--mass stars and brown dwarfs, depending on the mass/age of the ultracool dwarf (UCD--see \citealt{Burrows2001} and references therein).  

Distances provide a direct means for calculating the luminosity and (if there is a reliable radius estimate) the effective temperature of a star or brown dwarf.  Using luminosity, color,  and a probe of effective temperature, brown dwarf color-magnitude diagrams can be populated and used to investigate physical and chemical parameters such as gravity, metallicty, or dust properties.  Moreover, parallaxes to a significant number of objects are required to create relations with spectral type that can be used to estimate distances to the majority of brown dwarfs lacking astrometric measurements.  

There are now nearly 1000 spectroscopically confirmed field L and T dwarfs\footnote{According to the dwarfarchives website maintained at http://dwarfarchives.org} that define observational near-IR color trends.  Spectroscopic standards have been designated at each subtype that exhibit characteristic features of the changing brown dwarf spectral energy distribution. The physical parameter that drives the major changes in the observable photometric and spectroscopic features of the brown dwarf population is a decreasing effective temperature (T$_{eff}$).  However, with an ever-growing list of brown dwarfs observed in the field, a number of outliers have emerged that exhibit how secondary parameters such as age, metallicity, and cloud variability can change observable properties.  There are a handful of brown dwarfs that have halo kinematics, exhibit blue near-IR colors, and have enhanced metal hydride bands along with weakened metal oxide absorption bands indicating that they are old and metal-poor subdwarfs (\citealt{2003ApJ...592.1186B, 2007ApJ...657..494B},\citealt{Cushing2009}, \citealt{Kirkpatrick2010}).  A number of objects have red near-IR colors, weak alkali lines, enhanced metal oxide absorption bands in the optical, and appear to be juvenile-aged members of nearby moving groups such as AB Doradus, $\beta$ Pictoris, etc (\citealt{2006ApJ...639.1120K,Kirkpatrick2010}, \citealt{2009AJ....137.3345C}, \citealt{2010ApJ...715..561A}, \citealt{2010ApJ...715L.165R}).  There are also objects that do not exhibit the extreme spectral features of subdwarfs or low-gravity dwarfs, but are nevertheless near-IR photometric outliers, whose photometric properties might be attributable to dust, subtle variations in age or metallicity, or photometric variability (\citealt{2004AJ....127.3553K}, \citealt{2008ApJ...674..451B}, \citealt{2009AJ....137....1F}, \citealt{2010AJ....139.1808S}, \citealt{Kirkpatrick2010},\citealt{2008ApJ...686..528L}, Radigan et al. in prep).   

With the early parallax programs of \citet{2002AJ....124.1170D}, \citet{2003AJ....126..975T} and \citet{2004AJ....127.2948V} as well as subsequent astrometric programs, the spectral-type/absolute magnitude relations and color-magnitude diagrams for brown dwarfs were first investigated (\citealt{2003PASP..115.1207T}, \citealt{2006AJ....132.1234C}, \citealt{2008ApJ...689L..53B}, \citealt{Lucas2010}, \citealt{Artigau2010}, \citealt{2009A&A...493L..27S}, \citealt{2010A&A...524A..38M}).   One of the remarkable features of UCD color-magnitude diagrams is the significant scatter in luminosity found among objects with similar spectral types.  Increasingly complex atmospheric and evolutionary models have explained this as the result of variations in secondary parameters such as gravity, metallicity, sedimentation efficiency and/or binarity (e.g. \citealt{1996A&A...308L..29T,2003ApJ...585L.151T}; \citealt{2006ApJ...640.1063B}; \citealt{2008MNRAS.391.1854H}; \citealt{2008ApJ...689.1327S}).  The models disagree as to which  parameter has the largest effect on the emergent spectra and color trends, and small numbers of objects with independently measured secondary parameters has hindered progress (\citealt{2004AJ....127.3553K}; \citealt{2006ApJ...651..502P}; \citealt{2010ApJ...710.1627L}).  Only by increasing the number of well-characterized UCDs with distance measurements can we hope to understand the source and extent of the variation in the color-magnitude and HR diagrams. 

Another major anomaly of brown dwarf color magnitude diagrams is an intriguing brightening (or bump) in the $J$-band (up to 1.5 magnitude;  \citealt{2004AJ....127.2948V}, \citealt{2003AJ....126..975T}), and to some extent $H$ and $K$, as objects transition between the warmer L dwarf and cooler T dwarf spectral classes.  In the past decade, several L/T transition objects have been confirmed as flux reversal binaries with resolved components straddling the bump.  These objects  confirm that the brightening is an intrinsic feature of brown dwarf evolution (e.g. \citealt{2008ApJ...685.1183L}; \citealt{2003AJ....125.3302G}; \citealt{2006ApJ...647.1393L}; \citealt{2006ApJS..166..585B, 2007AJ....134.1330B}), however the full extent and physical explanation remain mysterious.   One possible explanation is that opacity from condensate clouds is especially influential on the 1~$\micron$ region of L and early-type T dwarfs, so changes in cloud properties can explain the unusual brightening (e.g. \citealt{2002ApJ...571L.151B}; \citealt{2004AJ....127.3553K}). The small numbers of L/T transition objects with parallax measurements has hindered progress in understanding the degree and variation in brightening exhibited at this interesting phase of brown dwarf evolution.  

In late 2006 we initiated the Brown Dwarf Kinematics Project (BDKP) in order to address persistent questions of brown dwarf evolution and atmospheric properties using astrometric measurements of proper motion, parallax, and radial velocity. Proper motion analysis of the population was reported in \citet{2009AJ....137....1F, 2010AJ....139..176F, 2011AJ....141...71F}. In this work we report parallaxes for 70 UCDs.   Section ~\ref{obser} describes the target list as well as the data acquisition and reduction.  Section ~\ref{parallax} describes the parallax pipeline used to determine distances.  Section ~\ref{discuss} uses all parallax measurements reported in this work in combination with published values  and photometric information obtained from various catalogs to investigate spectral type/absolute magnitude relations and color-magnitude diagrams for the brown dwarf population.  In section ~\ref{Evol} an updated brown dwarf near-IR color magnitude diagram is examined using evolutionary models.  Sections ~\ref{LowG} and ~\ref{subdwarfs} discuss the absolute magnitude vs. spectral type relation for low-surface gravity  and subdwarfs respectively. Section ~\ref{kinematics} reviews the kinematics for an ensemble of all known brown dwarfs with parallax measurements and section ~\ref{individual} investigates both known and suspected binaries within the sample.  Conclusions are reported in section ~\ref{conclu}.

\section{Observations\label{obser}}
\subsection{Target List}
We compiled the parallax target list from the BDKP astrometric sample reported in \citet{2009AJ....137....1F}.  Instrumental limitations precluded measuring parallaxes to the faintest, most distant L and T dwarfs so we focused primarily on objects within 20pc.  However we were also interested in subsets of the population which included low-surface gravity dwarfs (potentially young sources) and subdwarfs (potentially old sources).  For these scientifically interesting subsets we relaxed our astrometric constraint to include sources whose predicted spectroscopic parallax was up to 50pc.   Our full target list consisted of 70 dwarfs, including 11 M, 32 L, and 27 T dwarfs (see Table ~\ref{targetlist}).  Among this sample there were 14 low-surface gravity dwarfs, two subdwarfs, six binary systems, and nine calibrators with previous parallax measurements.  

\subsection{Data Collection and Reduction}
\subsubsection{ANDICAM}
We obtained parallax frames with the ANDICAM (A Novel Double-Imaging CAMera- DePoy 2003) imager between November 2006 and March 2010 ($\sim$ 500 hours of observations).  All data were acquired through queue observing with telescope time allocated through the SMARTS (Small and Moderate Aperture Research Telescope System) consortium.  ANDICAM is a dual channel near-IR and CCD imager located on the 1.3m telescope at Cerro Tololo InterAmerican Observatory (CTIO).  The optical detector is a 1024 x 1024 CCD and the near-IR channel uses a Rockwell 1024 x 1024 HgCdTe HAWAII array.  The near-IR field of view is 2.4 arcminutes with an unbinned plate scale of 0.137 $\arcsec$/pixel. The optical CCD field of view is larger,  6.2 arcminutes, with a plate scale of  0.369 $\arcsec$/pixel.  The optical and near-IR channels operate independently with a dichroic filter directing light to the two independent cameras. Therefore, we were able to take a set of near-IR images while integrating in the optical. 

  To ensure the same reference stars for each parallax frame, we required the target star to always be placed in the same X,Y position on the detector.  We also required all observations to be made within $\pm$30 minutes of meridian crossing to minimize the corrections for differential color refraction (DCR--see section ~\ref{extract}  below).  Typical seeing was 1$\arcsec$ and useable conditions for our parallax program were up to 2 $\arcsec$.

In the optical we observed in the $I_{c}$ band with integration times that ranged from 265s for our brightest targets to 610s for our faintest (typical S/N $>$ 100 for all targets).   In the near-IR we observed in the $J$ band with integration times that ranged from 20s with 5 coadds for our brightest targets to 130s with 8 coadds for our faintest.  We acquired 5-7 near-IR images in a 10 $\arcsec$ dither pattern.  

The optical ANDICAM data was processed with overscan subtraction and flat-fielding, prior to distribution.   Initially we intended to use the near-IR data for parallaxes and the CCD data as a check on the astrometric quality.  However, we quickly realized that the optical images were far superior to the near-IR, which were plagued with imaging artifacts, an occasional elongated PSF, and a smaller field of view (therefore fewer reference stars).  As a result we report parallaxes in this paper based only on the optical imaging.

\subsubsection{Infrared Side Port Imager (ISPI)}
We collected parallax data for our faintest targets with the Infrared Side Point Imager (ISPI) on the CTIO 4m Blanco telescope (\citealt{2004SPIE.5492.1582V}).  ISPI is a 2048x2048 HgCdTe HAWAII-2 array with an $\sim$ 8 arcminute field of view and nominal plate scale of 0.303$\arcsec$/pixel.  Observations were conducted over a period of just under 2 years (from early March 2008 through late April 2010) on 15 observing runs.  After the spring of 2010, ISPI was replaced on the Blanco 4m telescope with the NEWFIRM infrared detector.  However it was put back on the telescope in October of 2011 and we obtained three more epochs over six nights of observing in November 2011, January 2012, and February 2012.  These new parallax frames are also included in this work.

As opposed to ANDICAM, ISPI data were collected classically.  All observations were carried out in the $J$ band under seeing conditions up to 2$\arcsec$ full width half maximum (FWHM) with typical conditions between 0.8-1.1$\arcsec$.  Most of the parallax observations were made when the target was within $\pm$30 minutes of the meridian to minimize the corrections for DCR (see section ~\ref{extract}  below).  However, due to observational constraints (weather, instrument issues, etc) some targets were observed within $\pm$1 hr of meridian crossing.  

In order to minimize the effects of distortion and to ensure the same reference stars in each frame, we placed the target star on the same X,Y pixel position for each parallax frame.   On the first observing run for a target, the frame was initially offset from the center of the chip to avoid the four-quadrant seam along the detector.  This initial frame was used in all subsequent observing runs as a reference for determining where to place the parallax star.  

Integration times were set by the magnitude of the target and the conditions at the telescope.   They ranged from 30 to 60s with 5-10 coadds and 5-10 images in a 10 $\arcsec$ dither pattern (typical S/N $>$ 100 for all targets).  Depending on the weather and seeing conditions, the typical integration time  per target was 15-40 minutes.

Dark frames and lights on/off dome flats were obtained at the start of each evening.   Reduction procedures were based on the prescriptions put together by the ISPI team\footnote{$http://www.ctio.noao.edu/instruments/ir\_instruments/ispi/$} utilizing a combination of IRAF routines.  $J$ band flats were created by median-combining the lights on and lights off images then subtracting the two.  Bad pixel masks were created from a dome flat image.  Individual parallax frames were flat-fielded and corrected for bad pixels with the calibration images.  All images were flipped to orient North up and East to the left using the IRAF routine $\it{osiris}$ in the $\it{cirred}$ package.  Finally, the IRAF routine $\it{xdimsum}$ was used to perform sky subtractions and mask holes from bright stars.\footnote{We note that during observations the primary mirror would occasionally vibrate causing the PSF to appear elongated.  When this occured we would halt and restart an integration.  The problem was sporadic but did not affect any of the final images.}  
  
\section{Parallax Pipeline\label{parallax}}
\subsection{Source Extraction\label{extract}}
Once all images were reduced, we used the Carnegie Astrometric Planet Search (from here-on ATPa) software to extract all point sources and solve for relative parallaxes and proper motions (\citealt{2009PASP..121.1218B}). Images were not co-added, rather sources were extracted on every image at every dither position.

The processing of each night of observation (or epoch) was treated separately and consisted of extracting the position of all point sources. This required the specification of (1) a guess of the Full-Width Half Maximum (FWHM) of each set of images (2) the plate scale for the instrument and (3) a high quality image to be used as the template used to generate a preliminary list of sources to be extracted on all other frames in a given epoch. 

The precise centroids of the stars were measured by binning the stellar profile in the X and Y directions using a box of $\sim 2\arcsec$ around the pixel with maximum flux. Each one-dimensional profile was precisely centroided by finding the zero of the profile convolution using \textit{Tukey's biweight function} (Tukey 1960). The width of this function depends on a scale parameter to compute the centroid which was varied and the average taken (FWHM-0.5 pix, FWHM-0.25 pix, FWHM, FWHM+0.25 pix, FWHM-0.5 pix) to mitigate systematic effects caused by poor sampling of the stellar PSF.  We tested a variety of centroiding methods using different profiles but decided that the convolution approach described above provided the best centroid accuracy and robustness \citep[see][for further details]{2009PASP..121.1218B}. Once all sources were extracted on all the frames in a given night, the 10 stars showing the best formal centroid uncertainties were used to define a preliminary reference frame and their standard deviation assigned as the precision per image. 

ATPa generates a text file (from now on referred to as a plate file) containing X, Y positions with corresponding uncertainties (in pixels), and a rough flux measurement of all the successfully extracted stars. Plate files were used in the next step of the processing to calibrate the field distortions and measure the motion of each star as a function of time.  Objects displaying positional uncertainties larger than 5 times the median precision were automatically removed. These were usually spurious sources caused by unfiltered detector artifacts or cosmic rays. The typical centroiding uncertainty for parallax targets in ISPI and ANDICAM was $\sim$ 0.01 pixels. 

DCR corrections are typically required because the parallax star and reference stars have very different colors.  As a result, their positions shift relative to one another due to different amounts of atmospheric refraction.  The effect is wavelength, weather, and zenith-distance dependent.  \citet{1996PASP..108.1051S,2002PASP..114.1070S} presents a theoretical method for determining DCR effects.  That work demonstrates that by maintaining small zenith distances, DCR effects in $I$ and longer wavelengths (such as $J$) are minimal, typically $<$ 1 mas.  Similar results were found using the empirical methodology proposed by \citet{1992AJ....103..638M}.  The low-mass star optical parallax program of \citet{2005AJ....129.1954J} and the brown dwarf optical parallax program of \citet{2002AJ....124.1170D} also found negligible $I$ band DCR corrections as did the near-IR T dwarf parallax program of \citet{2003AJ....126..975T}.   Therefore, DCR corrections are not applied to the positions in our pipeline. To ensure that even this small effect was minimized, we observed targets (with few exceptions) within $\pm$ 30 minutes of meridian crossing.

\subsection{Astrometric solution and Parallax\label{solution}}
Using the \textit{Astrometric Iterative Solution} included in the ATPa package, the extracted X,Y pixel positions were mapped to a local tangent plane in order to solve for the astrometric parameters. The highest quality image obtained for a given parallax target was designated as the initial catalog to which everything was matched in the first astrometric iteration. The solution needed to be initialized with the right ascension (RA) and declination (DEC) coordinates of the brightest star in the field; however, the coordinates were only required to compute the projected paralactic motion in the tangent plane. Therefore a precision of a few arcseconds on this bright star was more than sufficient to initialize the solution. In the first
iteration, the position of the stars in the initial catalog were still approximate, so only a linear transformation was applied to each plate to correct for telescope pointing, field rotation and changes in the plate scale. This matching provided the position of each star as a function of time in the reference frame defined by the initial catalog. The apparent trajectory of each star was then fit to a linearized astrometric model

\begin{eqnarray}\label{eq:model}
x(t) &=& x_0 + \mu_{\alpha} \left(t-t_0\right) - \Pi p_{\alpha}(t)\, ,\\
y(t) &=& y_0 + \mu_{\delta} \left(t-t_0\right) - \Pi p_{\delta}(t)\, ,
\end{eqnarray}

\noindent where $x_0$ and $y_0$ are positional offsets at the first epoch of
observation $t_0$, $\mu_\alpha$ and $\mu_\delta$ are the proper motion in RA
and DEC respectively, $\Pi$ is the parallax, and $p_{\alpha}$ and $p_\delta$
are the parallax factors in RA and DEC respectively. At this point, all the
quantities are given in milliarcseconds (mas) and the time $t$ is measured in
years. The parallax factors are computed using the Earth geocenter as obtained
from the DE405 Ephemeris \footnote{http://ssd.jpl.nasa.gov/}. This linearized
model is based on the prescriptions laid out in the HIPPARCOS
(\citealt{1997A&A...323L..49P}) and Tycho
Catalogues\footnote{$http://www.rssd.esa.int/SA/HIPPARCOS/docs/vol1\_all.pdf$}
(\citealt{2000yCat.1259....0H}). 

After the astrometric solution of the field was obtained, a subset of well-behaved stars (RMS $<$ 5 mas and at least 4-5 observations) was chosen and the second iteration begun using those as reference sources. The whole process (crossmatching, field distortion fit and astrometric solution) was repeated a number of times. After the first iteration, ATPa allowed fitting more detailed field distortion corrections using higher order polynomials in X and Y. We tested up to 3rd order polynomials but these yielded negligible improvement in both ANDICAM and ISPI images.  Consequently, we ran all targets using a second order polynomial.

We set the number of total iterations to 3-5. The convergence was monitored by checking the average RMS of the well-behaved reference sources. This iterative process was automated by ATPa and required little supervision. The selection of reference stars was done automatically, but ATPa allowed the user to flag problematic reference stars. As a result, after a first solution was obtained, we checked the final catalog and re-ran the whole astrometric process eliminating the following objects from the reference frame : (1) the
target parallax star, (2) other high proper motion stars in the field, (3) saturated stars, and (4) elongated or extended sources (e.g. galaxies). The whole iterative process was repeated a second time. The final catalog that was output by ATPa contained the five free parameters defined above for each star in the field. It also contained formal uncertainties derived from the covariance matrix of the least squares solution for the equations above, information about the number of observations employed, and the RMS of the residuals per epoch.

In Table ~\ref{targetlist} we list the data on each target including the baseline between the first and last observations, the number of nights a target was observed (epochs), and how many parallax frames were included in the solution.  In the case of a number of targets, parallactic sampling was low (e.g. 5-7 epochs); therefore, to ensure realistic uncertainties, we applied a Monte Carlo method to the solution. For each target, we independently measured the standard deviation of the RA and DEC residuals. However, since many of the targets had a small number of epochs, we used the median standard deviation from the reference stars instead. These more realistic residuals were used to add random Gaussian noise to simulated observations with the parallax and proper motion from the final ATPa catalog. We repeated this experiment 1000 times and solved for the five astrometric parameters each time. The standard deviation around the mean value of the parallax over the Monte Carlo runs is the final uncertainty listed in Table ~\ref{astrometry}. As long as the standard deviations used in the Monte Carlo experiment are realistic, this approach automatically accounts for the correlation between the parallax and proper motion and offsets the issue of under sampled astrometry.  We flag sources in Table ~\ref{targetlist} that have $<$ 9 epochs and potentially under sampled parallaxes. 

\begin{figure}[!ht]
\begin{center}
\epsscale{1.0}
\plotone{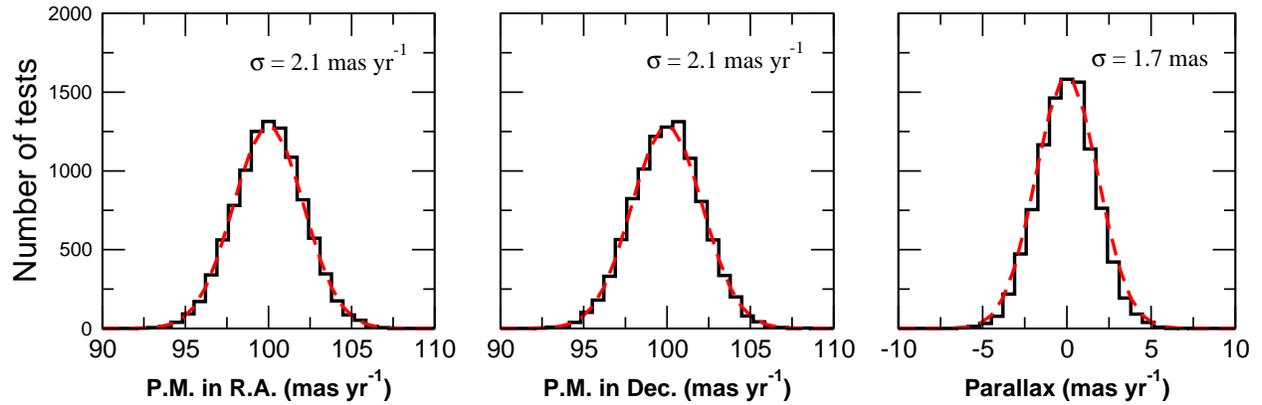}
\end{center}
\caption[Monte Carlo Simulation]{Histograms (black) of the obtained distributions from 10$^5$ synthetic datasets obtained using the epochs for calibrator 2M0746+2000 . Each simulated dataset assumes a 0 parallax and +100 mas yr$^-1$ in both RA and DEC. The red dashed line is a Gaussian distribution with $\sigma$ equal to the standard deviation measured on the Monte Carlo generated datasets, illustrating the perfect agreement between the two, and validating our approach to determine empirical uncertainties. } 
\label{fig:MC}
\end{figure}

We note that parallax and proper motion are linear parameters of the model. As a consequence, one does not depend on the particular value of the other in anyway. Still, it is true that they can be correlated due to the sampling cadence. One of the advantages of the Monte Carlo approach used here is that such a correlation is automatically accounted for in the process of generating synthetic datasets at the same observing epochs and deriving the empirical standard deviations over the repetitions of the same measurement.   To illustrate this, we use the observing epochs of 2M0746+2000 (13 epochs over 2.5 years) and simulate astrometric measurements assuming 0 parallax and a proper motion of +100 mas yr$^{-1}$ on both RA and DEC. We introduce random Gaussian noise with a standard deviation of 5 mas in both RA and DEC and solve for all the astrometric parameters. We repeat this experiment 10,000 times and produce the numerical distributions of the obtained $\mu_{\alpha}$, $\mu_{\delta}$ and parallax. As expected from a linear model, the obtained distributions are non-biased and the shape of the marginalized distributions is Gaussian with the same $\sigma$ we measure on the scatter of the 10$^5$ Monte Carlo obtained parameters (see Figure ~\ref{fig:MC}).

\subsection{Correction from relative to absolute parallax}
The final parallaxes from the astrometric solution are relative to the motion of the background stars chosen as references.  A correction is required based upon the true parallaxes of the reference stars to convert to an absolute measurement.   

Typically there are three ways to convert from relative to absolute parallaxes:  (1) using statistical methods which rely on a well-determined model of the Galaxy and is most relevant for faint distant reference stars, (2) spectroscopic parallaxes which rely on spectral data obtained for every reference star, or (3) photometric parallaxes which rely on colors for all reference stars.   We determined the parallax corrections using the third method because the reference stars are primarily the brightest in the field and spectral data are not available.

In order to measure photometric parallaxes for the reference stars we assume that all sources are main-sequence dwarfs.  Following a similar procedure described in \citet{2004AJ....127.2948V}, we obtained 2MASS photometry for all reference stars.  We compared with the intrinsic colors described in \citet{1983A&A...128...84K} so we first converted $J$,$H$,$K_{s}$ values to the Koornneef photometric system using the transformations detailed in \citet{2001AJ....121.2851C}.   The near-IR $J$-$H$ and $H$-$K$ colors were used to estimate spectral types and $V$-$K$ colors of the background stars based on the relations detailed in \citet{1983A&A...128...84K} for main sequence dwarfs.   Absolute V magnitudes were taken from color spectral type relations described in \citet{2004AsNow..18d..24K} and then converted to M$_{J}$ and M$_{K}$ values. Distances to reference stars were determined by averaging (m-M)$_{J}$ and (m-M)$_{K}$ values which were in good agreement. 

Each reference star was given equal weight in the astrometric solution. As a result we averaged the photometric parallaxes to calculate the distance correction and used a standard deviation of the mean for the correction uncertainty.  We added the distance correction to our relative parallax and added the correction uncertainty in quadrature with our parallax uncertainty to obtain the final absolute parallax.   The average correction to absolute parallax for the full list of targets was 1.5$\pm$0.5~mas ranging from  0.8 to 2.9 mas.   The final parallaxes with absolute corrections  are shown in Table ~\ref{astrometry}.

\subsection{Comparison of Calibrators}
\begin{figure}[!ht]
\begin{center}
\epsscale{1.2}
\includegraphics[width=.55\hsize]{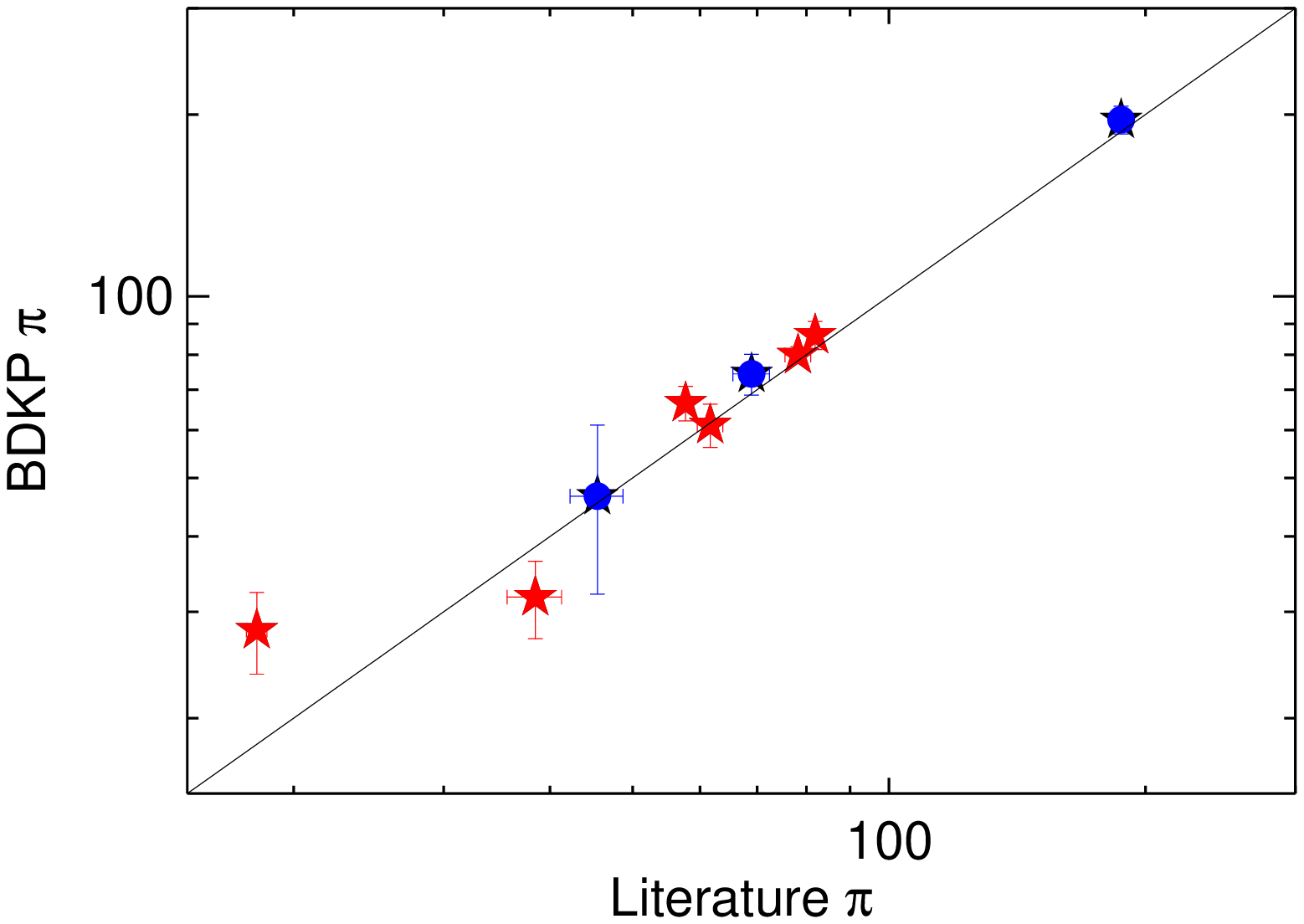}
\plottwo{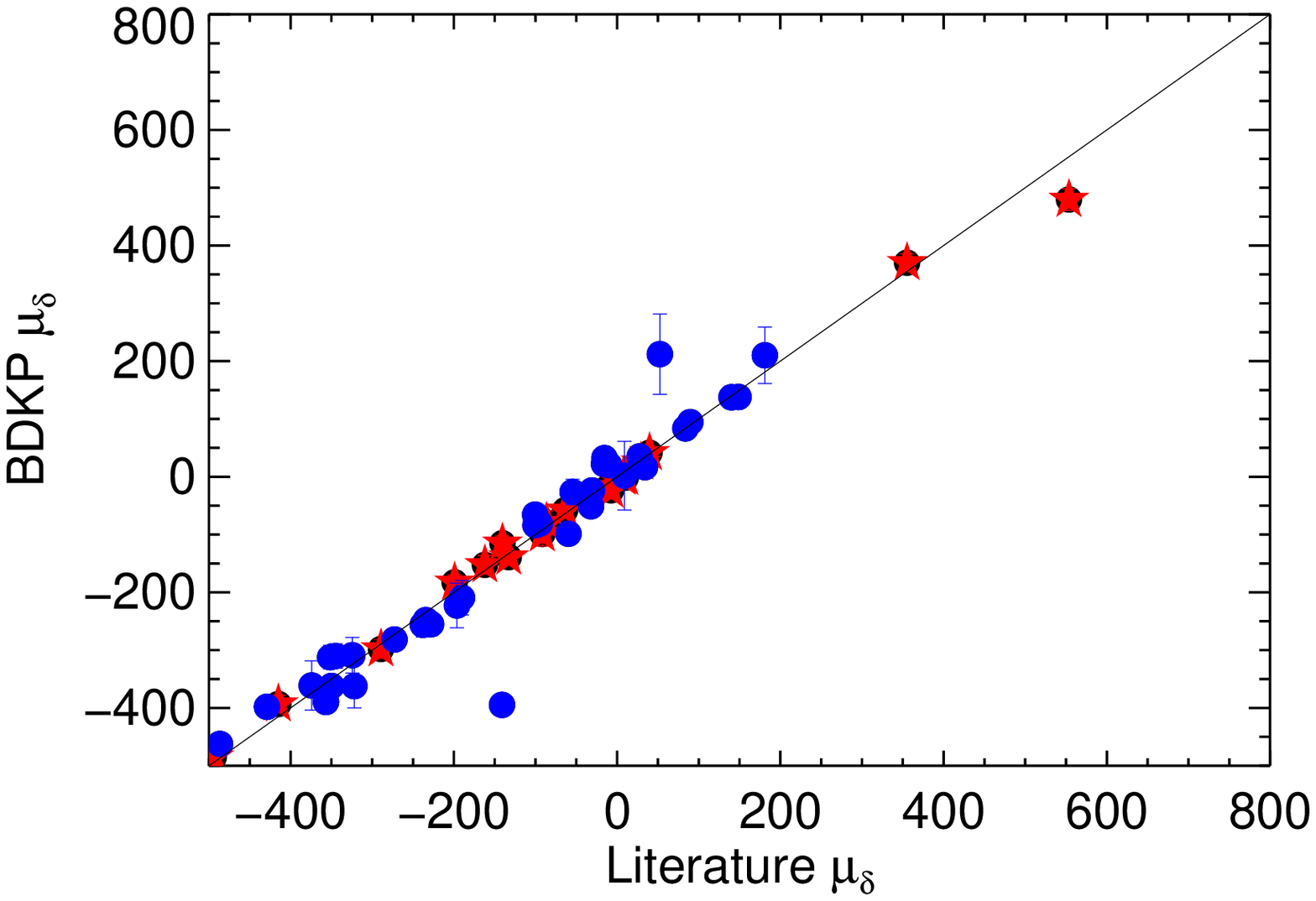}{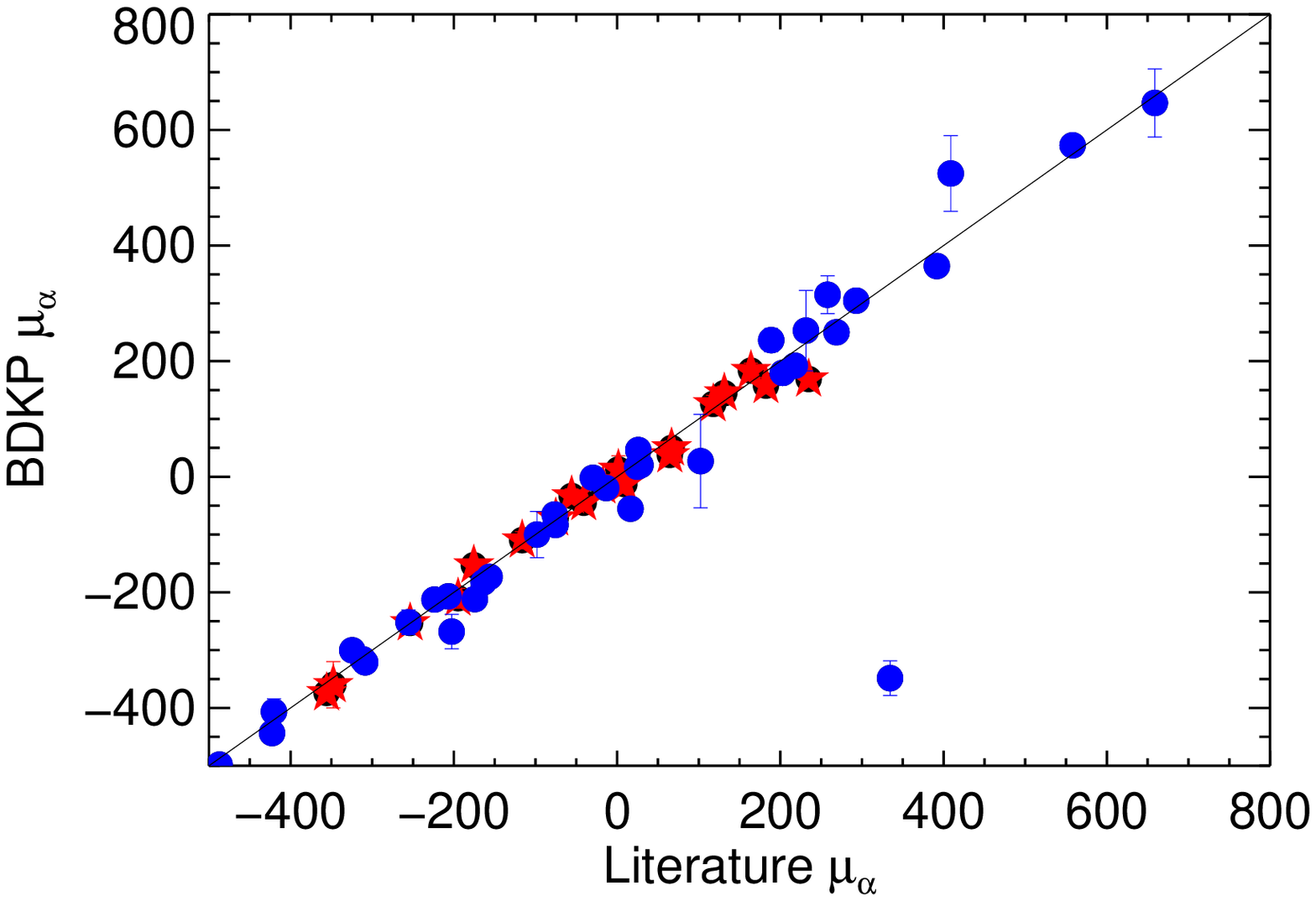}
\end{center}
\caption[Parallax Calibrators]{ Top plot:  The parallax measurement comparison of the nine calibrator dwarfs.  All but two match previous measurements within 1$\sigma$.  Middle and lower plots:  Comparison of literature proper motion components to those measured in this work.  Three objects are discrepant by more than 4$\sigma$ and listed in Table~\ref{discrepant}.  In each plot blue filled circles represent dwarfs that were measured with ISPI and red five point stars were measured with ANDICAM.  } 
\label{fig:Check}
\end{figure}

There are nine calibrator stars in our full astrometric sample (three imaged with ISPI and six imaged with ANDICAM) that we obtained as a check on the reliability of our methods.  Table ~\ref{calibs} lists the astrometry for the calibrators measured in this work and compares those values with results reported in the literature (see Figure ~\ref{fig:Check}).     Of the nine calibrators, seven match within 1 $\sigma$ and all but 1 (TWA 28) match within 2$\sigma$ of published values.   No systematic trends were detected in the parallax measurement differences.  The mean difference and scatter between literature and BDKP values is  -4.0$\pm$ 5.0~mas.  This indicates that our parallax pipeline produces reliable results.  

As a check on non-calibrator stars we examined how well the proper motion values match published values for all sources in our sample.  The two lower panels in Figure ~\ref{fig:Check} show plots of $\mu_{\alpha}$ and $\mu_{\delta}$  from the literature versus our calculated values with uncertainties.  We find that 42\% of the sample match both components within 1$\sigma$, 76\% match within 2$\sigma$, 87\% match within 3$\sigma$ and all but 3 match within 4$\sigma$.  This is different from a Gaussian error distribution, possibly indicating underestimated uncertainties in either published values or our own astrometric solution.  The mean difference between $\mu_{\alpha}$ and $\mu_{\delta}$ values was 4$\pm$32 mas and -4$\pm$30 mas respectively. No systematic trends were detected in the proper motion component differences.  We list objects whose $\mu_{\alpha}$ or $\mu_{\delta}$  components were discrepant by more than 4$\sigma$ in Table ~\ref{discrepant}. 2M1404-3159 is among the discrepant proper motion objects yet we believe there was a sign reporting error in Looper et al. 2007 as we calculate the same magnitude of motion in the opposite direction.

\section{Absolute Magnitude Relations\label{discuss}}
As of September 2011 there were 106 L and T dwarfs with published parallax measurements.   We have added 59 to this list, doubling the number of measurements in some spectral bins.  The precision on parallaxes in this work as well as within the literature varies greatly.  To ensure that the analysis that follows was not biased by uncertain parallaxes or photometry, we required all sources to have M$_{JHK}$ uncertainty $<$ 0.5 magnitude.  We list photometry, parallax measurements, and references for all known and new UCDs in Table ~\ref{Luminosity}.  With a substantial increase in the number of objects, we can re-evaluate the color-magnitude and spectral type-luminosity trends originally defined by \citet{2002AJ....124.1170D}, \citet{2003AJ....126..975T}, and \citet{2004AJ....127.2948V}, particularly across the poorly sampled L/T transition region. We list all L and T dwarfs with parallaxes and their corresponding magnitudes in optical through mid-IR wavelengths in Table ~\ref{Luminosity}.  The apparent MKO $JHK$ magnitudes listed in Table ~\ref{Luminosity} were used to calculate the absolute MKO M$_{JHK}$ magnitudes used throughout the analysis.

Figure ~\ref{fig:ABS_Mag2} shows the M$_{JHK}$ magnitude versus spectral type for all late-type M, L, and T dwarfs with parallax measurements.  Seemingly normal field objects (from here-on ''normal" is defined as excluding tight binaries unresolved in 2MASS, young sources with low-surface gravity features, and subdwarfs) provide a guideline for how near-IR intrinsic brightness changes with spectral type.  Absolute magnitudes of brown dwarf subgroups (unresolved binaries in red, subdwarfs in blue, and low-surface gravity or companions to young field stars in green and purple) allow us to investigate how secondary parameters such as binarity, metallicity, and age influence brown dwarf observables.  In this section we investigate what can be extrapolated for the population using the sequence of normal field dwarfs including resolved L/T transition binaries.  In section  ~\ref{LowG}, \ref{subdwarfs}, and \ref{individual} we discuss in detail the differences among the subdwarfs, low-surface gravity, and binary dwarfs respectively.  

Brown dwarfs have highly structured spectral energy distributions, and magnitudes in $JHK$ are extremely sensitive to the exact filter bandpass used.  Therefore, we converted all magnitudes onto the Mauna Kea Observatory filter set (MKO; \citealt{2002PASP..114..180T}), whose narrow bandpasses are less affected by atmospheric absorption than the CIT and 2MASS filter sets (particularly at $J$).  If required, the transformations from Stephens \& Leggett (2004) were used to convert from 2MASS to MKO magnitudes.  

Most of the L dwarfs in our sample were classified spectrally from red optical data following the scheme of \citet{1999ApJ...519..802K}, while the T dwarfs were classified in the near-IR (\citealt{2006ApJ...637.1067B}).  An optical spectral type was used for any object classified as an L dwarf and a near-IR spectral type was used for any object classified as a T dwarf.  For any L dwarf lacking optical data we used its near-IR spectral type.  

\subsection{Brown Dwarf HR Diagram\label{nearab}}
\begin{figure}[!ht]
\begin{center}
\epsscale{1.2}
\includegraphics[width=.55\hsize]{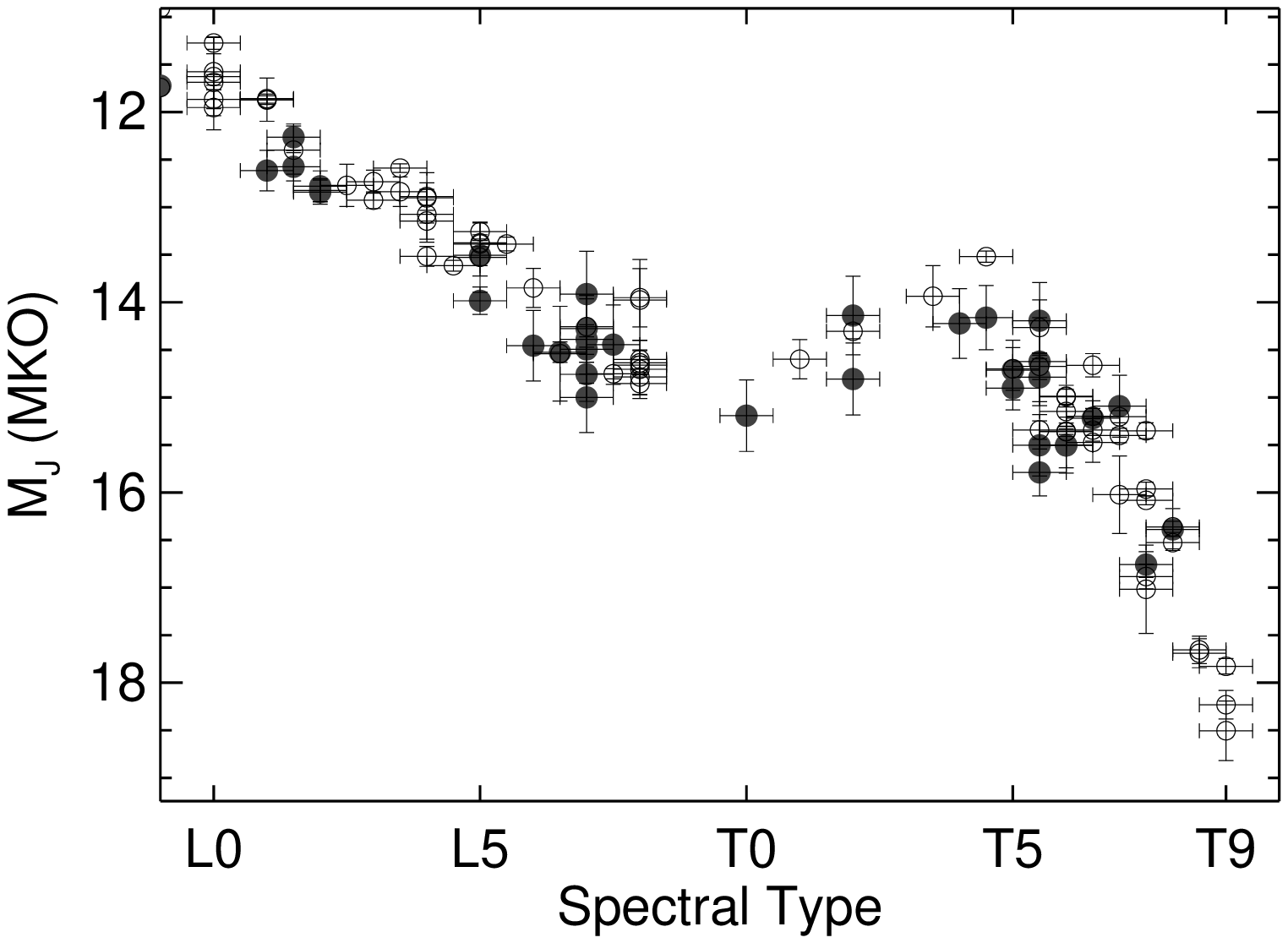}
\plottwo{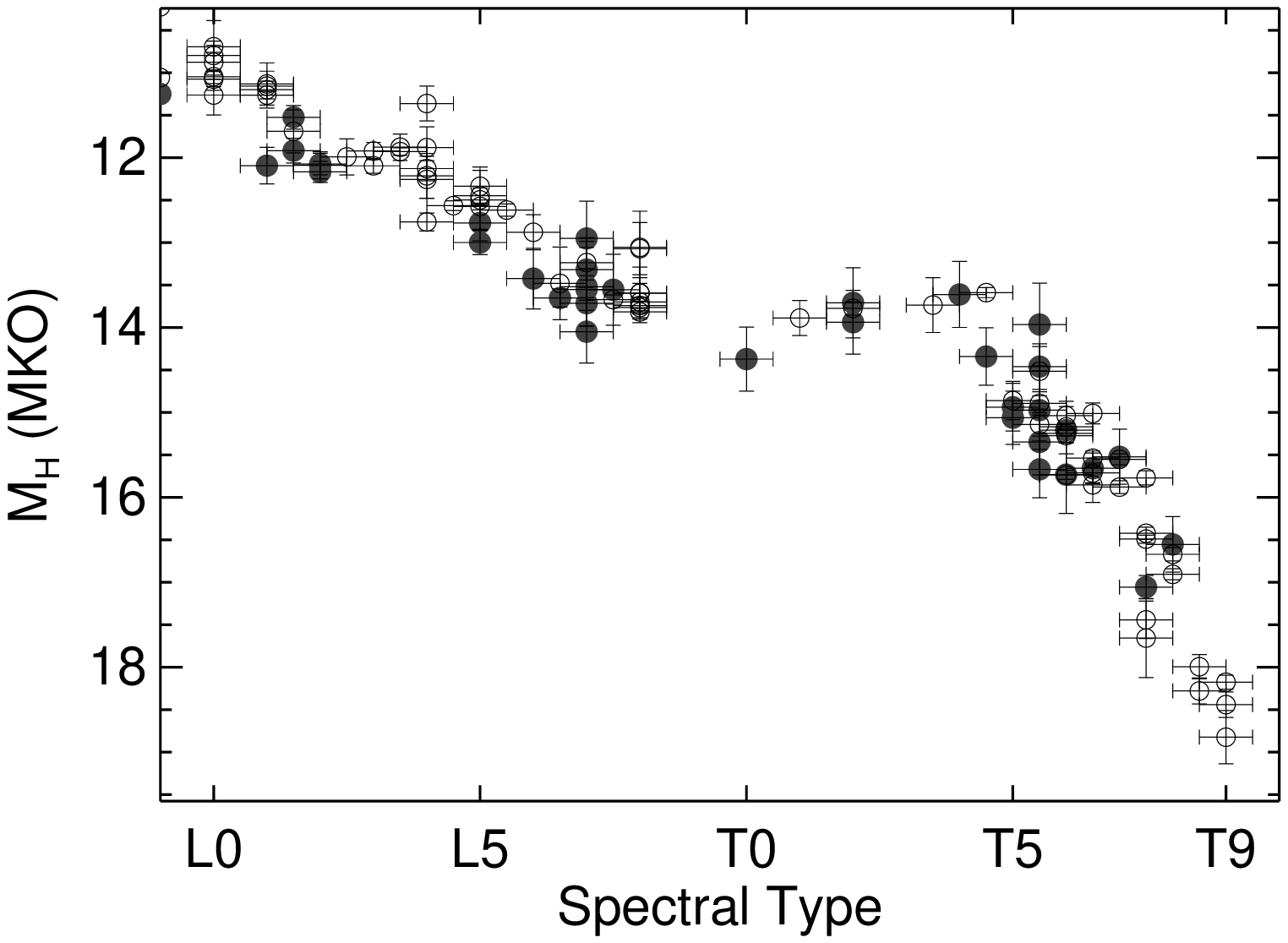}{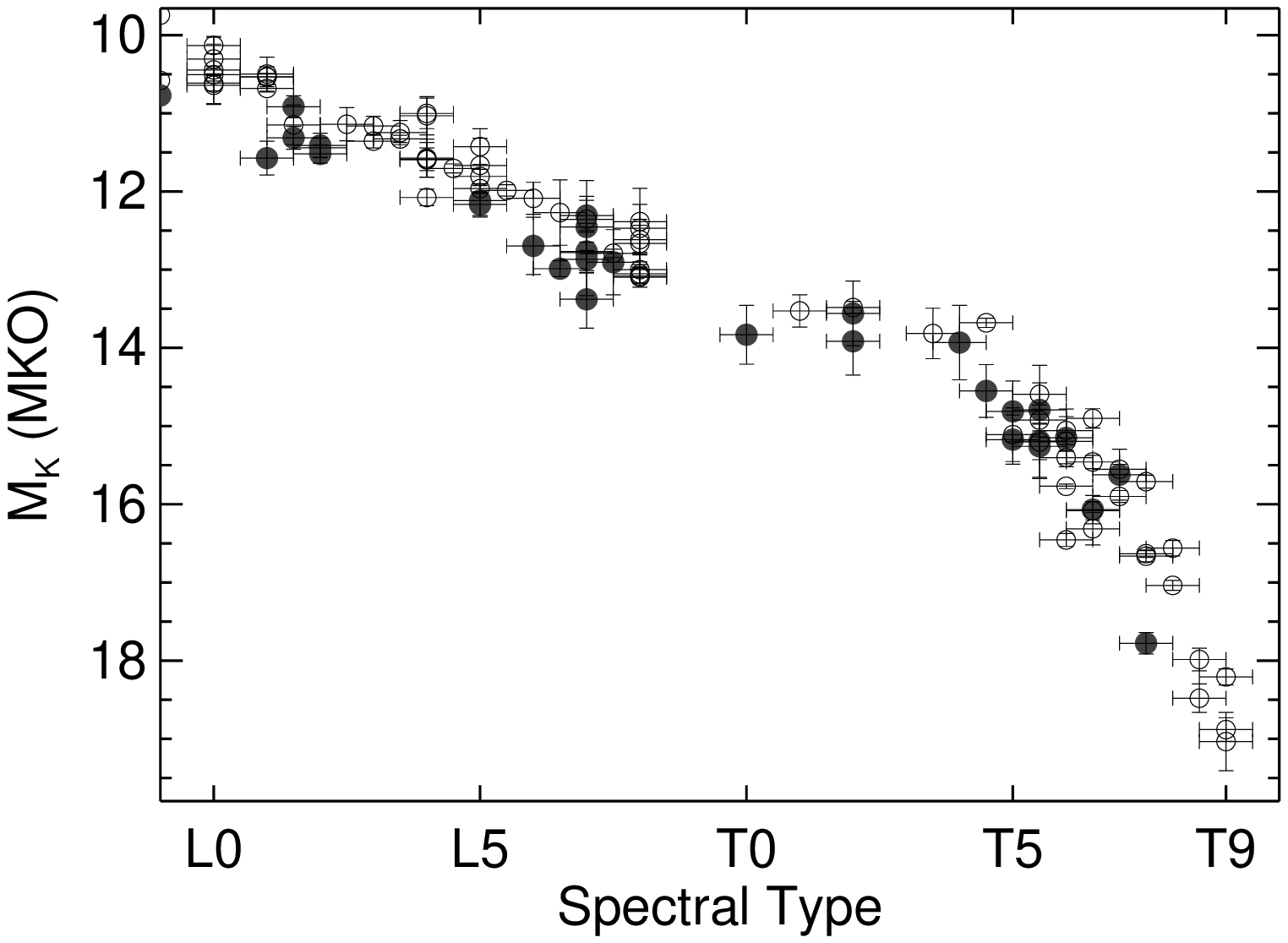}
\end{center}
\caption{ Spectral type versus absolute magnitude in the MKO $JHK$ filters for normal L through T dwarfs.  Unfilled circles are ultracool dwarfs with parallax measurements gathered from the literature and filled circles are those reported in this work. Low-surface gravity, subdwarf, binary, and young companion brown dwarfs are not shown.  Only objects with absolute magnitude uncertainties $<$ 0.5 mag are displayed. } 
\label{fig:ABS_Mag}
\end{figure}

\begin{figure}[!ht]
\begin{center}
\epsscale{1.2}
\includegraphics[width=.55\hsize]{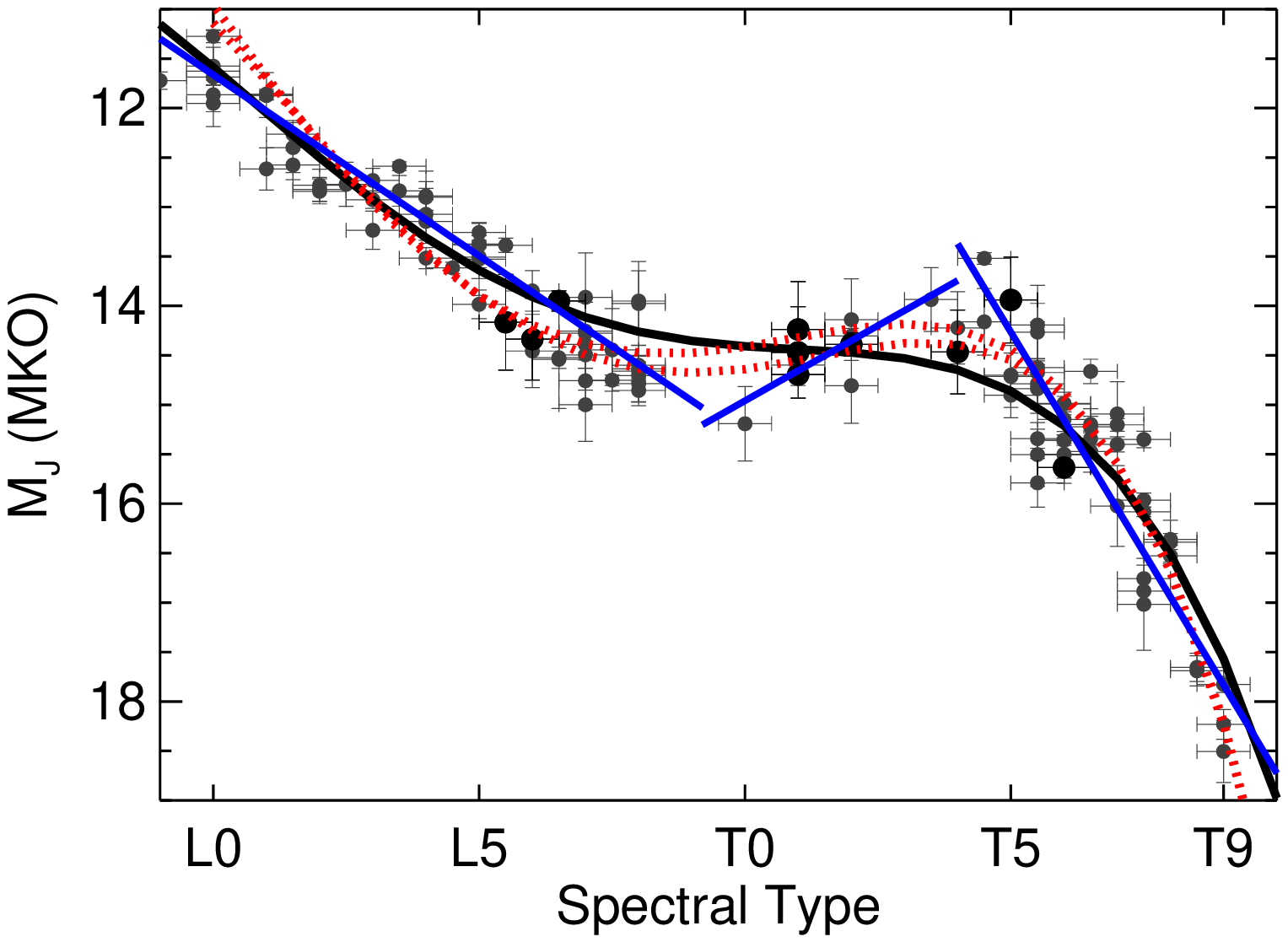}
\plottwo{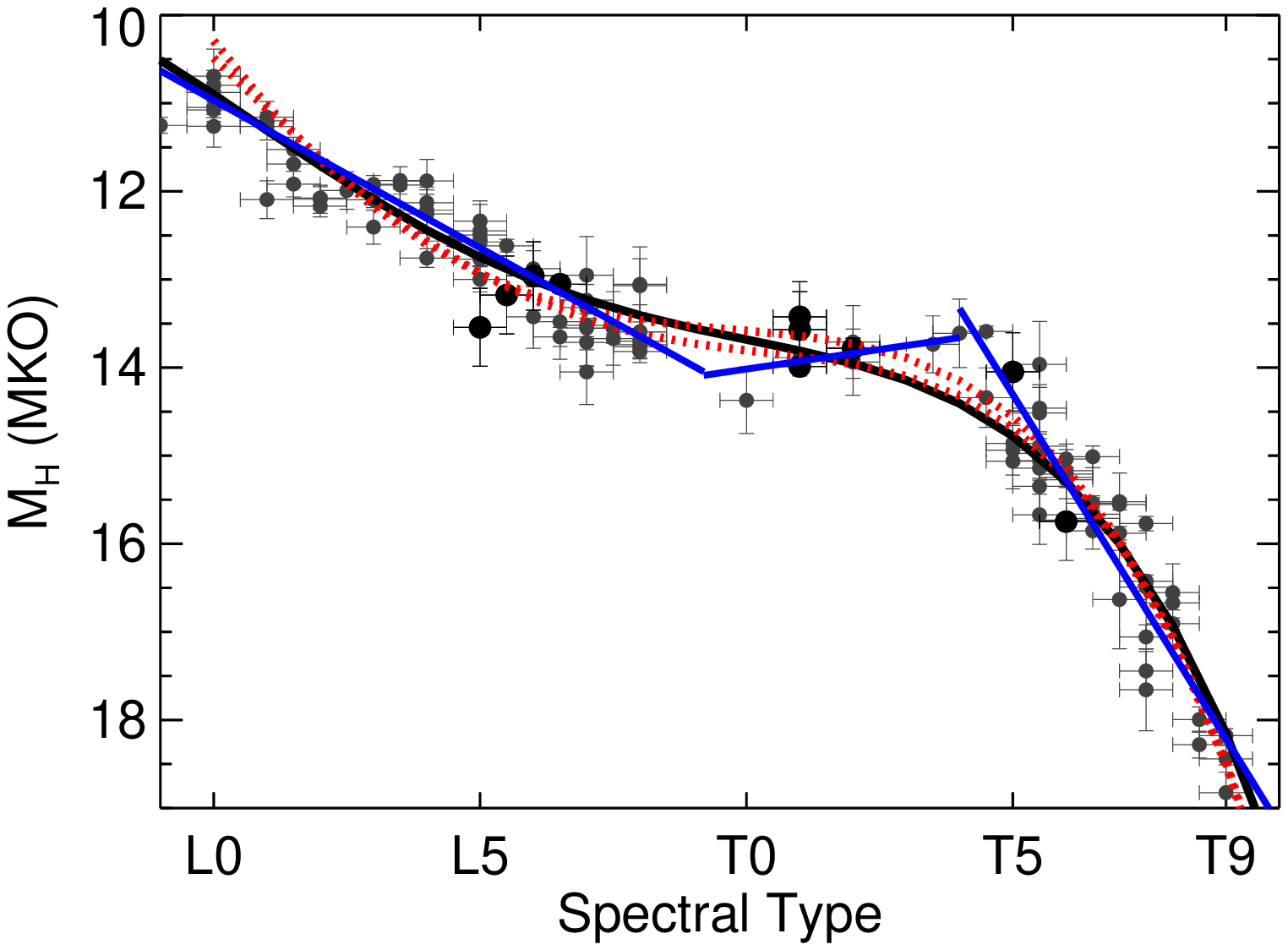}{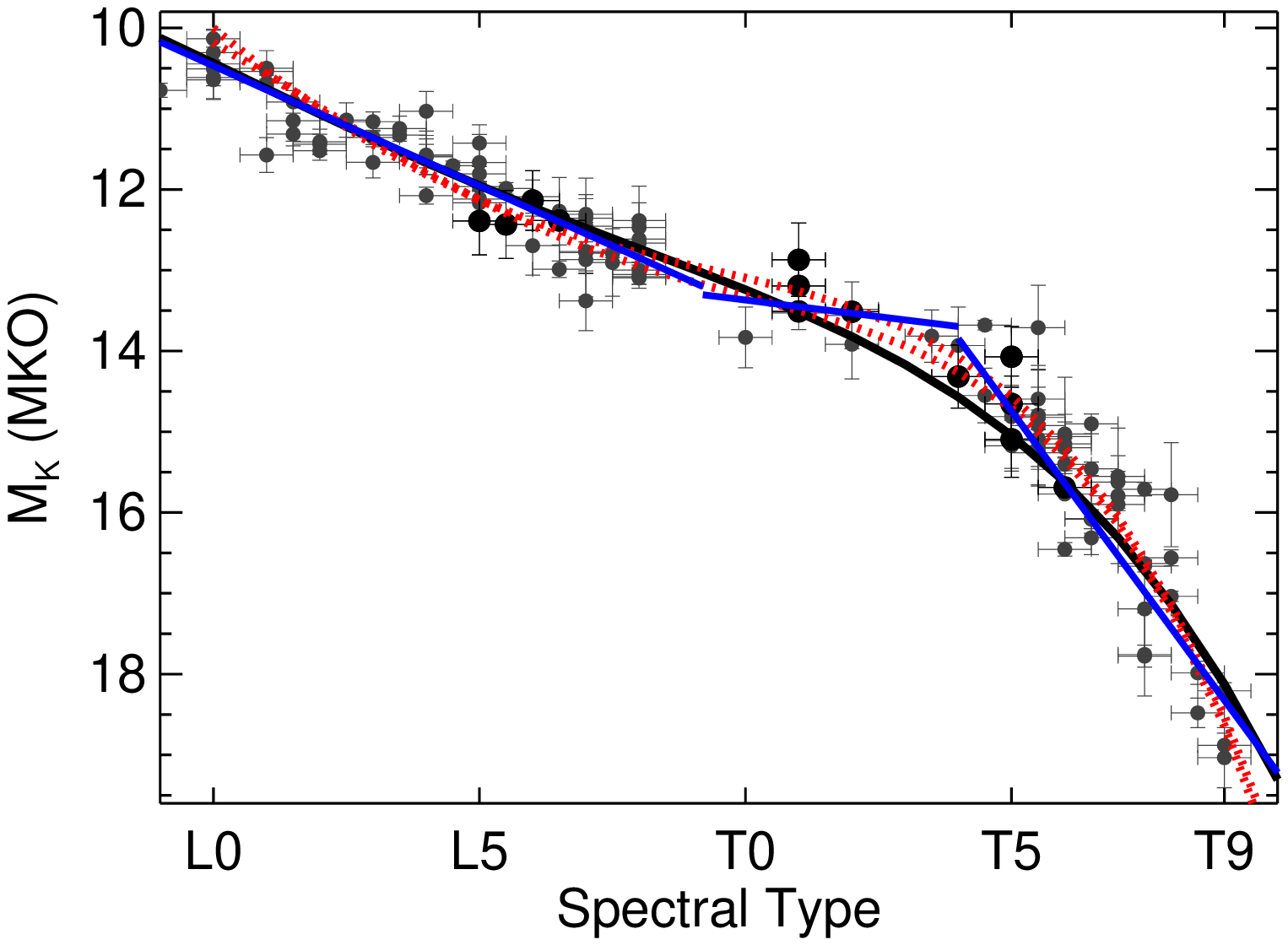}
\end{center}
\caption {Spectral type versus absolute magnitude in the MKO $JHK$ filters for normal L and T dwarfs and L/T transition binaries with resolved components.  Light grey filled circles represent normal field dwarfs.  Black filled circles are the resolved absolute magnitudes of the binary components.  Over-plotted on each panel is the best fit polynomial for L0-T9 dwarfs (see Table ~\ref{coeffs}) in solid black, the bright/faint polynomials in red-dashed from \citet{2010A&A...524A..38M}, and the best fit linear functions to the ranges L0-L9, T0-T4, and T5-T9 in solid blue.  Only objects with absolute magnitude uncertainties $<$ 0.5 mag are shown.} 
\label{fig:Binaries}
\end{figure}

Figure ~\ref{fig:ABS_Mag} shows the M$_{JHK}$ sequence for normal L and T dwarfs with uncertainties $<$ 0.5 mag.  We have distinguished published parallaxes (open circles) from those reported in this work (filled grey circles) to demonstrate the impact of these new measurements.   We have augmented the field dwarf sample in Figure ~\ref{fig:Binaries} with resolved photometry for 9 L dwarf/T dwarf transition binaries for a more detailed look at brightness trends across this transition (\citealt{2010ApJ...710.1142B}; \citealt{2008ApJ...685.1183L}; \citealt{2004A&A...413.1029M}; \citealt{2010ApJ...722..311L}; \citealt{2011A&A...525A.123S}).  Four known binary sources, 2MASS J0518-2828, 2MASS J1209-1004, 2MASS J1404-3159, and SDSS J2052-1609 had parallax measurements reported in this work as did one suspected binary SDSS J1511+0607 (\citealt{2010ApJ...710.1142B}, Gelino et al. in prep).  We list the component magnitudes of the binaries (including all known binary L/T transition objects with parallaxes) in Table ~\ref{LTBinaries2}.  

We have used the full parallax sample of normal objects with M$_{JHK}$ uncertainties $<$ 0.5 mag to re-examine commonly used near-infrared absolute magnitude/spectral type relations.  To determine the best fit between these parameters, we applied an F-test to a polynomial with increasing coefficients.  The F-test allows us to distinguish the false-alarm probability of a decreasing $\chi^2$ due to additional degrees of freedom.  In this manner, we converged on a fourth order fit to spectral type versus MKO M$_{JHK}$ for normal field dwarfs.  The coefficients are listed in Table ~\ref{coeffs} and the solution is over-plotted in Figure ~\ref{fig:Binaries}.  For comparison, \citet{2010A&A...524A..38M} recently reported 11 mid to late-type T dwarf parallaxes and also converged upon a 4th order polynomial fit to MKO M$_{JHK}$ for L0-T9 dwarfs.  In that work, near-IR spectral types were used for L0-T9 likely leading to a steeper fit for L dwarfs than our fit.   In the same manner as \citet{2006ApJ...647.1393L}, \citet{2010A&A...524A..38M} report two absolute magnitude/spectral type polynomials (over-plotted as red dashed lines on Figure ~\ref{fig:Binaries}): the first excluding known binaries and the second  excluding known binaries as well as five binary candidates across the L/T transition (as selected by Liu et al).  \citet{2006ApJ...647.1393L} found a 0.6 magnitude difference between the peak of the L/T transition when excluding binary candidates but with the addition of the  \citet{2010A&A...524A..38M} parallax sample this was reduced to 0.2 magnitude. Adding our new parallaxes, we find no difference in a polynomial fit which excludes or does not exclude the five binary candidates, therefore we report one fit for the entire sample.  

As the data indicates in Figure ~\ref{fig:Binaries}, the brightening across the L/T transition is more pronounced than demonstrated by a polynomial fit from L0-T9.  By eye, the L dwarfs, L/T transition objects and T dwarfs appear to follow distinct and independent linear trends.  Therefore, as an alternative to a full range L0-T9 polynomial, we have split the normal, single source objects on Figure ~\ref{fig:ABS_Mag} into three ranges of L0-L9, T0-T4, and T5-T9 and fit a linear polynomial to each.  We obtained uncertainties on the coefficients by randomly shifting the known objects in each range within the given uncertainty 10,000 times and fitting a gaussian to the range of parameters.  The best fit piecewise lines are over plotted on Figure ~\ref{fig:ABS_Mag} and reported in Table ~\ref{coeffs2}.    For the L/T transition between T0-T4, we find a brightening at J between [1.2 - 1.4] magnitudes, a brightening at H between [0.3 - 0.5] magnitudes, and a plateau or dimming at K between [-0.2- -0.3] magnitudes.  The increased sampling across the L/T transition demonstrates that the brightening is a real and prominent feature, however further investigation of individual objects is necessary to disentangle the role that dust and unresolved binarity play in creating the effect.  We discuss the L/T transition objects in relation to model predictions in section ~\ref{LTtransition}.

\subsection{Color-Magnitude Trends for L and T dwarfs\label{trends}}
\begin{figure}[!ht]
\begin{center}
\epsscale{1.0}
\plotone{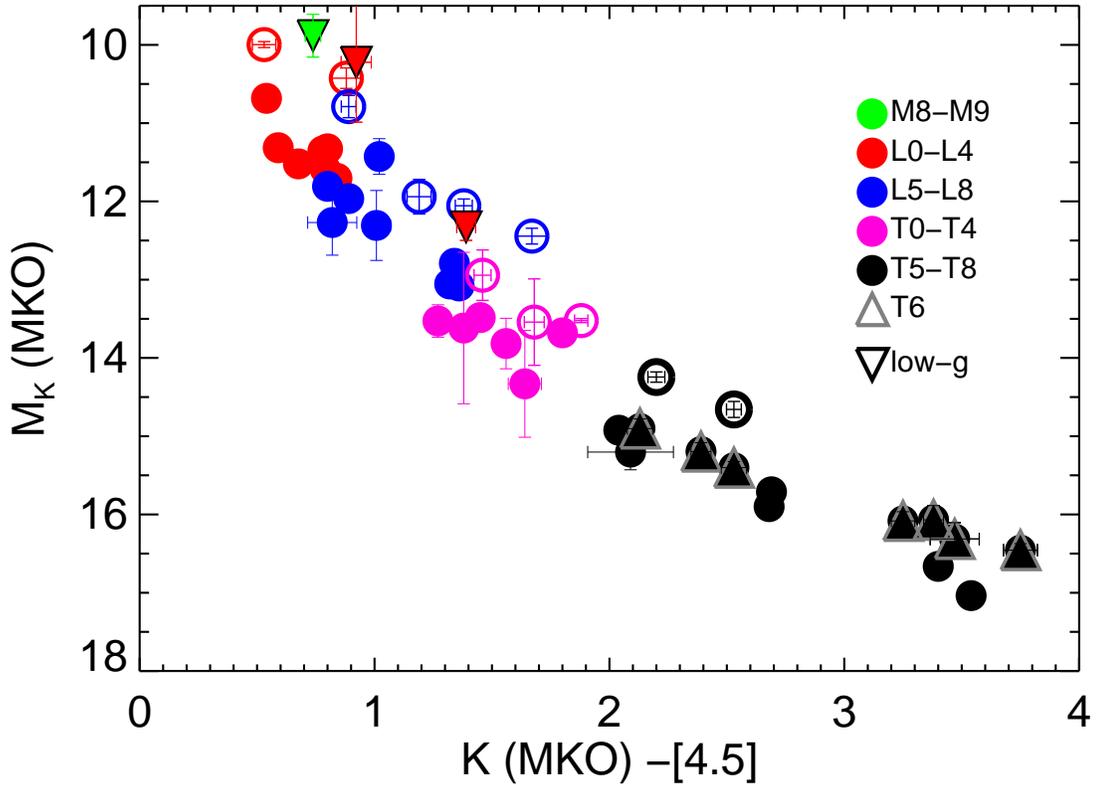}
\end{center}
\caption{The tightest linear relationship between color and luminosity identified for brown dwarfs. L dwarfs smoothly transition into T dwarfs although the Figure is non-linear for spectral subtypes (as depicted by the T6 dwarfs in grey upward-facing triangles).  Unresolved binaries are shown as open circles.  Low-surface gravity dwarfs are shown as downward facing triangles and appear slightly overluminous for their color. } 
\label{fig:color_mag2}
\end{figure}

\begin{figure}[!ht]
\begin{center}
\epsscale{1.2}
\includegraphics[width=.55\hsize]{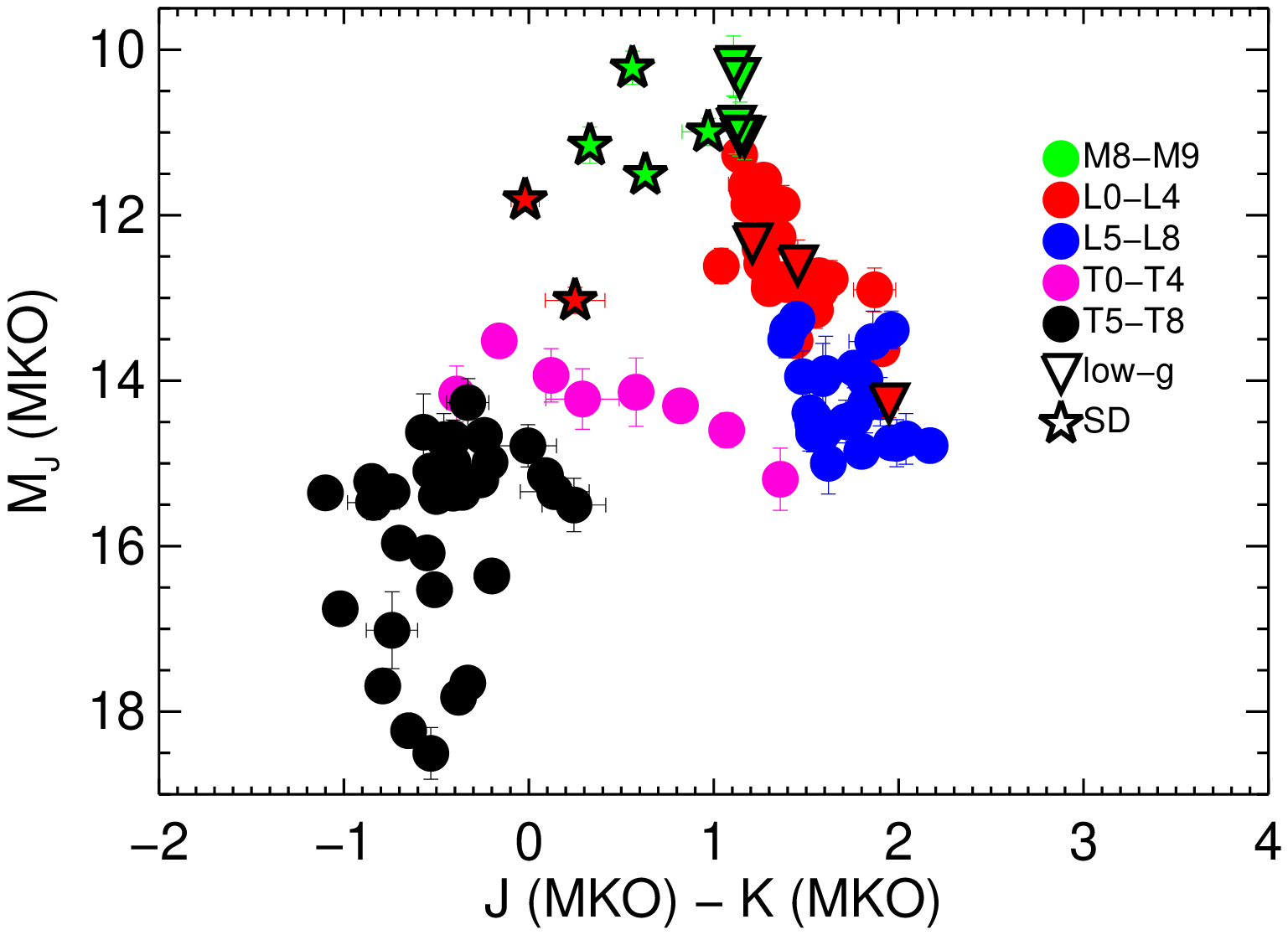}
\plottwo{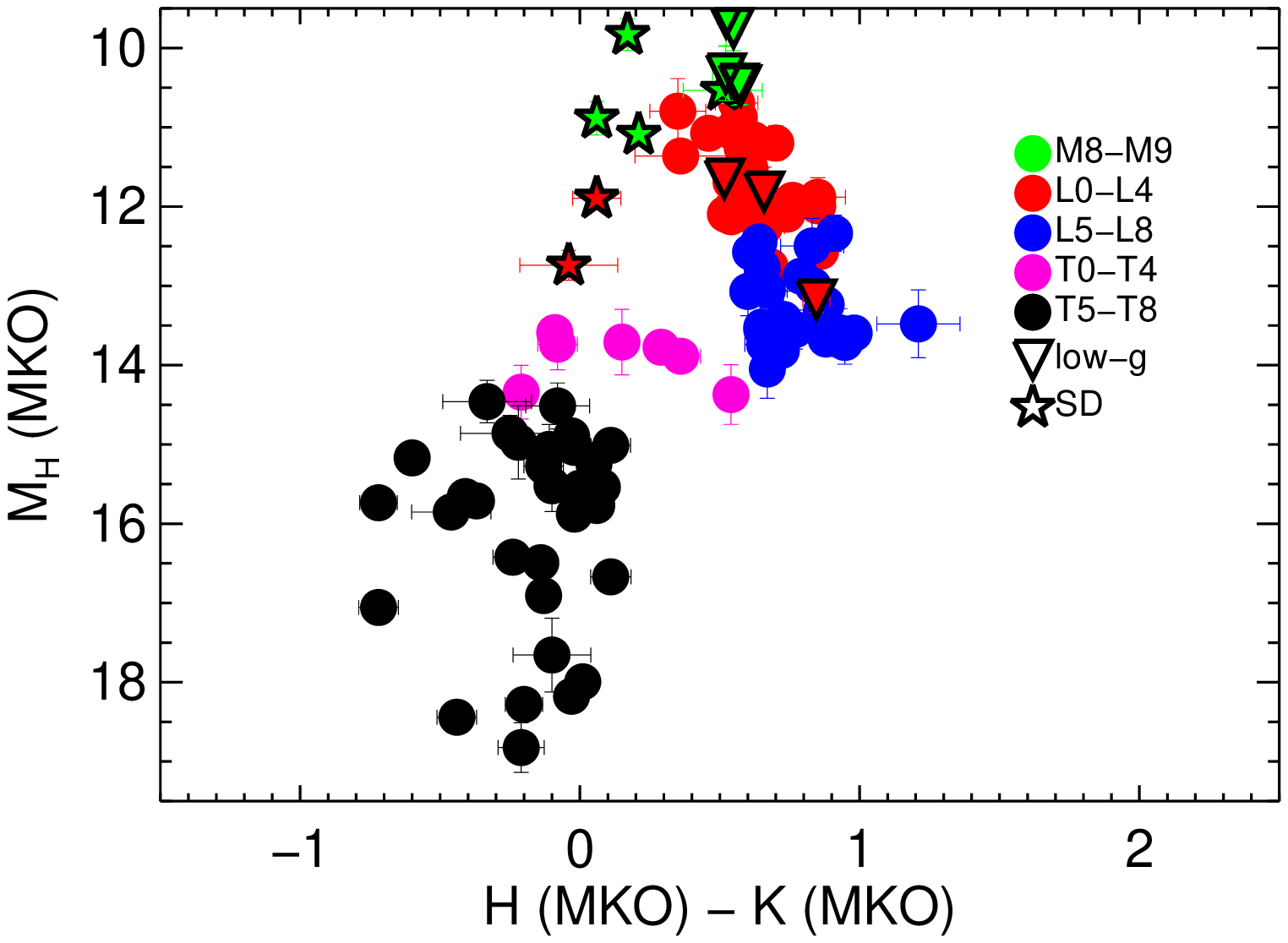}{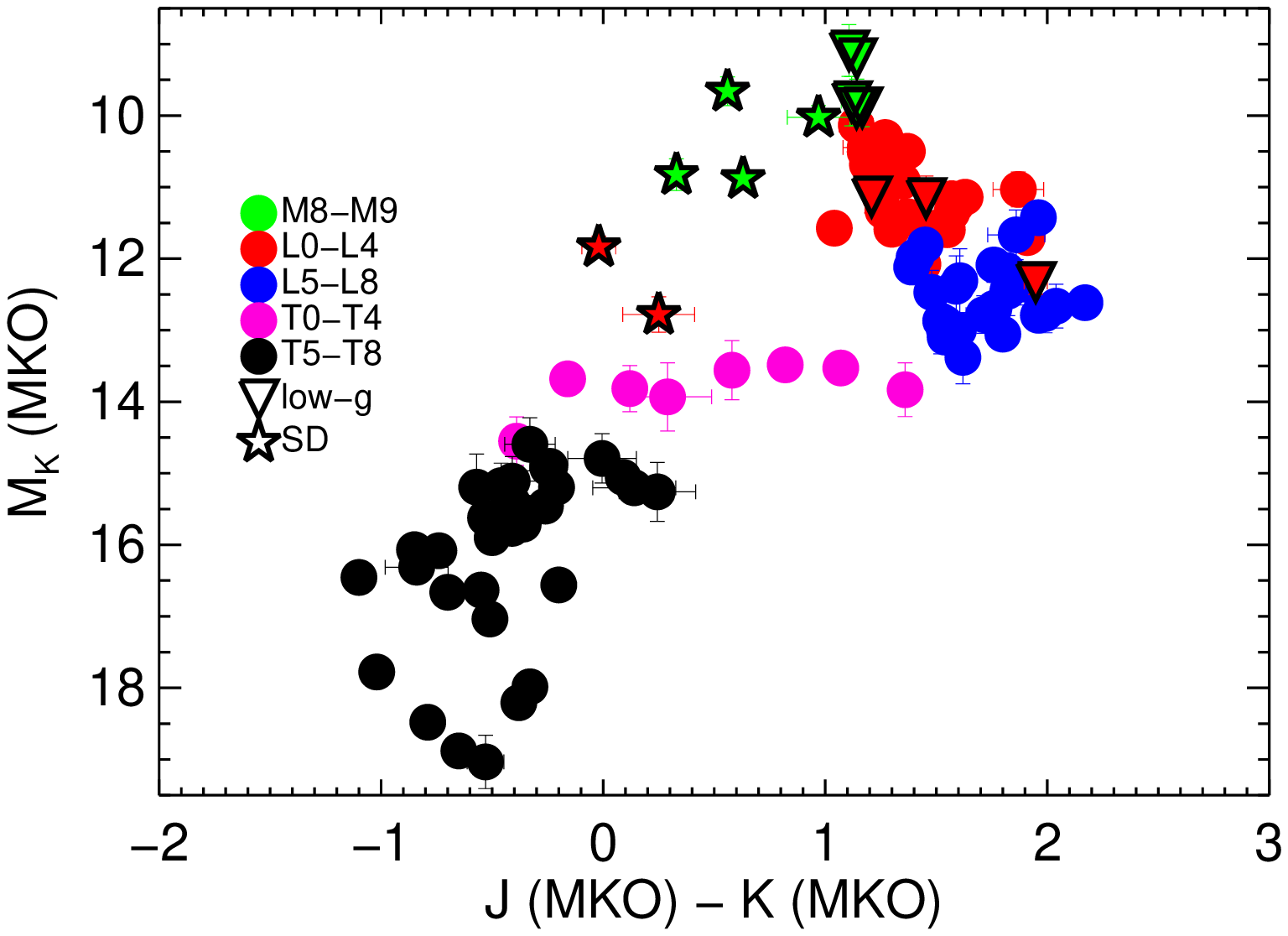}
\end{center}
\caption{The near-IR color  vs. absolute magnitude diagrams in a combination of MKO $JHK$ filters.  Spectral subtypes are color-coded and low-gravity as well as subdwarfs are depicted as downward facing triangles and five point stars respectively.  } 
\label{fig:color_mag}
\end{figure}

We have collected photometric information (from optical to mid-IR) for all known L and T dwarfs with parallaxes (106 literature and 59 added in this work), totaling 165 (see Table ~\ref{Luminosity}).  We examined various combinations of optical, near-IR and mid-IR colors to find correlations on color-magnitude diagrams that provided the strongest insight into differentiating brown dwarf spectral types and/or physical properties of the  population.  Figures ~\ref{fig:color_mag2} - ~\ref{fig:color_mag} show representative color-magnitude diagrams using only objects whose absolute magnitude uncertainties are $<$ 0.5 mag.  

Figure ~\ref{fig:color_mag2} shows the most striking linear relationship between color and absolute magnitude.   \citet{2006ApJ...651..502P} first showed a relatively smooth progression of M through T dwarfs in the M$_{K_{s}}$ versus $K_{s}$-$[4.5]$ diagram.  We verify that this relation separates the early and mid L dwarfs (0.5 $<$ $K$-$[4.5]$ $<$ 1.4) from the early and mid T dwarfs  (1.4 $<$ $K$-$[4.5]$ $<$ 4.0).  However, it does not show a linear progression with spectral subtypes.  The degeneracy is most clearly depicted for the T6.0 and T6.5 dwarfs (designated by grey triangle symbols) which have a nearly 2.0 mag spread in $K$-$[4.5]$ color.  

Interestingly the low-surface gravity dwarfs, depicted as filled downward-facing triangles, appear slightly overluminous for their color in comparison to normal field dwarfs.  This is in contradiction to findings discussed in section ~\ref{LowG} where objects are underluminous in M$_{JHK}$ for their spectral type. The discrepancy between the behavior  of low gravity objects on the mid-IR color magnitude diagram and the spectral-type/absolute magnitude diagrams suggests that spectral type is a poor gauge for effective temperature of young objects.   

The most prevalent change on these color magnitude diagrams is a color reversal.  Figure ~\ref{fig:color_mag} shows the three most studied versions of this effect.  Condensate cloud opacity is likely the dominate contributor to the increasingly red colors of L dwarfs and the onset of CH$_{4}$ is responsible for the change to blue as temperatures drop into the T dwarf regime.  As objects transition from L into T dwarfs the brightening (discussed in section ~\ref{nearab}) is most clearly seen in $J$ band although it is seen to a lesser effect in $H$ and plateaus in $K$.   Subdwarfs occupy their own space on this diagram with extremely blue near-IR colors.  Low-surface gravity dwarfs fall within the color space of normal field dwarfs however they are redward and underluminous compared to normal dwarfs with the same spectral type.  In Section~\ref{Evol} below, we discuss the M$_{K}$ vs $J$-$K$ color magnitude diagram in context against models to disentangle subtle effects that drive much of the scatter seen in  Figure ~\ref{fig:color_mag}.

\section{Comparison to Evolutionary Models\label{Evol}}
\begin{figure}
\centering
\begin{tabular}{cc}
\epsfig{file=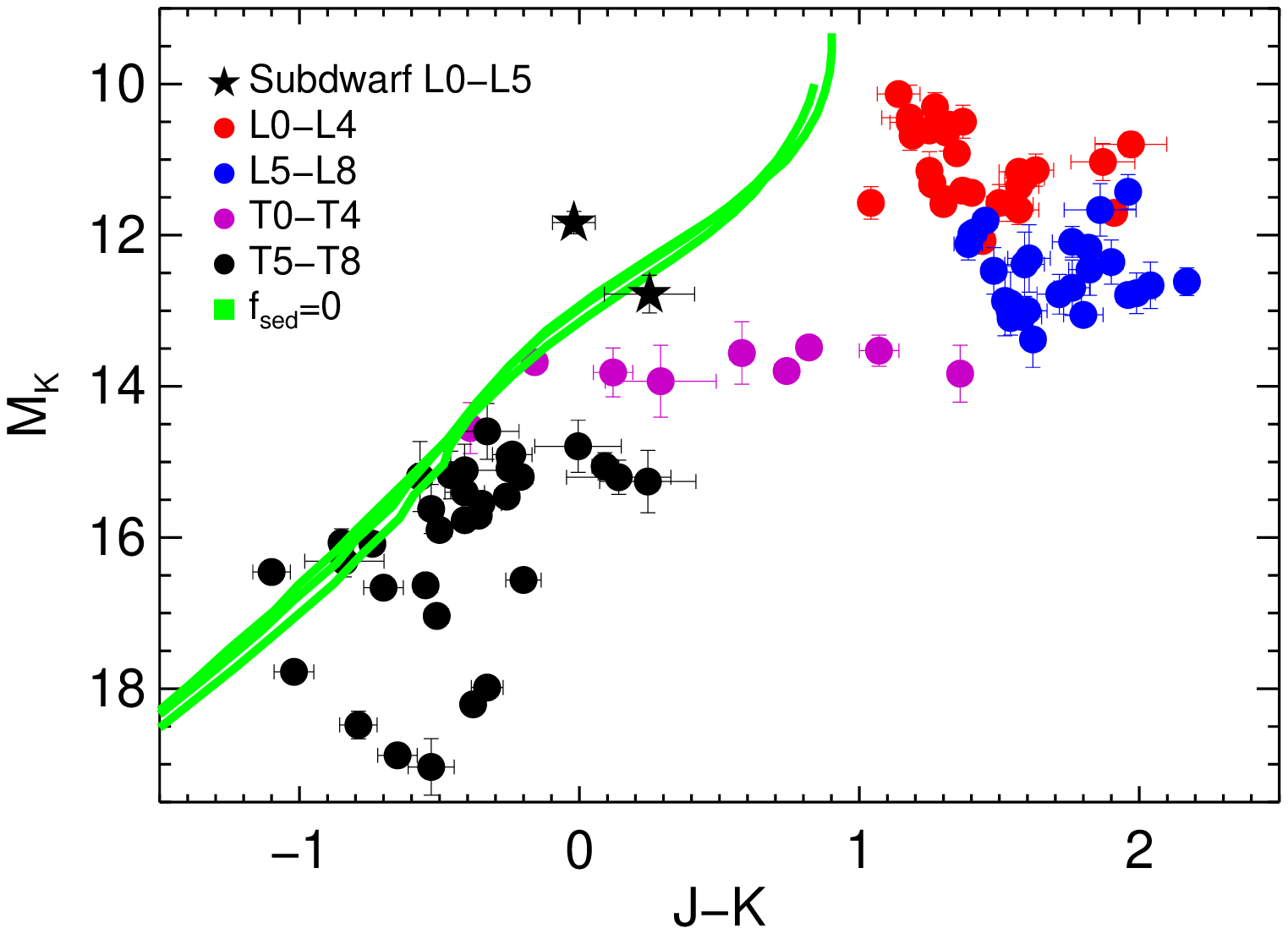,width=0.5\linewidth,clip=}&
\epsfig{file=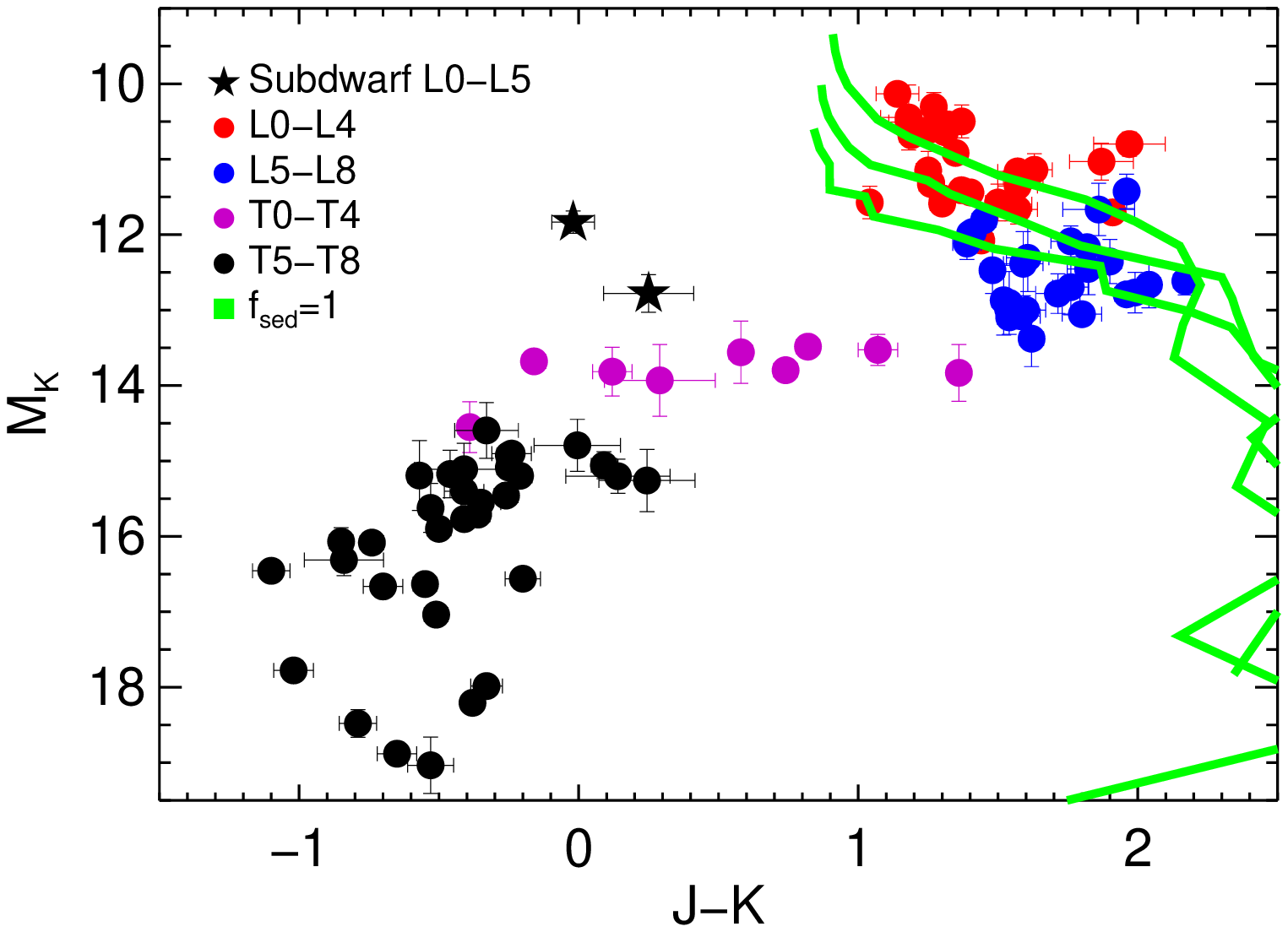,width=0.5\linewidth,clip=} \\
\epsfig{file=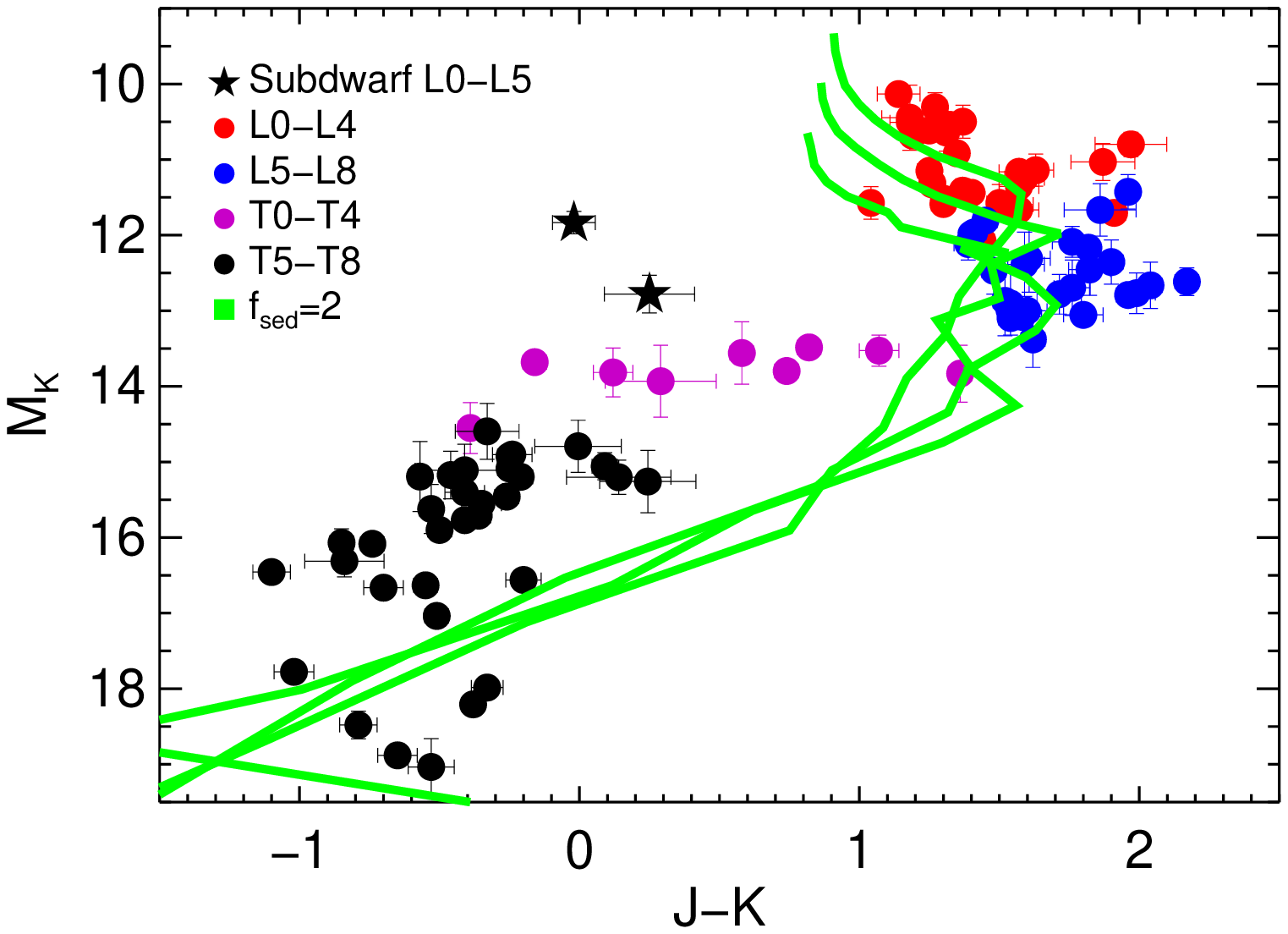,width=0.5\linewidth,clip=} &
\epsfig{file=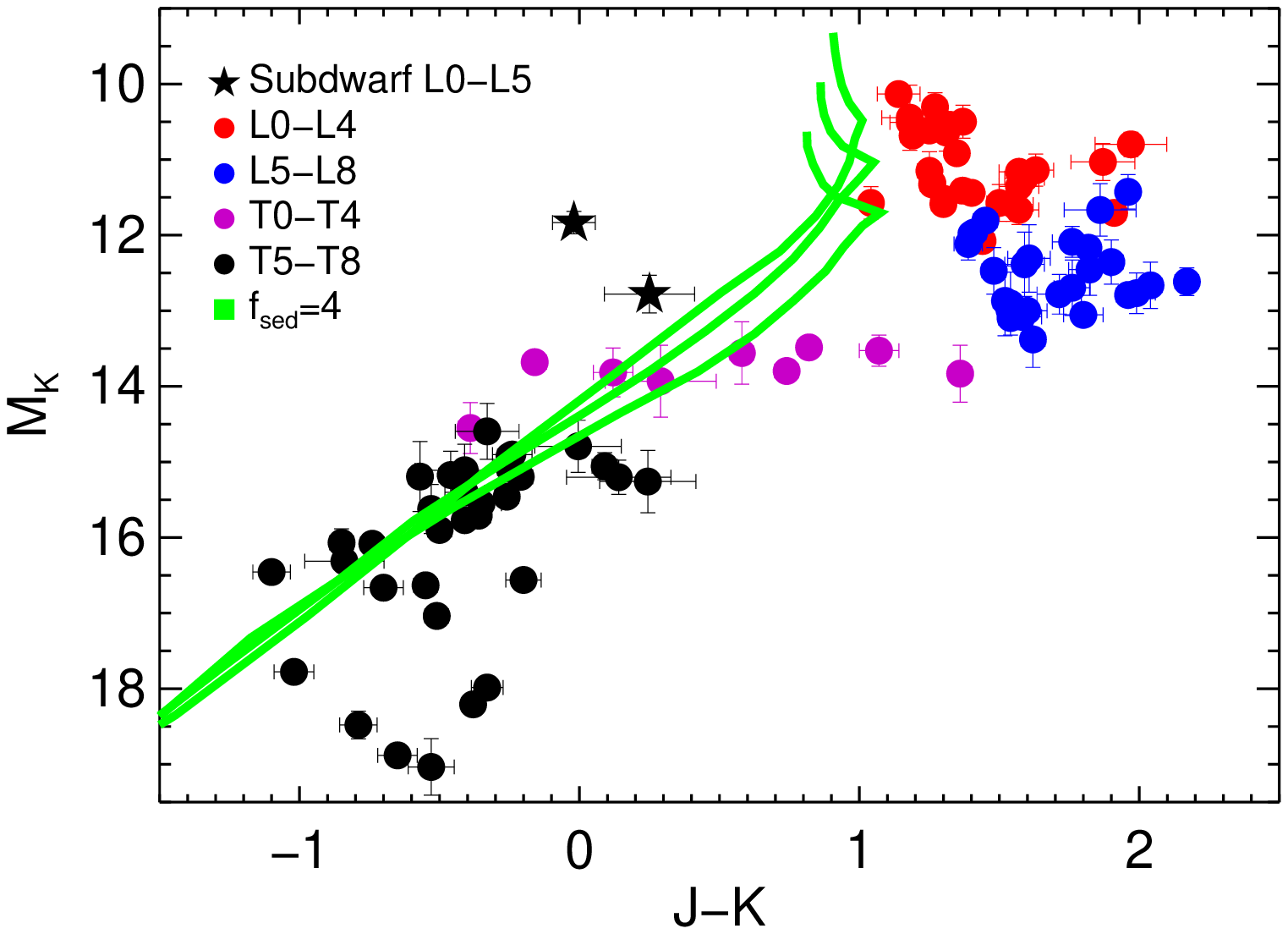,width=0.5\linewidth,clip=} 
\end{tabular}
\caption{The $J$-$K$ vs. M$_{K}$ diagrams for L and T dwarfs with the evolutionary models of \citet{2008ApJ...689.1327S}  over-plotted.  Varying the cloud thickness parameter, f$_{sed}$, from thick (f$_{sed}$=1) to thin (f$_{sed}$=4) to cloudless (denoted as f$_{sed}$=0), fits the L and T dwarf sequences with varying degrees of accuracy.  The three different tracks in each plot represent different gravities (log(g)=[4.5,5.0,5.5]).}
\label{fig:fsed}
\end{figure}

\begin{figure}
\centering
\begin{tabular}{cc}
\epsfig{file=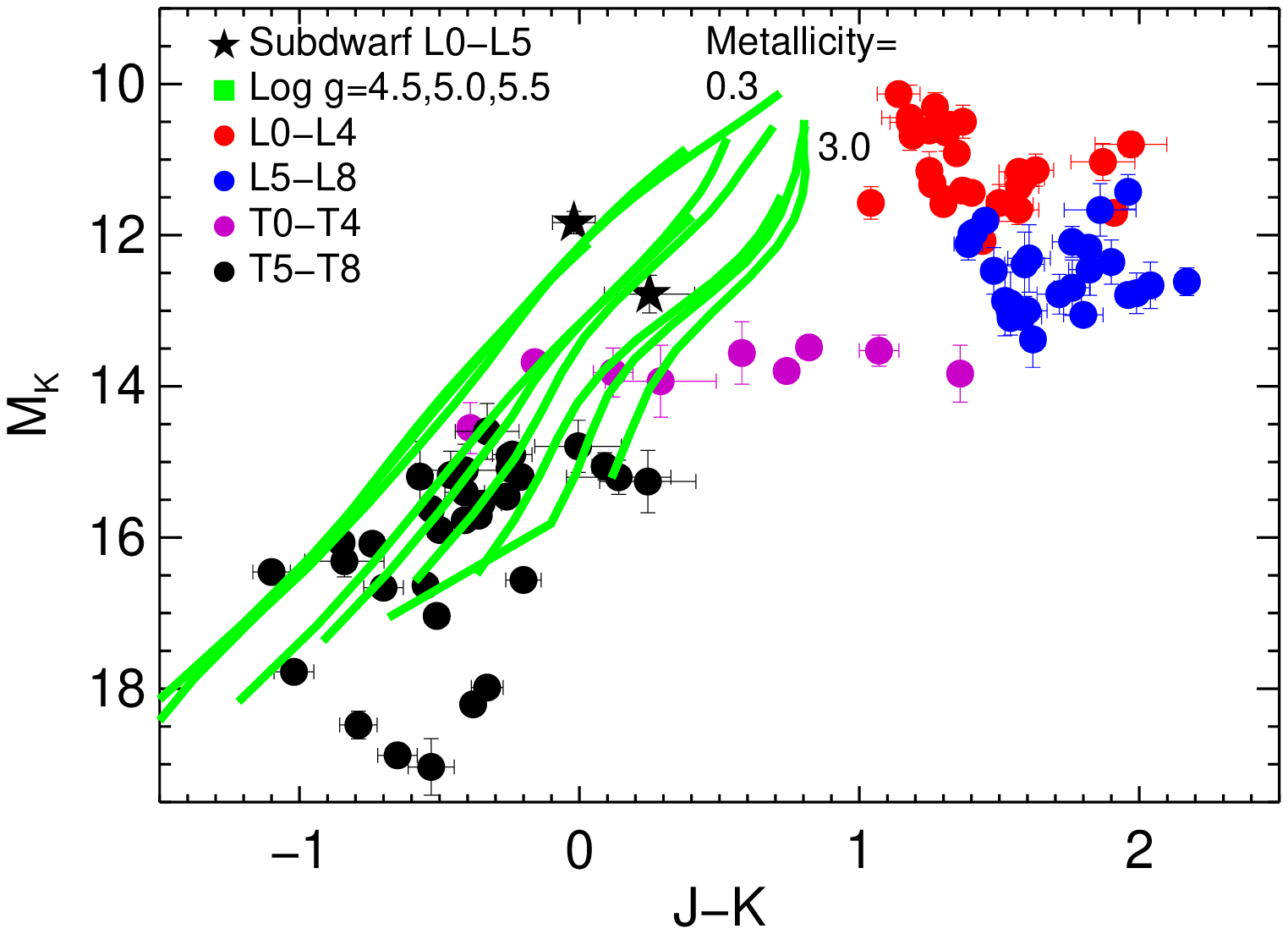,width=0.5\linewidth,clip=} &
\epsfig{file=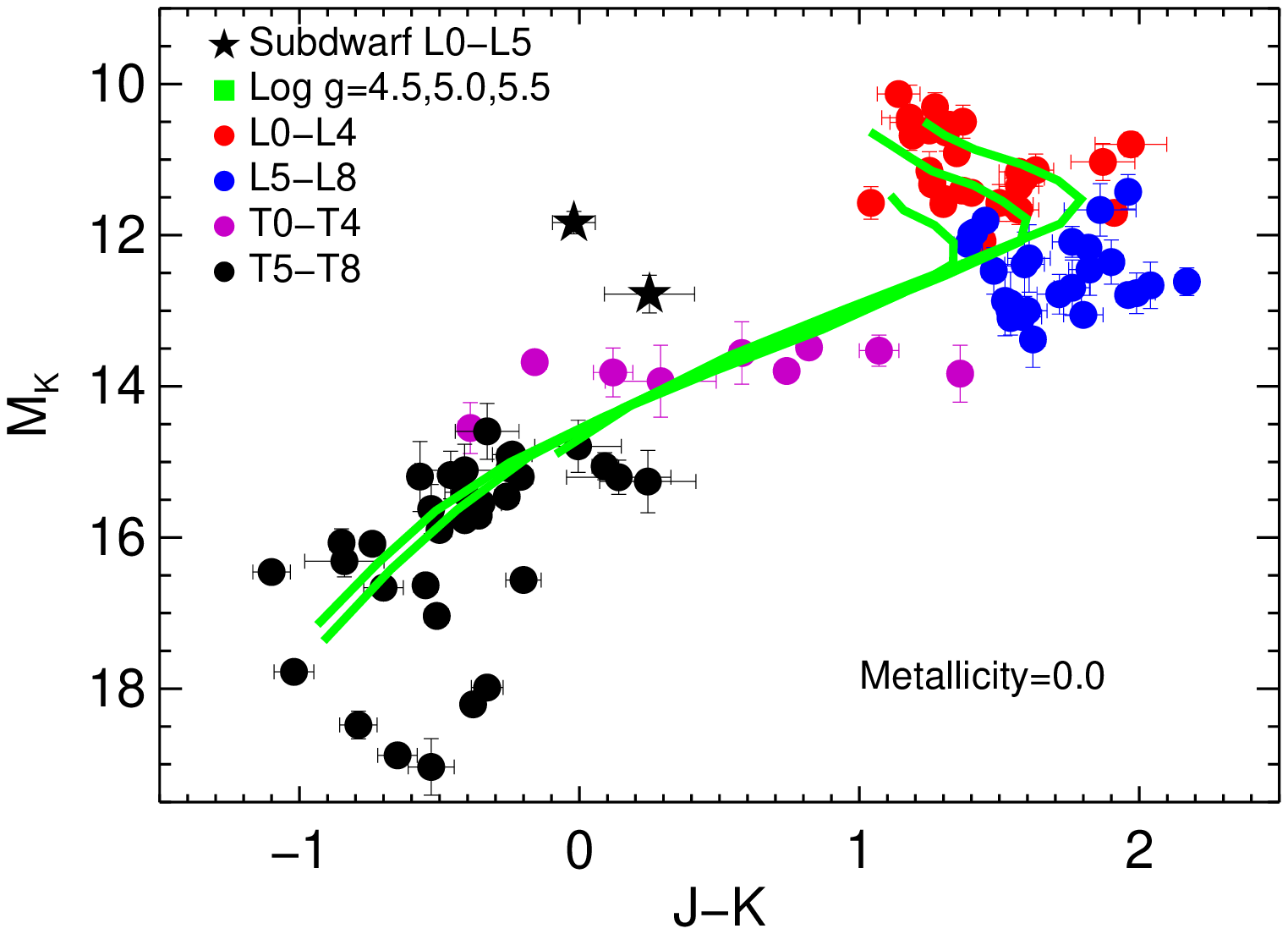,width=0.5\linewidth,clip=} \\
\epsfig{file=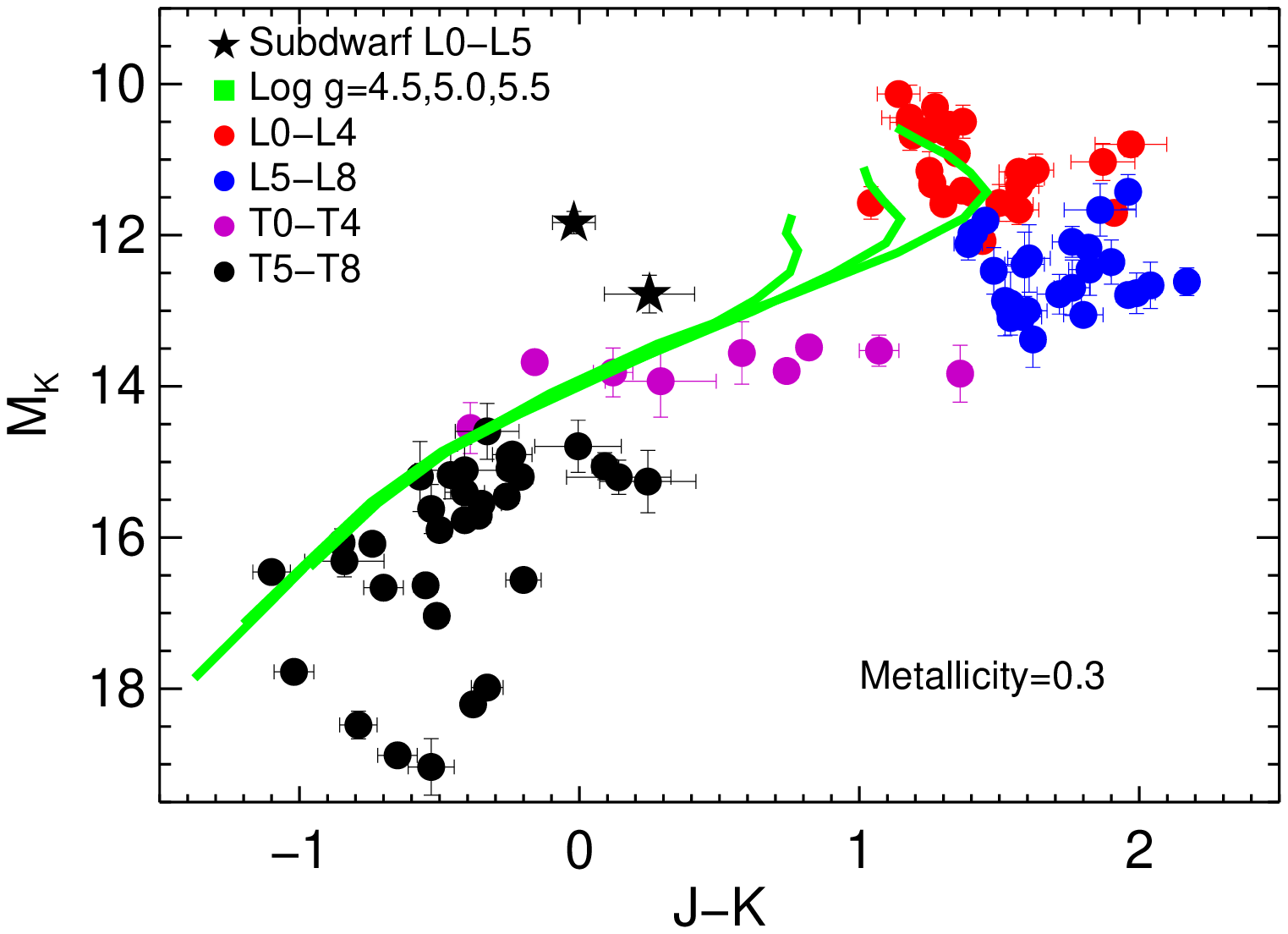,width=0.5\linewidth,clip=} &
\epsfig{file=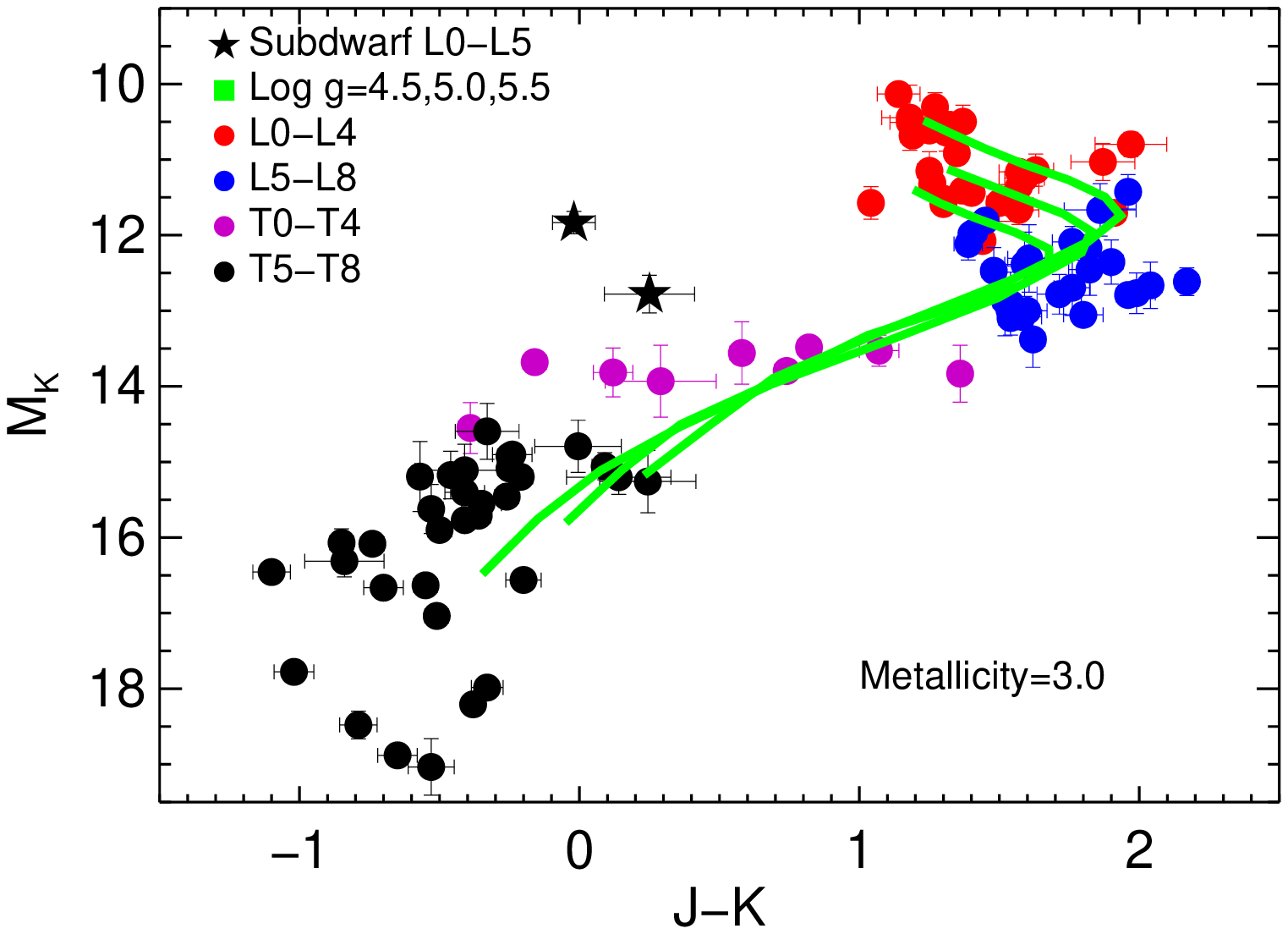,width=0.5\linewidth,clip=}
\end{tabular}
\caption{The $J$-$K$ vs. M$_{K}$ diagrams for L and T dwarfs with the evolutionary models of  \citet{2006ApJ...640.1063B} over-plotted.  The top left panel shows the cloudless model with all metallicities and the remaining three panels show the cloudy model.  Varying the metallicity from subsolar (0.0) to super-solar (3.0) and gravity from low (log(g)=4.5) to high (log(g)=5.5) fits the L and T dwarf sequences with varying degrees of accuracy.} 
\label{fig:burrows_met}
\end{figure}

To put the observed trends on color-magnitude diagrams in context, we compared the data to two sets of evolutionary models.  \citet{2008ApJ...689.1327S} present a set of models that include a cloud sedimentation parameter and three gravity choices (log(g)=[4.5,5.0,5.5]) which can be varied to explain (with different levels of accuracy) the near-IR color magnitude diagram for L and T dwarfs.  \citet{2006ApJ...640.1063B} present a model that includes refractory clouds as well as a completely cloudless model with varying gravity and metallicity parameters.  In Figures ~\ref{fig:fsed} and ~\ref{fig:burrows_met} we examine the M$_{K}$ vs. $J$-$K$ diagram for L and T dwarfs using the full sample with trigonometric parallaxes and the respective evolutionary models.  In Figure ~\ref{fig:fsed} the sedimentation parameter (f$_{sed}$) from \citet{2008ApJ...689.1327S} is shown with increasing value and varying gravity to represent decreasing cloud thickness.  In Figure   ~\ref{fig:burrows_met} metallicity and gravity variations are examined using the models of \citet{2006ApJ...640.1063B}.  All early and late-type L and T dwarfs (including subdwarfs) with trigonometric parallaxes ($\sigma_{M_{K}}$ $<$ 0.5 mag) are over-plotted.  In the following subsections, we discuss how variations of the models fit different spectral types.  The low-surface gravity dwarfs are discussed in detail in section ~\ref{LowG}.

\subsection{L Dwarfs\label{Ldwarfs}}
Varying gravity and metallicity within the cloud model of \citet{2006ApJ...640.1063B} encompasses the majority of early L dwarfs (top right and bottom two panels of Figure ~\ref{fig:burrows_met}); however, late-type L dwarfs are still poorly represented.  Compared to the highest gravity, super solar metallicity track, there are a number of late-type L dwarfs that are fainter and redward of predictions.  In Figure ~\ref{fig:fsed}, the L dwarf sequence is best modeled with the f$_{sed}$=1,2 parameters (top right and bottom left plots of Figure ~\ref{fig:fsed}).  However, there are a handful of red or potentially ``ultra-cloudy" objects that are not fit by either model. Significant outliers include 2MASS J1442+6603, which is a close ($\sim$30 AU) companion to the moderately young M1.5 dwarf G239-25 (\citealt{2004A&A...427L...1F}), and 2MASS J0619-5803 which is a companion ($\sim$260 AU) to the young K2 star AB Pic (\citealt{2005A&A...438L..29C}).  The independent assessment that these objects are young (or moderately young) yet redward of the cloudy model implies a connection between youth and a dusty photosphere (further discussed in section ~\ref{LowG}).

\subsection{T Dwarfs\label{Tdwarfs}}
The mid to late-type T dwarfs are best fit by the thin clouds (f$_{sed}$=4) track from \citet{2008ApJ...689.1327S} and the clear model from \citet{2006ApJ...640.1063B}.  In the case of the latter model (bottom right panel of Figure ~\ref{fig:fsed}) the predicted range in both M$_{K}$ and $J$-$K$ shows very little spread whereas empirical measurements show significant scatter.  There are a handful of T dwarfs including those reported in this work, 2MASS J1114-2618, and 2MASS J1754+1649, as well as previously reported ULAS J0034-0052, ULAS J1335+1130, CFBDS J0059-0114, ULAS J0722-0540, and ROSS 458C which are notably under-luminous and red compared to the f$_{sed}$=4 model predictions.  The colors of these late-type T dwarfs are better (or equally) fit by the f$_{sed}$=2 parameter (thicker clouds) which also encompasses the majority of mid to late-type L dwarfs. Comparing the spread to the \citet{2006ApJ...640.1063B} clear model (top left panel of Figure ~\ref{fig:burrows_met}) shows similar red, under-luminous outliers.   Inconsistencies with both models suggests that thick condensate clouds continue to play a role in the photospheres of some cooler dwarfs (see discussion in \citealt{2010arXiv1009.5722B}, \citealt{2010ApJ...723L.117M}).  At least one of the ultra-cloudy T dwarf outliers, Ross 458C, is known to be young (\citealt{2010arXiv1009.5722B}; \citealt{2011MNRAS.414.3590B}).  A connection between youth and dust in the photospheres of T dwarfs is consistent with the L dwarf results discussed in sections ~\ref{Ldwarfs} and ~\ref{LowG} of this work as well as recent modeling work of young exoplanets (\citealt{2011ApJ...733...65B}, \citealt{2011ApJ...729..128C}).

\subsection{L/T Transition Dwarfs\label{LTtransition}}
The L/T transition objects are not fit by a single f$_{sed}$ parameter using the \citet{2008ApJ...689.1327S} models nor by any single combination of gravity and/or metallicity on the \citet{2006ApJ...640.1063B} models\footnote{\citealt{2006ApJ...640.1063B} also note that variations in cloud particle size can not account for the transition objects.}.  The steady and significant decrease in $J-K$ color with near constant M$_{K}$ for the objects in the transition region has been attributed to the clearing of clouds or a change in the atmospheric dynamical state (e.g. overall cloud thickness) as temperatures cool into the T dwarfs (see \citealt{2002ApJ...571L.151B};  \citealt{2004AJ....127.3516G};  \citealt{2004AJ....127.3553K}).  Without clouds to provide a significant, nearly gray opacity, flux can emerge through molecular opacity windows in $J$ and $H$ bands explaining the significant brightening discussed in section ~\ref{nearab}.

\begin{figure}[!ht]
\begin{center}
\epsscale{1.2}
\includegraphics[width=.55\hsize]{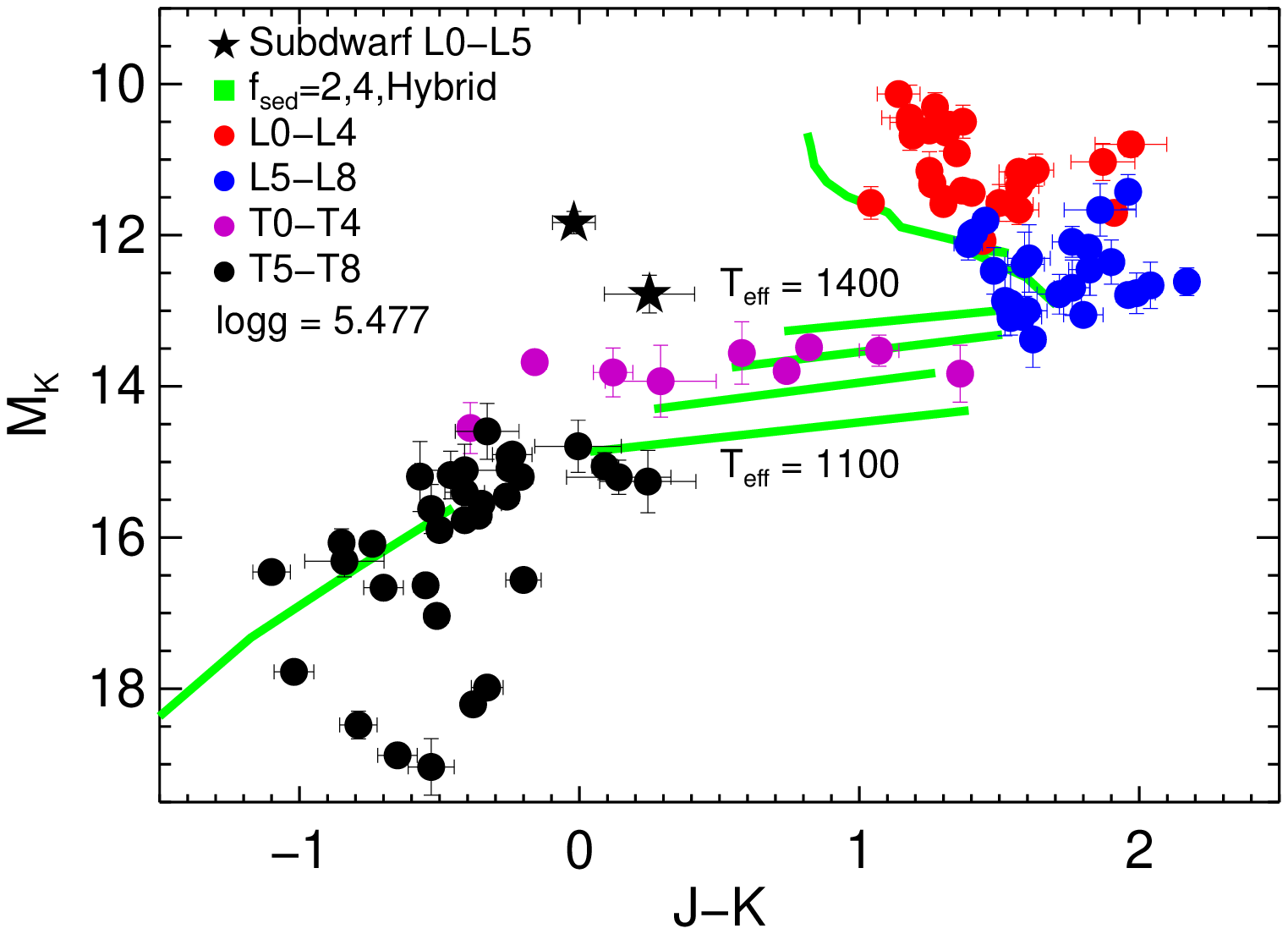}
\plottwo{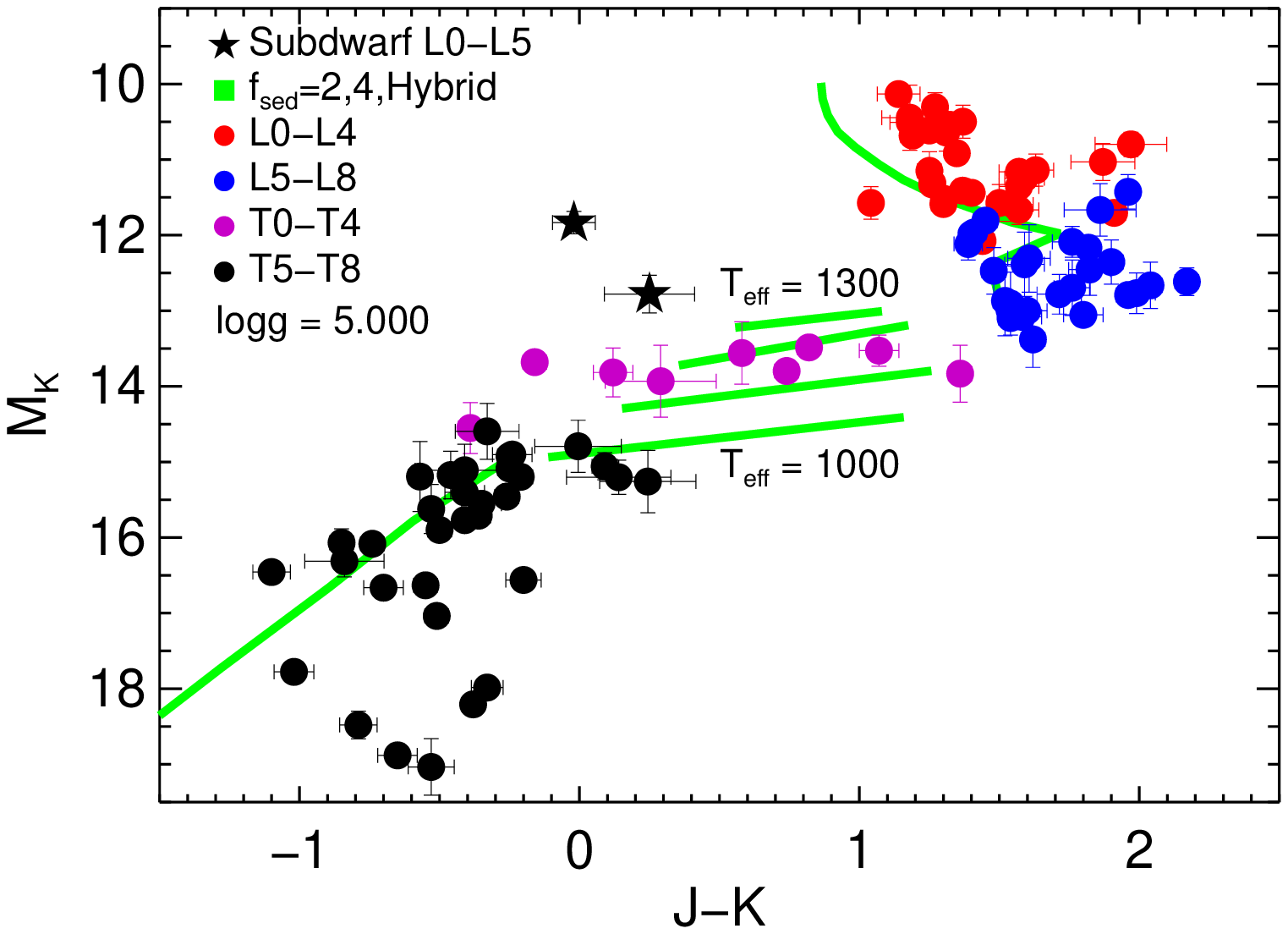}{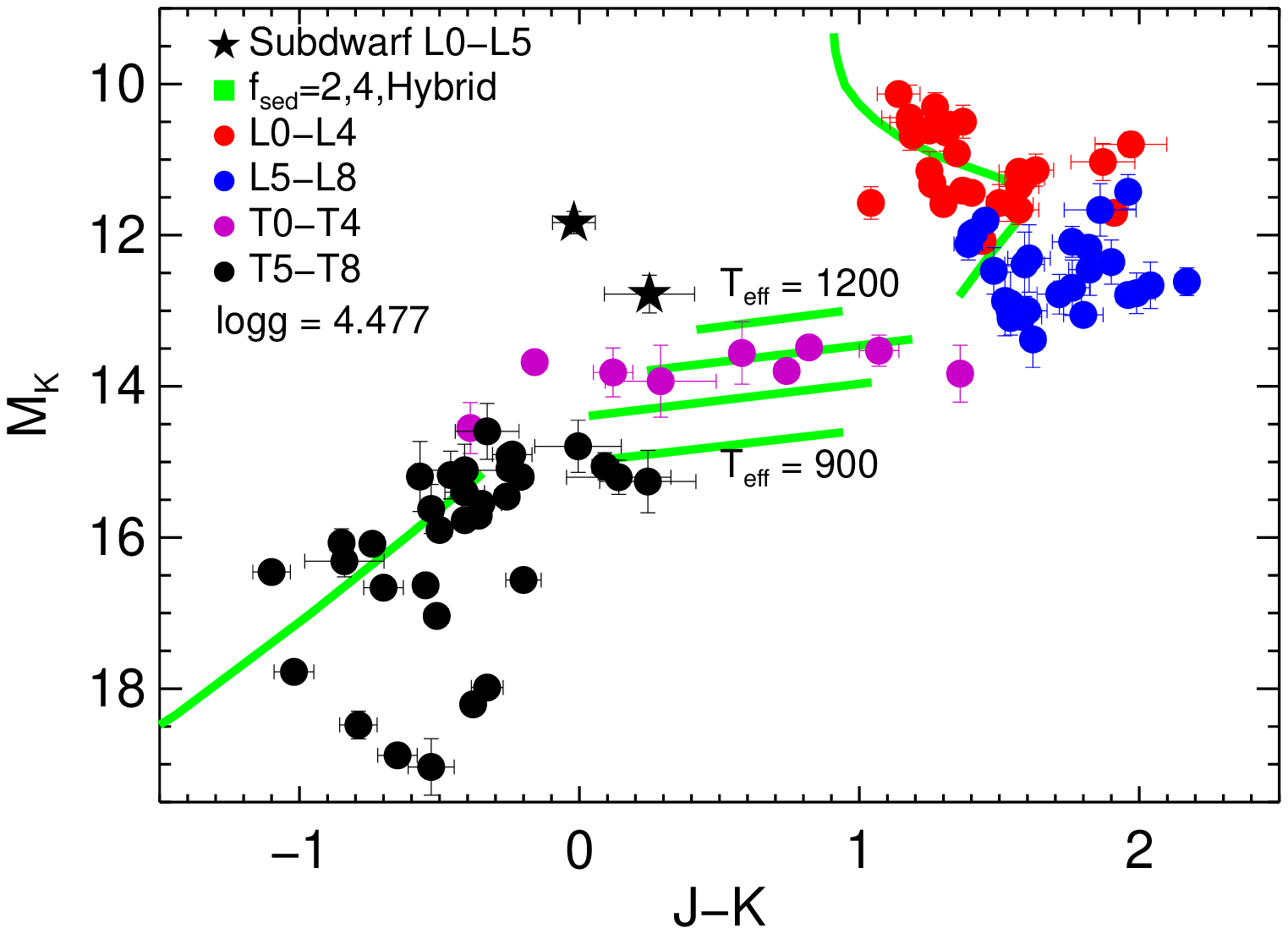}
\end{center}
\caption{The $J$-$K$ vs. M$_{K}$ diagram for L and T dwarfs with the evolutionary models of \citealt{2008ApJ...689.1327S} over-plotted to demonstrate the best fit for L/T transition objects.  For late-type L dwarfs (M$_{K}<$13.0) we have over-plotted the f$_{sed}$=2 tracks and for late-type T dwarfs (M$_{K}>$15.0) we have over-plotted the f$_{sed}$=4 tracks.   For the L/T transition we created a hybrid model between the two by adding the predicted model magnitudes of the latter to the former in 10\% increments from 13.0 $<$ M$_{K}$ $<$ 15.0. }\label{fig:hybrid}
\end{figure}

To investigate whether our expanded parallax sample supports a rapid cloud clearing, we created a hybrid model using the Saumon \& Marley tracks.  Similar to the work of \citet{2002ApJ...571L.151B} we varied the sedimentation parameter between the f$_{sed}$=2 and f$_{sed}$=4 models across the region between 13.0$<$ M$_{K}$ $<$ 15.0.  We started with the f$_{sed}$=2 color, then added the f$_{sed}$=4 color in 10\% increments across the transition.  The result is plotted in Figure ~\ref{fig:hybrid}.  The L/T transition objects lie within an absolute magnitude range corresponding to model temperatures spanning $\pm$150 K from a mean T$_{eff}$ that depends on the gravity chosen.  

For an intermediate surface gravity of log(g)=5.0, which is consistent with field age objects ($\sim$ 3 Gyr), the mean T$_{eff}$ of our model is $\sim$ 1200$\pm$100 K for T0-T4 dwarfs.  \citet{2004AJ....127.3516G} empirically measured the T$_{eff}$ of 11 L7-T4 dwarfs and reported a near constant T$_{eff}$ of $\sim$1450 K. This warmer constant temperature across a broader range of spectral types was strong evidence for unresolved binarity across the transition. Indeed 2 of the 11 objects in that sample have since been resolved into near-equal mass binaries.  The splitting of spectral-type/absolute magnitude polynomials into bright and faint samples (see section ~\ref{nearab}) was largely driven by this result.  However, our result is in agreement with the outcome of other toy and sophisticated models (e.g  \citealt{2002ApJ...571L.151B};    \citealt{2008ApJ...689.1327S}; \citealt{2010ApJ...723L.117M}) which converged upon a similar cooler T$_{eff}$.  Our comparatively lower temperature than found by \citet{2004AJ....127.3516G} across the transition is consistent with our finding that the brown dwarf temperature plateau occurs across a narrower spectral subtype range (T0-T4 rather than L7-T4)  than previously thought.

\section{Low Surface Gravity Dwarfs\label{LowG}}
A subset of our parallax sample are the low surface gravity dwarfs including seven M and seven L dwarfs.   Their optical spectra are characterized by unusually weak FeH absorption, weak Na I and K I doublets  and very strong vanadium oxide bands (\citealt{2009AJ....137.3345C};  \citealt{2006ApJ...639.1120K}).   They have extreme red near-IR colors and small tangential velocities relative to the rest of the brown dwarf population (\citealt{2009AJ....137....1F}).   \citet{2006ApJ...639.1120K} and \citet{2009AJ....137.3345C} have suggested that a number of the low-surface gravity dwarfs are candidate members of nearby moving groups such as $\beta$ Pictoris, Tucana-Horlogium, and AB Doradus, implying ages roughly spanning 10- 50 Myr (e.g. \citealt{2010ApJ...715L.165R}) . 

Younger ultracool dwarfs have been examined on color-magnitude diagrams, but as only a handful have reported parallaxes, reports of under- or -over luminosity have been speculative and in some cases contradictory. There is evidence that young dwarfs at the L/T transition such as  HD 203030B and HN Peg B, are underluminous or have a lower T$_{eff}$ than equivalent spectral type dwarfs. This has been explained as a gravity dependent temperature/spectral type relation at the transition (\citealt{2007ApJ...654..570L}; \citealt{2006ApJ...651.1166M}).  HR 8799b as well as the early L dwarf companion to AB Pic are also underluminous while some earlier L and M dwarfs such as HD 130948B and CD-35 2722 B are overluminous on color magnitude diagrams (\citealt{2010ApJ...723..850B}; \citealt{2005A&A...438L..29C}; \citealt{2008arXiv0807.2450D}; \citealt{2011ApJ...729..139W}). 

\begin{figure}[!ht]
\begin{center}
\epsscale{1.2}
\includegraphics[width=.55\hsize]{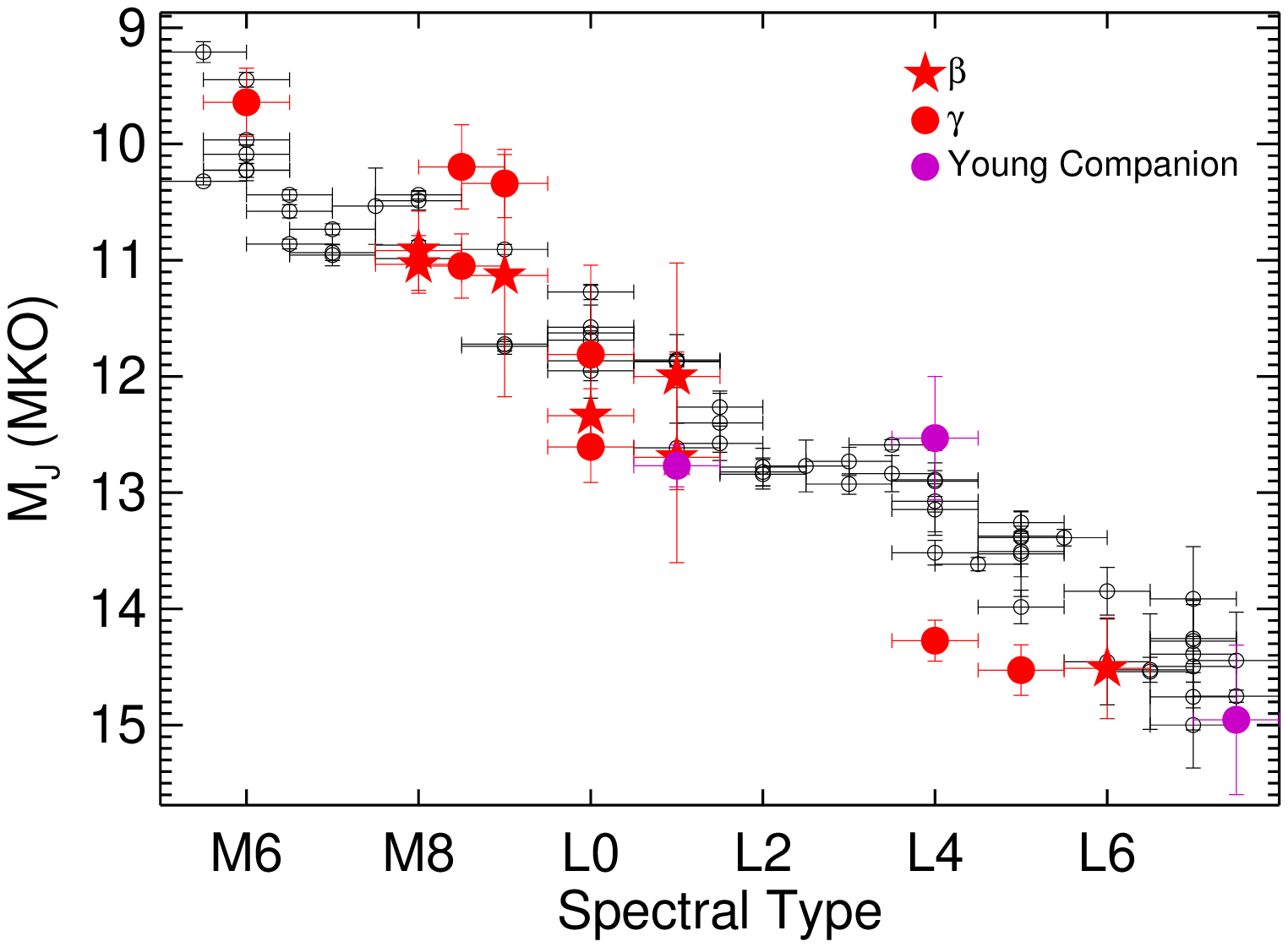}
\plottwo{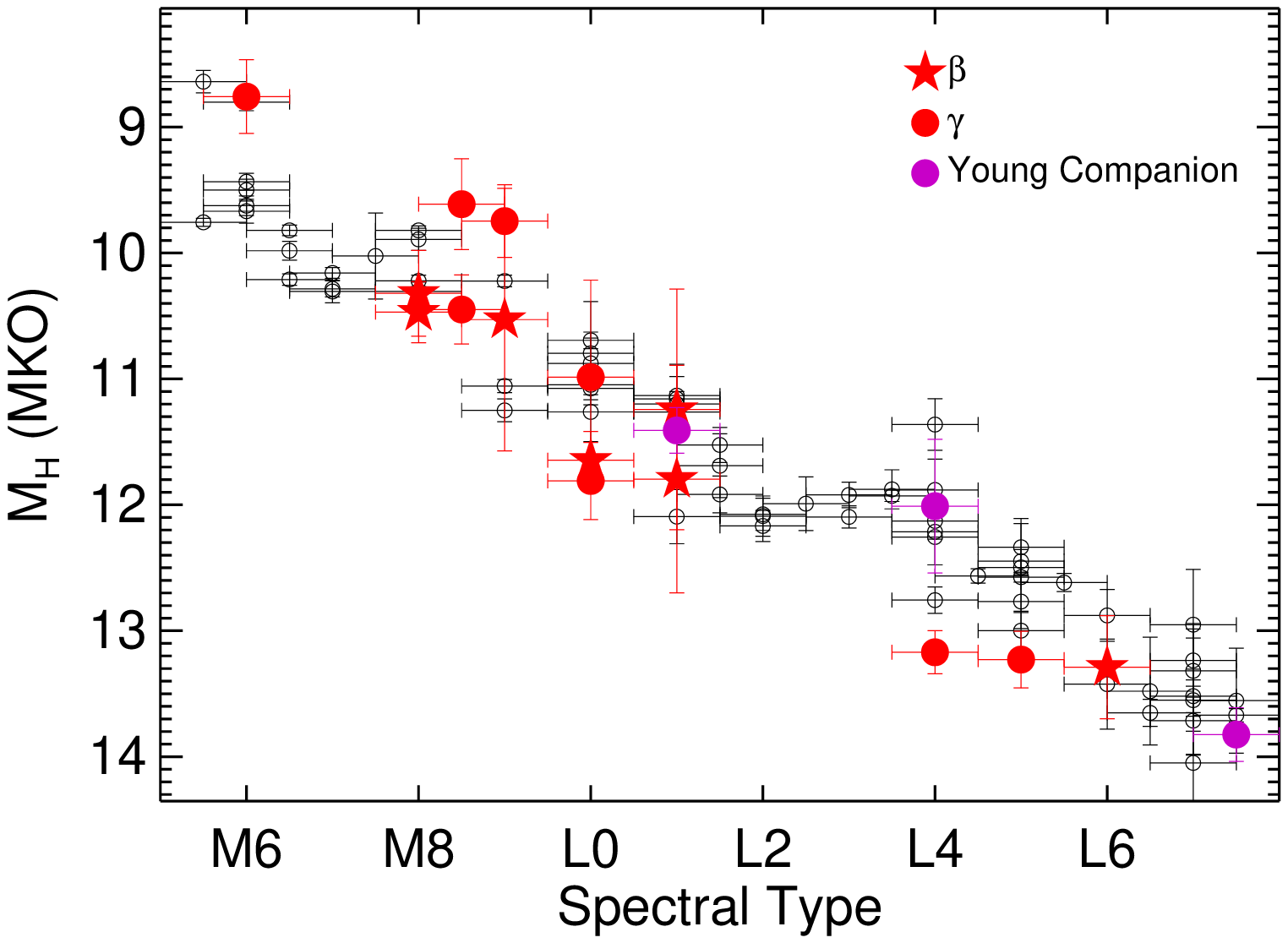}{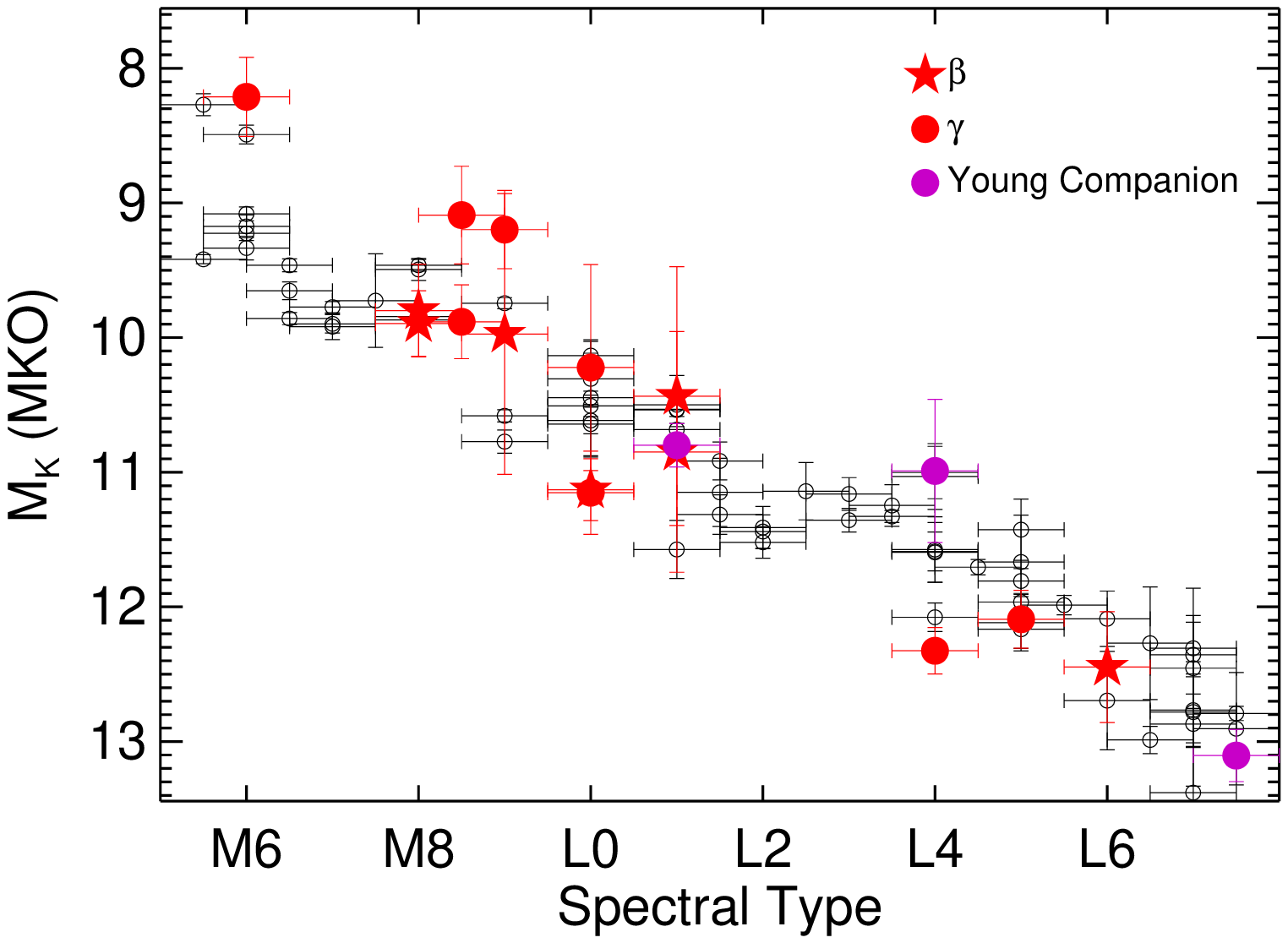}
\end{center}
\caption{Spectral type versus absolute magnitude in the MKO $JHK$ filters for late-type M through mid-L dwarfs.  Unfilled circles are normal dwarfs with parallax measurements.  Red five point stars and filled circles are intermediate ($\beta$) and low ($\gamma$) surface gravity dwarfs, and purple filled circles are young ($<<$ 1 Gyr) companions to nearby stars. In Table ~\ref{lg} we report the difference in magnitude for each source from the M$_{JHK}$ value calculated from the polynomials in Table ~\ref{coeffs}. } 
\label{fig:ABS_LG}
\end{figure}

Figure ~\ref {fig:ABS_LG} shows the near-IR absolute magnitude vs. spectral type diagrams for normal mid-type M through late-type L dwarfs with the low-surface gravity dwarfs and young companions over-plotted.  Table ~\ref{lg} lists the absolute magnitude for each object in $JHK$ as well as the deviation from the M$_{JHK}$ values for each spectral bin calculated from the polynomial in Table ~\ref{coeffs}.  Of the seven M dwarfs in this sample, three are $>$ 0.5 mag overluminous for their spectral type as might be expected for a young object which has not contracted to its final radius.  Indeed two of the three overluminous M dwarfs are suspected members of the TW Hydrae association.  However, 8 out of the 10 low gravity or young companion L dwarfs ($\sim$80\%) are [0.2-1.0] magnitude underluminous for the average of their spectral subtype (compared to the polynomial described in Table ~\ref{coeffs}) in 1 or more near-IR bands.  On Figure ~\ref {fig:ABS_LG} we have plotted the $\beta$ and $\gamma$ designations assigned for each object to indicate intermediate and low-gravity respectively (see discussion in \citealt{2005ARA&A..43..195K}; \citealt{2009AJ....137.3345C}).  Within this sample, there does not appear to be a correlation between $\Delta_{M_{JHK}}$ (defined as the difference in M$_{JHK}$ between the source and the predicted polynomial value) and the strength of low-surface gravity features.   

\begin{figure}[!ht]
\begin{center}
\epsscale{1.2}
\includegraphics[width=.55\hsize]{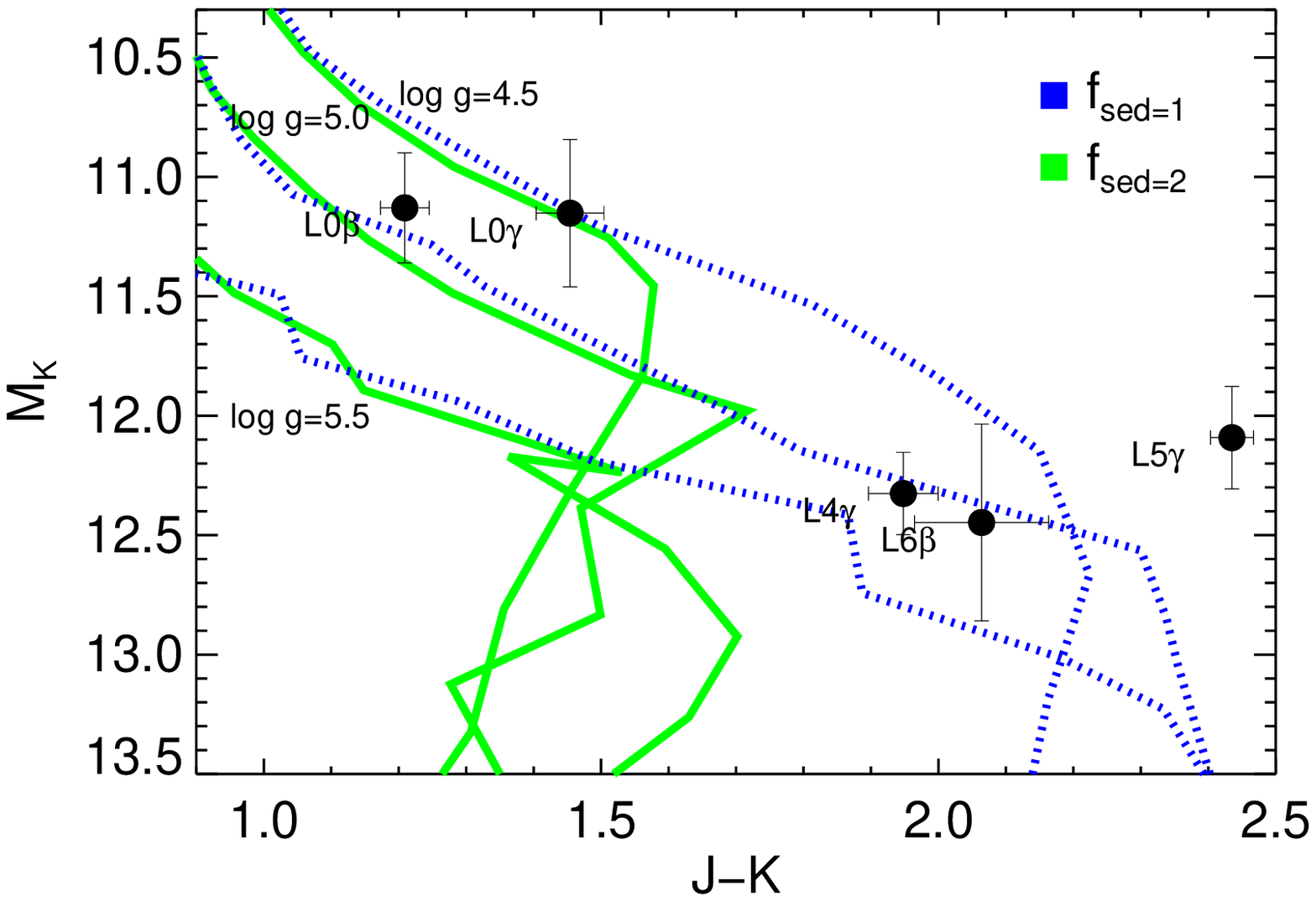}
\includegraphics[width=.55\hsize]{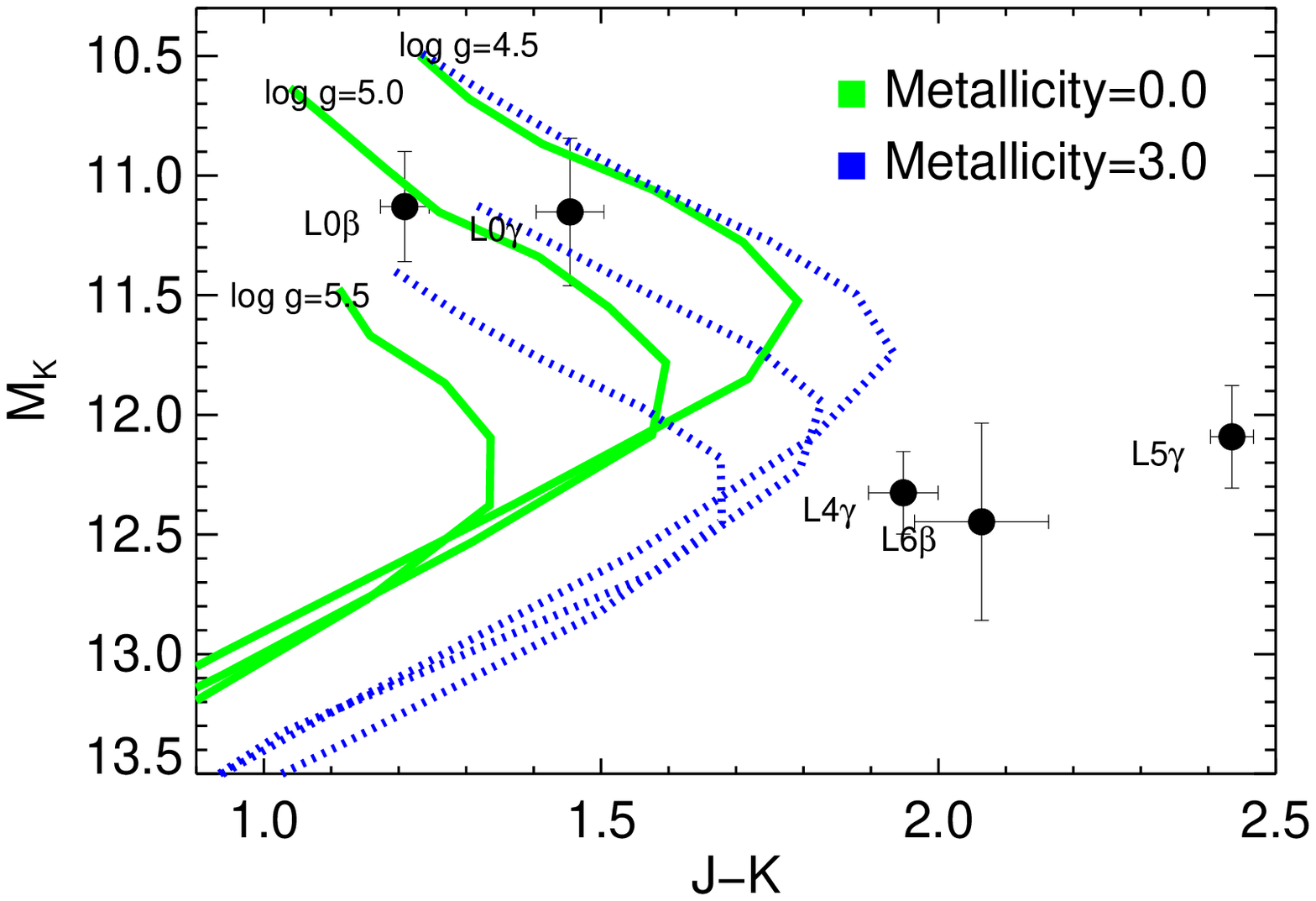}
\end{center}
\caption{The $J$-$K$ vs. M$_{K}$ diagram with the evolutionary models of \citealt{2008ApJ...689.1327S} (top panel) and Burrows et al (2006-bottom panel) over-plotted along with 4 low-surface gravity L dwarfs. The log(g)=[4.5,5.0,5.5], and f$_{sed}$=1,2 parameters of the former are shown as are the log(g)=[4.5,5.0,5.5], Metallicity=[0.0,3.0] parameters of the later.} 
\label{fig:fsed_lowg}
\end{figure}

The trend of low-surface gravity ultracool dwarfs appearing under-luminous for their spectral type is surprising given that young M dwarfs, such as those in TW Hydrae, are 1-2 mag overluminous (Looper et al. in prep).  While we find that the TW Hydrae M dwarfs in our sample are indeed overluminous, the L dwarfs show a different trend.  According to the evolutionary tracks of \citet{1997ASPC..119....9B}; 10 Myr objects with masses ranging from 10-75 M$_{Jup}$ have radii which are 25-75\% larger than 1- 3 Gyr dwarfs with equivalent temperatures.  This translates into an over luminosity of  0.5-1.2 mag.  For 50 Myr objects radii can be 13-50\% larger and would be 0.3-0.9 mag overluminous. We speculate that there are at least two factors that could contribute to the under-luminosity:  First, the low-gravity spectral classification scheme may have a different temperature relation than the \cite{1999ApJ...519..802K} classification scheme used for normal field dwarfs. For example, an L0$\gamma$ dwarf might have a significantly different temperature (and luminosity) than a normal L0 dwarf.  Second, young objects could have dustier photospheres than field-aged dwarfs thus making young objects appear both fainter and redder than field objects of similar temperature.  Observationally, both low-surface gravity dwarfs and dusty L dwarfs show red near-IR colors and similar spectral characteristics (\citealt{2008ApJ...686..528L}, \citealt{2010ApJ...715..561A}). Evolutionary models demonstrate that the lower gravity and dustier (lower f$_{sed}$) tracks have redder near-IR colors than intermediate, high gravity, or larger f$_{sed}$ tracks (see Figure ~\ref{fig:fsed_lowg}).  

In Figure ~\ref{fig:fsed_lowg} we isolate the low-surface gravity L dwarfs with M$_{K}$ uncertainties $<$ 0.5 mag on a color-magnitude diagram with the \citet{2008ApJ...689.1327S} and \citet{2006ApJ...640.1063B} evolutionary tracks over-plotted. In general the sources do not follow the low-gravity track.   Moreover, each model traces objects at temperatures which are several hundred degrees lower than expected for equivalent spectral type objects.  For example, temperatures for L4 dwarfs range from $\sim$1600-1900 K (Golimowski 2004); however the L4$\gamma$ dwarf, 2MASS J0501-0010, is traced by model temperatures of $\sim$1200-1300 K.

The two latest-type L dwarfs in our sample are significantly redward of any of the \citet{2006ApJ...640.1063B}  predictions but within the gravities explored on the cloudy model of \citet{2008ApJ...689.1327S}.  Recent work by \citet{2011ApJ...733...65B} investigating HR8799b's red near-IR color and relatively smooth near-IR spectrum concluded that thick photospheric dust cloud opacity could explain the planets observed luminosity and color.  The latest-type L dwarfs in our sample may be higher mass analogs to HR 8799b  (\citealt{2011ApJ...733...65B}).  

\section{Subdwarfs\label{subdwarfs}}
\begin{figure}[!ht]
\begin{center}
\epsscale{1.2}
\includegraphics[width=.55\hsize]{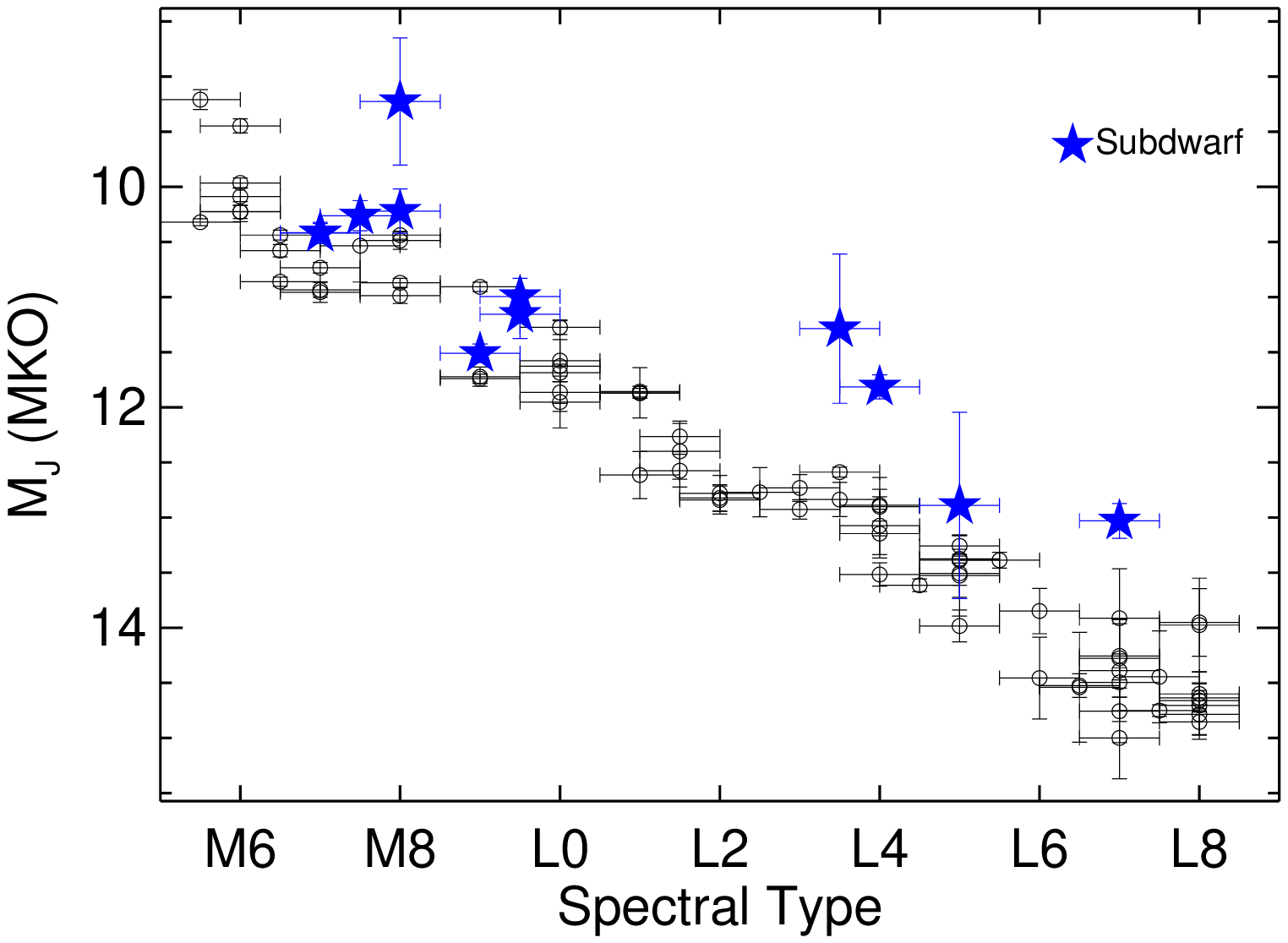}
\plottwo{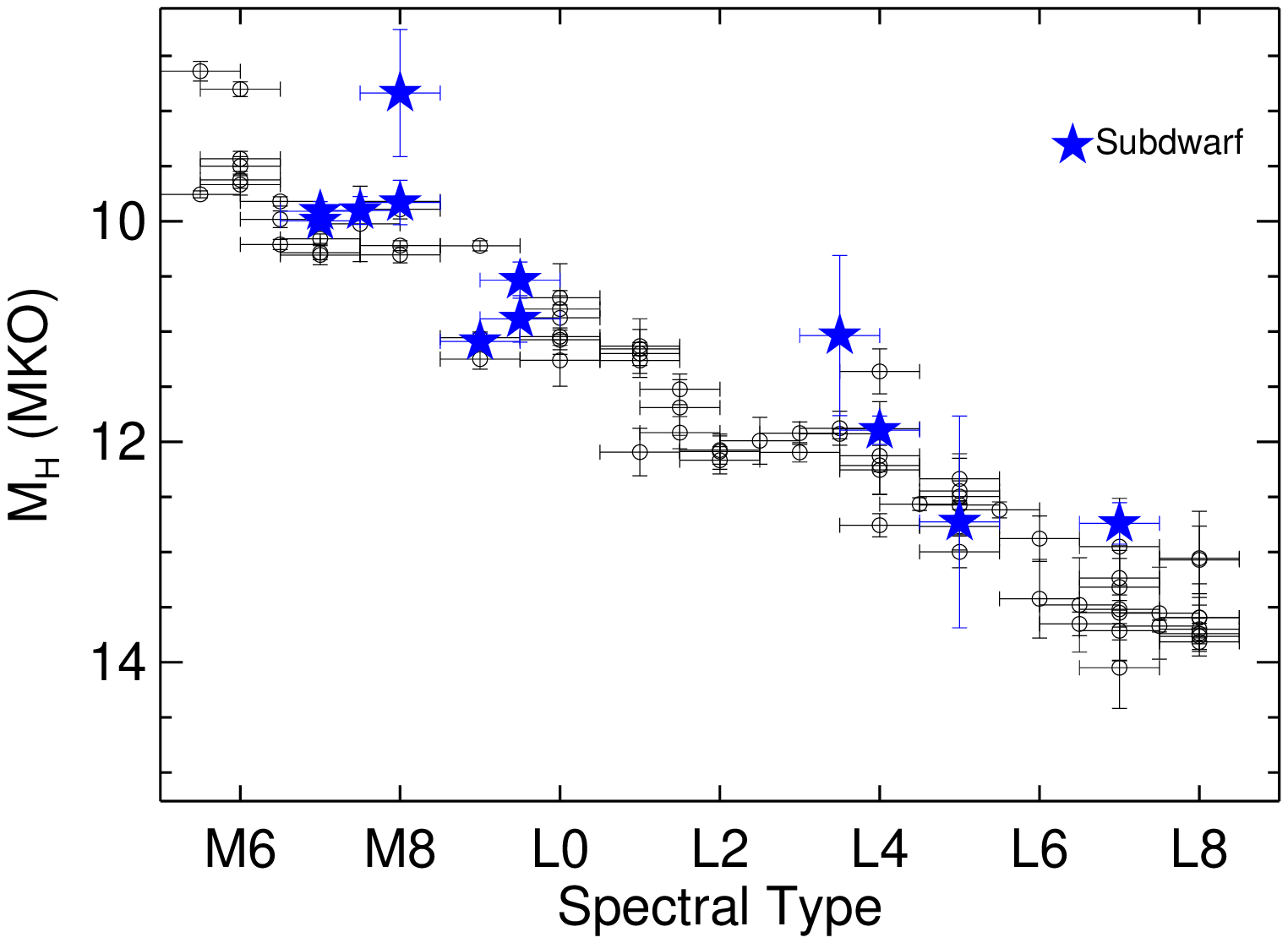}{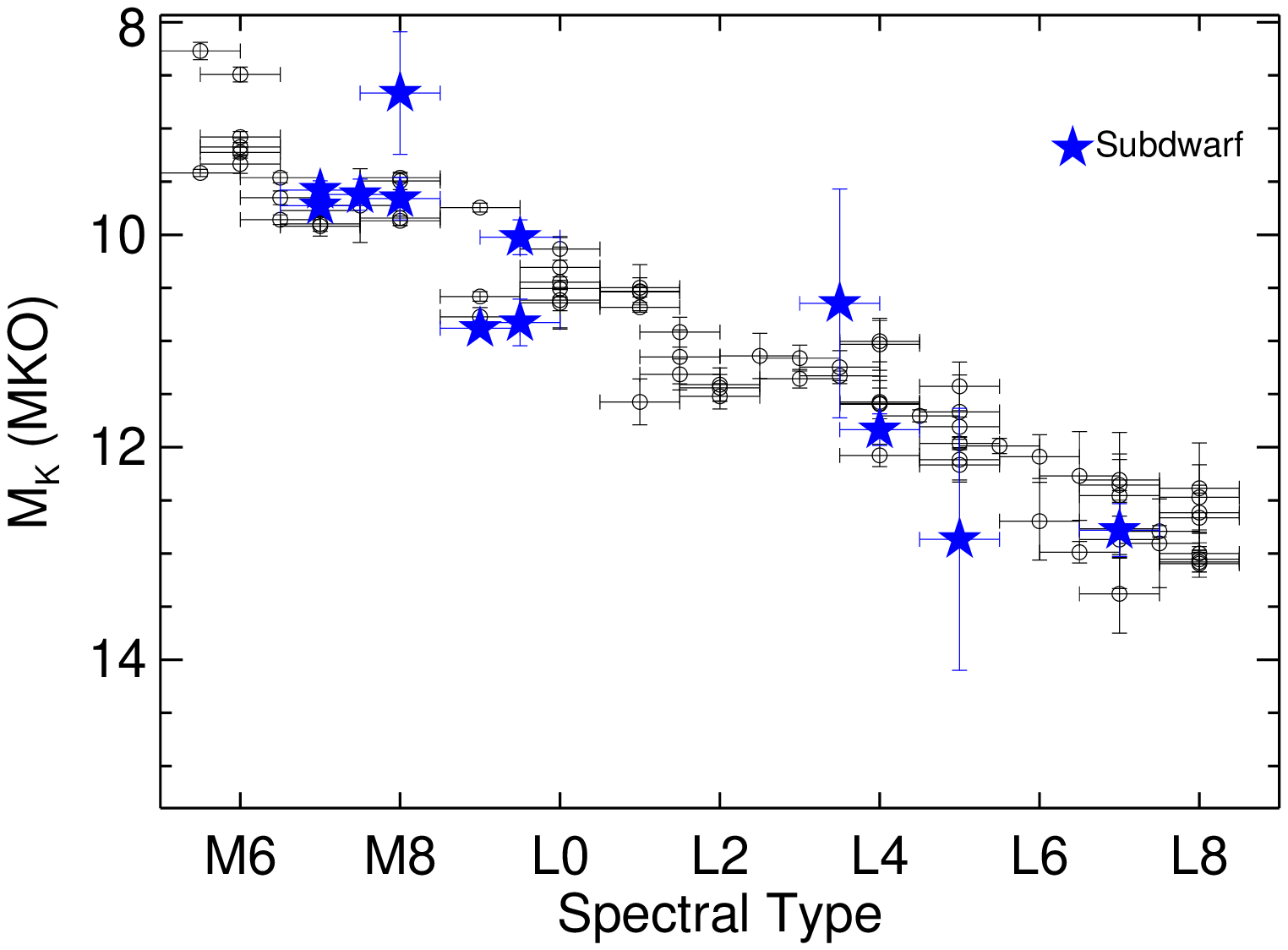}
\end{center}
\caption{Spectral type versus absolute magnitude in the MKO $JHK$ filters for late-type M through L dwarfs.  Unfilled circles are normal dwarfs with parallax measurements.  Blue five point stars are subdwarfs with parallax measurements.} 
\label{fig:ABS_SD}
\end{figure}

There are 12 ultracool subdwarfs with parallaxes including eight late-type M  and four L subdwarfs.  Figure ~\ref {fig:ABS_SD} shows the near-IR absolute magnitude vs. spectral type diagrams for normal mid-type M through late-type L dwarfs (excluding binaries and low-surface gravity dwarfs) with the subdwarfs over-plotted. We show all subdwarfs including objects with absolute magnitude uncertainty $>$ 0.5 mag.  As noted in \citet{2008ApJ...672.1159B}, the L subdwarfs are overluminous in M$_{J}$ but shift to normal or slightly underluminous by M$_{K}$.  This has been attributed to reduced condensate opacity, as evidenced by strong TiO, Ca I, and Ti I features; and enhanced collision-induced H$_2$ opacity at $K$-band (e.g. \citealt{2001ApJ...556..872A} ;\citealt{1996A&A...308L..29T}; \citealt{2003ApJ...592.1186B, 2007ApJ...657..494B}) .  The effect is not as pronounced for the late-type M subdwarfs which, with the exception of SSSPM J1256-1408, appear at most slightly overluminous in M$_{J}$ and normal or underluminous in M$_{H}$ and M$_{K}$.  

We compare photometry for the two L subdwarfs 2MASS J0532+8246 and 2MASS J1626+3925 on Figures ~\ref{fig:fsed} and ~\ref{fig:burrows_met}.  In section ~\ref{Ldwarfs} we find that early and mid-type L dwarfs are best fit by the cloudy tracks using both \citet{2008ApJ...689.1327S} and \citet{2006ApJ...640.1063B} evolutionary models. However, variations in the metallicity of the cloudy tracks in Figure ~\ref{fig:burrows_met} do not reproduce the colors of the L subdwarfs.  Instead, one must use the cloudless tracks supporting the idea that these objects are blue and overluminous at $J$ due to reduced cloud opacity.


\section{Kinematics\label{kinematics}}
Combining the absolute parallax with the relative proper motion gives the tangential velocity (V$_{tan}$) of a source (see Table ~\ref{astrometry}  for V$_{tan}$ values of objects studied in this work). As our full astrometric sample is composed of objects in the immediate solar vicinity, V$_{tan}$ values and their dispersions can be used as a rough indicator of age.  In general, older objects will have had enough time to interact with objects in the Galactic disk and have their orbits perturbed while younger objects will retain a motion consistent with that of the Galactic disk (i.e., co-moving with their nascent cloud).  The dispersion of a population is more informative than individual values for determining ages.  

The median $V_{tan}$ and $\sigma_{tan}$ values for the 71 normal and unresolved binary L dwarfs with absolute magnitude uncertainties $<$ 0.5 are 27 km~s$^{-1}$ and 20 km~s$^{-1}$ respectively.  For the 50 normal and unresolved binary T dwarfs we find similar values of 31 km~s$^{-1}$ and 20 km~s$^{-1}$ respectively.   Our results are in good agreement with earlier population analyses (\citealt{2009AJ....137....1F},\citealt{2010AJ....139.1808S}).  Compared to the kinematic results of Vrba et al. (2004), we do not find a significant difference between the kinematics of L and T dwarfs.  Based on the difference in dispersion between the L and T dwarfs, \citet{2004AJ....127.2948V} concluded from their much smaller sample of UCDs that the L dwarfs were a kinematically younger population than the T dwarfs.  In part, their conclusion was drawn from the fact that there were no T dwarfs with V$_{tan}$ values $<$ 20 km s$^{-1}$.  In our larger sample, we find 14 T dwarfs with V$_{tan}$ values $<$ 20 km s$^{-1}$. A two-sided KS test on the L and T velocities yields a significant probability (p $\sim$ 0.33) that the L and T dwarfs in our sample have identical kinematics hence ages.  

We isolated the low-surface gravity dwarfs and the subdwarfs and compared their kinematics to the overall sample (note that we have included late-type M dwarfs in each subset but only discuss objects whose absolute magnitude uncertainties are $<$ 0.5 mag). The former have significantly smaller V$_{tan}$ values and tighter dispersions than the overall population and the latter significantly larger values.  The median V$_{tan}$ and $\sigma_{tan}$ values for the 10 low-surface gravity dwarfs are 10 km s$^{-1}$ and 14 km s$^{-1}$ respectively.  For the 9 subdwarfs,  the median V$_{tan}$ and $\sigma_{tan}$ values are 241 km s$^{-1}$ and 68 km s$^{-1}$ respectively.  The considerable difference in values for each subset compared to the overall population further confirms expectations that they are younger (low surface gravity objects) and older (subdwarfs) than the overall ultracool dwarf population.

\section{Binaries\label{individual}}
\begin{figure}[!ht]
\begin{center}
\epsscale{1.2}
\includegraphics[width=.55\hsize]{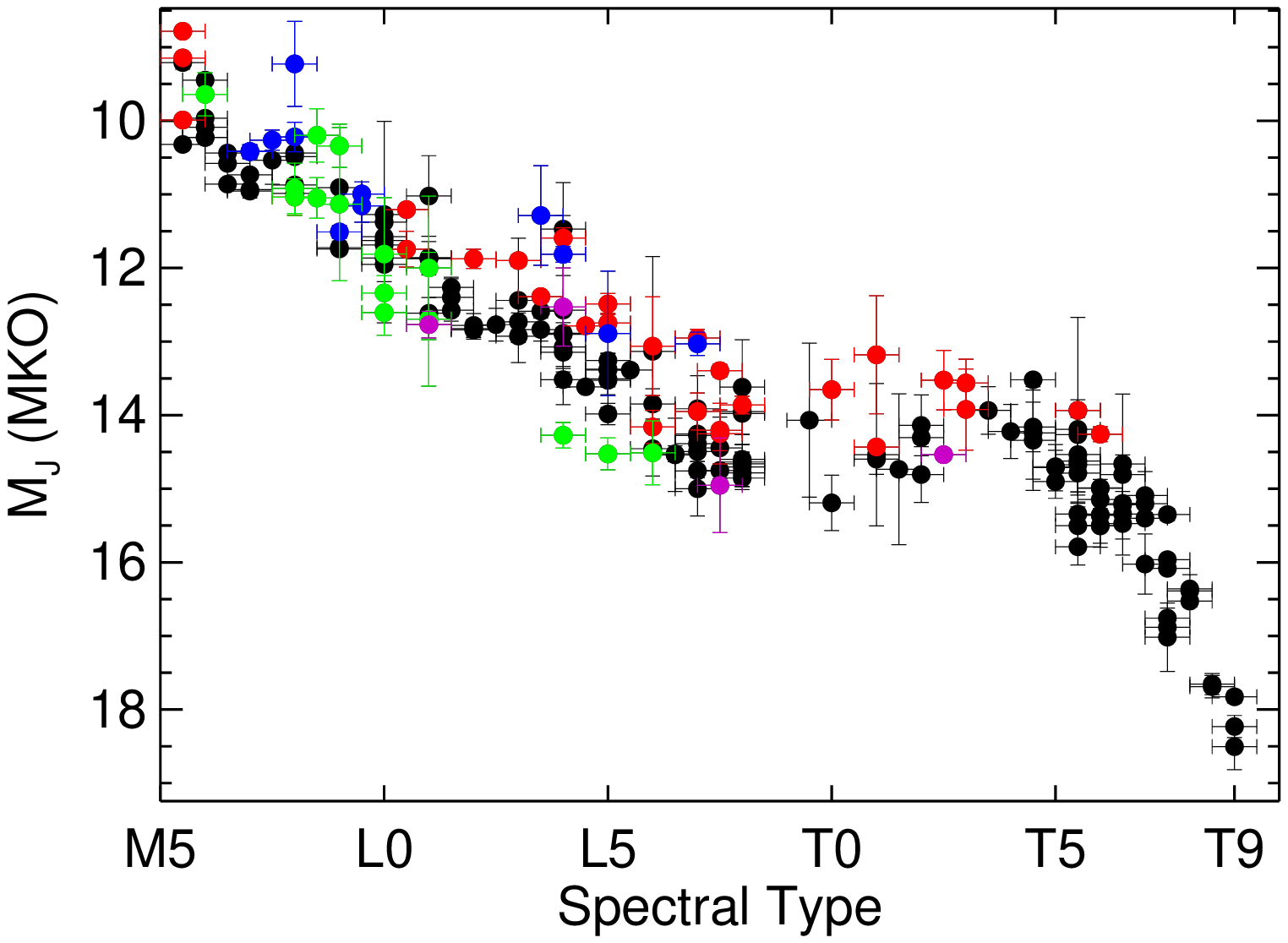}
\plottwo{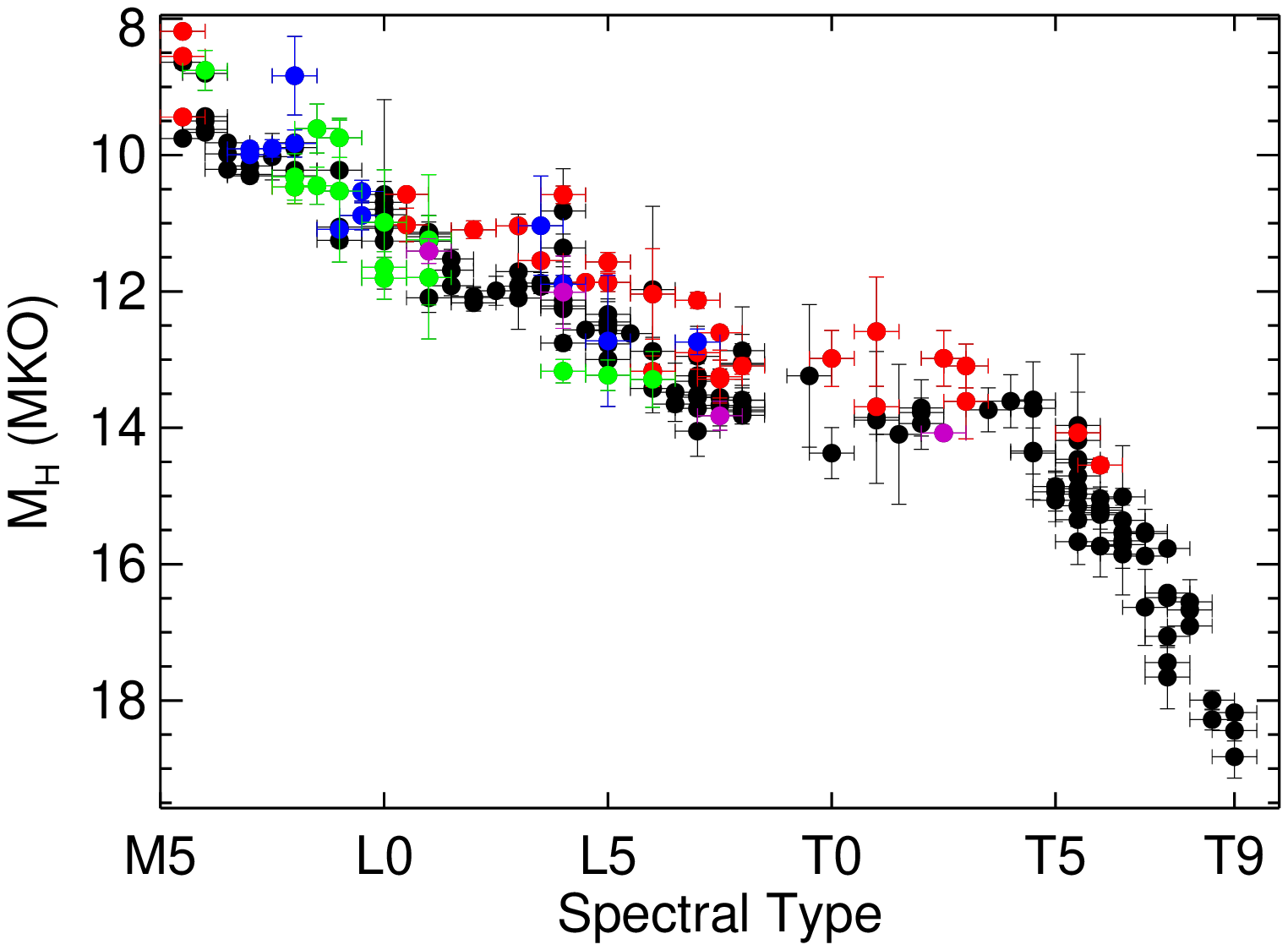}{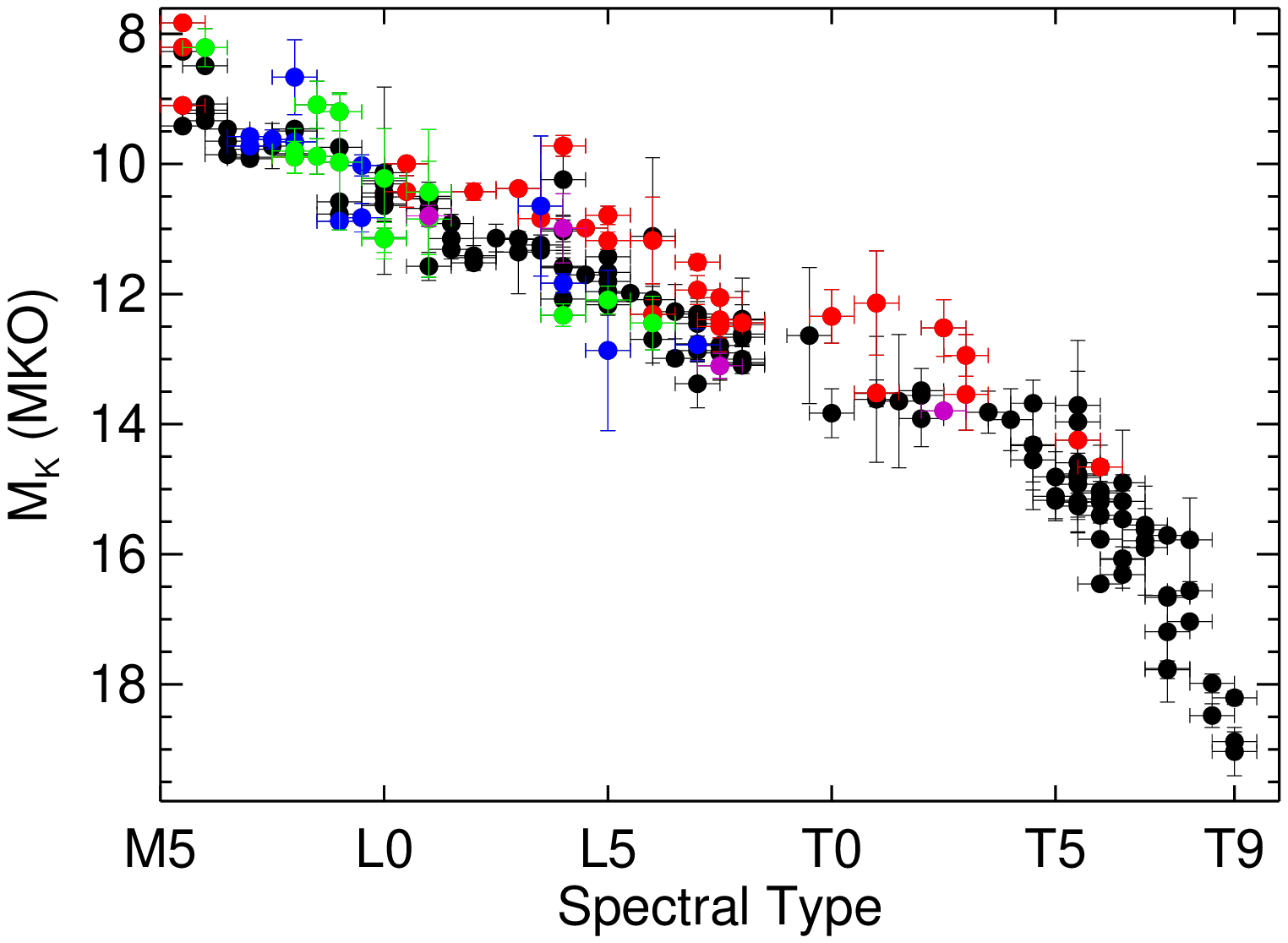}
\end{center}
\caption[Spectral type versus absolute magnitude in the MKO $JHK$]{ Spectral type versus absolute magnitude in the MKO $JHK$ filters for normal (black filled circles), combined-light binary (red filled circles), young (companions to known $<<$ 1 Gyr  stars; purple filled circles), low-surface gravity (green filled circles), and subdwarf (blue filled circles) M through T dwarfs. All UCDs with a parallax measurement in this work or within the literature are shown regardless of uncertainty.} 
\label{fig:ABS_Mag2}
\end{figure}

When found as companions, brown dwarfs are primarily tightly bound (separations $<$ 20 AU) with a mass ratio close to 1 (e.g. \citealt{2003ApJ...586..512B}; \citealt{2003ApJ...587..407C}).  As such, near-equal mass brown dwarf binaries can be identified on an HR diagram by their over-luminosity compared to equivalent spectral type objects.  Moreover, if component photometry has been acquired one can investigate the properties of co-evolving systems and measure dynamical masses (e.g \citealt{2009ApJ...706..328D,2008arXiv0807.2450D}; \citealt{2010arXiv1001.4800K}).  Among the parallax sample listed in Table ~\ref{Luminosity} there are 25 binaries unresolved in 2MASS. These are shown on Figure ~\ref{fig:ABS_Mag2} as filled red circles.  

There were four sources with parallaxes reported in this work with published high resolution imaging and predicted component spectral types from photometric data and/or spectral template fitting: 2M0518-2828, 2M1209-1004, 2M1404-3159, and 2M2052-1609.  These four sources were used in the analysis of section ~\ref{nearab} and their system properties are listed in Table ~\ref{LTBinaries2}. When compared to the absolute magnitudes calculated for each spectral subtype using the polynomial reported in Table ~\ref{coeffs}, both components of 2M1209-1004, 2M2052-1609, and 2M0518-2828  fit within uncertainties on the brown dwarf main sequence.  While the primary for 2M1404-3159 also fits within the brown dwarf main-sequence, the T5 secondary is $\sim$ 1 mag underluminous in M$_{JHK}$.  

Burgasser et al. 2010a used a template fitting technique to identify 17 new L-T transition binaries and three of the suspected binaries have parallaxes reported in this work:  SDSS1511+0607, 2MASSJ0949-1545, and SDSS1207+0244.  The first was regarded as a strong binary candidate while the latter two are weak binary candidates.  Examining the HR diagram in Figure ~\ref{fig:ABS_Mag2} we find SDSS1511+0607 to be nearly 1 magnitude over-luminous in $M_{JHK}$.  Gelino et al. (in prep) have followed up with Adaptive optics and resolved the two components.   We show both components in Figure ~\ref{fig:Binaries} and list the individual magnitudes in Table ~\ref{LTBinaries2}.  The proposed L5.5 and T5 components fit well within the brown dwarf main sequence.  

Using new parallax measurements, 2MASSJ0949-1545 and SDSS1207+0244 fit well on the HR diagram as single objects.  If decomposed into the binary components proposed by Burgasser et al. 2010a, the primary and secondary sources would be $>$ 1 mag underluminous in multiple bands.  We conclude that these sources are best treated as single sources.

\section{Conclusions\label{conclu}}
We have measured parallaxes for 11 M, 32 L and 27 T dwarfs in the local solar neighborhood.   Nine calibrator stars were included in the sample to verify the reliability of our pipeline. The focus of this project was on low surface gravity dwarfs, L/T transition objects, and late-type T dwarfs within 20 pc of the sun.   The 70 new parallaxes significantly increases the number of brown dwarfs with accurate distance measurements.  

We combined our sample with 115 literature measurements and used the full sample to re-define color-magnitude and spectrophotometric diagrams in $JHK$. Adding decomposed L/T transition binaries we find no reason to split spectral-type/absolute magnitude polynomials into  ``bright'' and ``faint'' trends to account for unresolved binarity as has been done in the past.   Isolating T0-T4 dwarfs to investigate the extent of the L/T transition brightening we find there is a [1.2 - 1.4] magnitude difference at J, [0.3-0.5] magnitude difference at H, and a plateau or dimming of [-0.2 - -0.3] magnitude at K.  In agreement with flux reversal binary studies, this confirms the brightening---and the physical mechanism that drives it---as an intrinsic feature of brown dwarf atmospheric evolution. 

We compared the  $J$-$K$ vs. M$_{K}$ data for the full parallax sample to the evolutionary models of \citet{2008ApJ...689.1327S} and \citet{2006ApJ...640.1063B}.  The f$_{sed}$=1,2 parameters best fit the L dwarf sequence and f$_{sed}$=4 (corresponding to a very thin cloud layer) best fit the late-type T dwarf sequence using the \citet{2008ApJ...689.1327S} models.  The cloud model with varying gravity and metallicity reproduce the L dwarfs and the clear model with similar variations fit the T dwarfs using the \citet{2006ApJ...640.1063B} models.   However comparisons of both models to empirical data show significant red or potentially ``ultra-cloudy" L dwarf outliers.  Similarly there is significant scatter seen in the latest type T dwarfs that is unaccounted for in the clear and f$_{sed}$=4 models indicating that condensate clouds continue to play a role in the photospheres of some low-temperature brown dwarfs.   Investigations of individual objects in the ``ultra-cloudy" sample reveals objects which are young/moderately young, implying a correlation between youth and enhanced photospheric dust.

No single f$_{sed}$ parameter, gravity, nor metallicity track in the evolutionary models can account for the L/T transition objects.  However, a hybrid model which smoothly transitions from f$_{sed}$=2 (the best fit for late-type L dwarfs) to f$_{sed}$=4 (the best fit for late-type T dwarfs) at a near-constant T$_{eff}$ = 1200$\pm$100K encompasses the majority of T0--T4 dwarfs. This temperature range is consistent with recent toy and sophisticated hybrid models but demonstrates that the range of spectral subtypes for which the temperature plateau is applicable is narrower than previously suspected (T0-T4 as opposed to L7-T4).

The low-surface gravity objects with parallax measurements in this work are not explained by varying gravity in the evolutionary models.  Rather they appear to be under-luminous compared to model color-magnitude diagrams and for their low gravity spectral type.   Among the 10 low-surface gravity or young companion L dwarfs investigated, 80\% appear [0.2-1.0] magnitude under-luminous for their spectral type in $J$, $H$ and/or $K$. Possible explanations for their underluminosity are that (1) the low-gravity and field dwarf spectral classification schemes are on different temperature scales with the low-surface gravity objects intrinsically cooler than field age objects of the same type; and/or (2) young objects could have dustier photospheres than field-aged objects making them appear both fainter and redder.

A kinematic analysis of the astrometric sample reveals similar velocity dispersions between the L and T dwarf populations.  A two-sided KS test verifies that the two kinematic distributions are likely drawn from the same population and hence have similar age distributions.  The low-surface gravity and subdwarf samples have distinctly different velocity dispersions and are likely significantly younger and older (respectively) than normal objects.

\acknowledgments{We acknowledge receipt of observation time through NOAO as well as the SMARTS consortium.  Stony Brook's participation in the SMARTS consortium is made possible by generous support by the Dean of Arts and Sciences, the Provost, and the Vice-President for Research of Stony Brook University.  We would like to thank 4.0m telescope operators C. Aguilera, M. Gonzalez, and A. Alvarez as well as 1.3m observers A. Miranda, J. Espinoza, and J. Velasquez.   Faherty gratefully acknowledges support from Hilary Lipsitz and from the AMNH and further acknowledges the encouragement and support made possible by Sahne Nuss during observing runs.    This publication has made use of the Carnegie Astrometric Program parallax reduction software as well as the VLM Binaries Archive maintained by Nick Siegler at http://www.vlmbinaries.org and the data products from the Two Micron All-Sky Survey, which is a joint project of the University of Massachusetts and the Infrared Processing and Analysis Center/California Institute of Technology, funded by the National Aeronautics and Space Administration and the National Science Foundation. This research has made use of the NASA/ IPAC Infrared Science Archive, which is operated by the Jet Propulsion Laboratory, California Institute of Technology, under contract with the National Aeronautics and Space Administration. }

\begin{deluxetable}{lllcccccccccccc}
\label{tab:tab1}
\tabletypesize{\scriptsize}
\rotate
\tablecaption{Target List\label{targetlist}}
\tablewidth{0pt}
\tablehead{
\colhead{Name} &
\colhead{SpT} &
\colhead{SpT} &
\colhead{Nights} &
\colhead{Frames\tablenotemark{a}} &
\colhead{Ref Stars\tablenotemark{a}} &
\colhead{$\Delta$t} &
\colhead{$J$} &
\colhead{Instrument\tablenotemark{b}} &
\colhead{Note\tablenotemark{c}}  &
\colhead{Flag\tablenotemark{d}} &
\colhead{Ref} \\
 &
 \colhead{OPT}&
 \colhead{IR}&
&
&
&
\colhead{(yr)}&
\colhead{(MKO)}&
&
\\
\colhead{(1)}  &
\colhead{(2)}  &
\colhead{(3)}  &
\colhead{(4)}  &
\colhead{(5)}  &
\colhead{(6)}  &
\colhead{(7)}  &
\colhead{(8)}  &
\colhead{(9)}  &
\colhead{(10)}  &
\colhead{(11)}  &
\colhead{(12)}  \\
}
\startdata

        2MASSJ00325584-4405058    &   L0.0$\gamma$    &             --    &   12    &   49    &   71    &      2.24    &     14.68$\pm$      0.04    &         A    &        LG    &                                                 &                      23,  22\\
        2MASSJ00345157+0523050    &             --    &           T6.5    &    6    &   27    &   38    &      2.91    &     15.11$\pm$      0.03    &         I    &              &                        F                        &                      24, 19   \\
         2MASSIJ0103320+193536    &    L6.0$\beta$    &             --    &    6    &   30    &   54    &      3.08    &     16.16$\pm$      0.08    &         I    &        LG    &                        F                        &                      15     \\
        2MASSJ02212859-6831400    &    M8.0$\beta$    &             --    &   15    &   60    &   78    &      2.04    &     13.89$\pm$      0.03    &         A    &        LG    &                                                 &                      17     \\
        2MASSJ03185403-3421292    &           L7.0    &             --    &    8    &   34    &   46    &      3.08    &     15.44$\pm$      0.05    &         I    &              &                        F                        &                      16     \\
       SDSSJ032553.17+042540.1    &             --    &           T5.5    &    6    &   33    &   21    &      3.08    &     15.90$\pm$      0.03    &         I    &              &                        F                        &                       2     \\
        2MASSJ03341218-4953322    &           M9.0    &             --    &   16    &   54    &   73    &      2.96    &     11.32$\pm$      0.02    &         A    &              &                                                 &                      77     \\
\enddata
\end{deluxetable}

\begin{deluxetable}{lcrrrrrrrrrrrrr}
\label{tab:tab2}
\tabletypesize{\scriptsize}
\tablecaption{Astrometry of Targets\label{astrometry}}
\tablewidth{0pt}
\tablehead{
\colhead{Name} &
\colhead{SpT} &
\colhead{$\pi_{rel}$}  &
\colhead{$\pi_{abs}$}  &
\colhead{$\mu_{\alpha}$ } &
\colhead{$ \mu_{\delta}$}  &
\colhead{v$_{tan}$} \\
&
&
\colhead{(mas)}&
\colhead{(mas)}&
\colhead{mas yr$^{-1}$} &
\colhead{mas yr$^{-1}$}&
\colhead{km s$^{-1}$} \\
\colhead{(1)}  &
\colhead{(2)}  &
\colhead{(3)}  &
\colhead{(4)}  &
\colhead{(5)}  &
\colhead{(6)}  &
\colhead{(7)}  \\
}
\startdata
        2MASSJ00325584-4405058    &     L0$\gamma$    &      36.6$\pm$       4.6    &      38.4$\pm$       4.8    &     117.8$\pm$       4.3    &     -91.6$\pm$       4.3    &      18.4$\pm$       2.8    &        \\
        2MASSJ00345157+0523050    &           T6.5    &     103.3$\pm$       7.2    &     105.4$\pm$       7.5    &     658.8$\pm$       6.2    &     181.0$\pm$       6.0    &      30.8$\pm$       1.2    &        \\
         2MASSIJ0103320+193536    &      L6$\beta$    &      45.2$\pm$       7.5    &      46.9$\pm$       7.6    &     293.0$\pm$       4.6    &      27.7$\pm$       4.7    &      29.8$\pm$       4.2    &        \\
        2MASSJ02212859-6831400    &      M8$\beta$    &      23.7$\pm$       3.3    &      25.4$\pm$       3.6    &      53.9$\pm$       4.4    &      13.7$\pm$       4.5    &      10.4$\pm$       2.9    &        \\
        2MASSJ03185403-3421292    &             L7    &      70.2$\pm$       7.3    &      72.9$\pm$       7.7    &     391.7$\pm$       6.4    &      34.1$\pm$       6.5    &      25.6$\pm$       2.0    &        \\
       SDSSJ032553.17+042540.1    &           T5.5    &      52.9$\pm$      10.8    &      55.6$\pm$      10.9    &    -163.7$\pm$       5.8    &     -59.6$\pm$       5.7    &      14.9$\pm$       3.3    &        \\
        2MASSJ03341218-4953322    &             M9    &     119.1$\pm$       3.4    &     120.6$\pm$       3.6    &    2329.5$\pm$       4.2    &     553.6$\pm$       4.3    &      94.2$\pm$       0.8    &        \\
       2MASSJ04221413+1530525:    &     M6$\gamma$    &      22.2$\pm$       2.9    &      24.8$\pm$       3.1    &     -17.2$\pm$       2.7    &       7.4$\pm$       2.6    &       3.6$\pm$       1.5    &        \\
\enddata
\end{deluxetable}
\clearpage

\begin{deluxetable}{lcrrrrrrrrrrrrr}
\label{tab:tab3}
\tabletypesize{\scriptsize}
\tablecaption{Astrometric Calibrators \label{calibs}}
\rotate
\tablewidth{0pt}
\tablehead{
\colhead{Name} &
\colhead{$\pi_{abs}$}  &
\colhead{$\mu_{\alpha}$ } &
\colhead{$ \mu_{\delta}$}  &
\colhead{$\pi_{abs}$}  &
\colhead{$\mu_{\alpha}$ } &
\colhead{$ \mu_{\delta}$}  &
\colhead{Inst.\tablenotemark{a}}  &
\colhead{ref}\\
&
\colhead{(mas)}&
\colhead{mas yr$^{-1}$} &
\colhead{mas yr$^{-1}$}&
\colhead{(mas)}&
\colhead{mas yr$^{-1}$} &
\colhead{mas yr$^{-1}$}&
\\
&
\colhead{(BDKP)}&
\colhead{(BDKP)}&
\colhead{(BDKP)}&
\colhead{(Lit)}&
\colhead{(Lit)}&
\colhead{(Lit)}&
\\

\colhead{(1)}  &
\colhead{(2)}  &
\colhead{(3)}  &
\colhead{(4)}  &
\colhead{(5)}  &
\colhead{(6)}  &
\colhead{(7)}  &
\colhead{(8)}  &
\colhead{(9)}  \\

}
\startdata
LHS 1777						&   79.9$\pm$2.5    &    -175.6$\pm$2.2    &     937.9$\pm$2.3    &     78.20$\pm$2.70&   -153.00$\pm$5.00&    957.00$\pm$5.00  &   A&    1 \\
					&							&						&						&	79.72$\pm$	1.89	&	 &	 & &2\\
						&							&						&						&	80.11$\pm$	2.34	&	 &	 & &3\\
\hline
 WT 248&     31.7$\pm$4.7    &   -1199.4$\pm$5.9    &    -140.5$\pm$5.8    &     38.44$\pm$2.83&  -1191.00$\pm$     13.00&   -115.00$\pm$     13.00  &     A&1  \\
\hline
2MASS J0746+2000&86.2$\pm$4.6    &    -355.9$\pm$5.1    &     -63.7$\pm$5.2    &     81.90$\pm$0.30&   -374.04$\pm$0.31&    -57.91$\pm$0.65  &     A&4   \\
\hline
2MASS J0939-2448 &    196.0$\pm$10.4    &     558.1$\pm$5.8    &   -1030.5$\pm$5.6   &    187.30$\pm$4.60&    573.40$\pm$2.30&  -1044.70$\pm$2.50  &I&6 \\
\hline
TWA28&  28.0$\pm$4.3    &     -55.7$\pm$3.9    &-7.4$\pm$    3.8    &   18.1$\pm$0.5 &  -67.2$\pm$0.6 & -14.0$\pm$0.6 & A&8\\
\hline
2MASS J1058-1548&66.5$\pm$4.4    &    -253.5$\pm$3.4    &39.7$\pm$3.2    &     57.70$\pm$1.00&   -252.93$\pm$0.50&     41.42$\pm$0.45  &     A&4   \\
\enddata
\tablenotetext{a}{A=ANDICAM and I=ISPI}
\tablerefs{1=\citet{1995gcts.book.....V}, 2=\citet{2005AJ....129.1954J}, 3=\citet{2005AJ....130..337C}, 4=\citealt{2002AJ....124.1170D}, 5=\citet{2004AJ....127.2948V}, 6=\citet{2008ApJ...689L..53B}, 7=\citet{2009A&A...493L..27S}, 8=\citet{2008A&A...489..825T}}
\end{deluxetable}

\begin{deluxetable}{r r r r c c c r r r r r|}
\label{tab:tab7}
\tabletypesize{\scriptsize}
\tablecaption{ Discrepant Proper Motion Values\label{discrepant}}
\tablewidth{0pt}
\tablehead{
\colhead{Name} &
\colhead{$\mu_{\alpha}$ } &
\colhead{$ \mu_{\delta}$}  &
\colhead{$\mu_{\alpha}$ } &
\colhead{$ \mu_{\delta}$}  &
\colhead{Reference} \\

 & 
\colhead{mas yr$^{-1}$} &
\colhead{mas yr$^{-1}$}&
\colhead{mas yr$^{-1}$} &
\colhead{mas yr$^{-1}$} \\
&
\colhead{(BDKP)} &
\colhead{(BDKP)} &
\colhead{(Lit)}&
\colhead{(Lit)}\\

}
\startdata

        2MASSJ06164006-6407194    &    1304.0$\pm$       3.9    &     -31.9$\pm$       3.7    &     1405$\pm$        8    &      -51$\pm$       18    &  1 \\
          2MASSJ14044941-3159329\tablenotemark{a}    &     334.5$\pm$       4.9    &     -15.5$\pm$       4.7    &     -348$\pm$       30    &       32$\pm$        3    &   2\\
        2MASSJ21321145+1341584    &      16.4$\pm$      11.6    &    -141.0$\pm$      11.2    &      -55$\pm$        9    &     -395$\pm$        9    &   3\\
\enddata
\tablecomments{Details on the discrepant proper motion objects.  We note only objects whose proper motion values were discrepant by more than 4$\sigma$.}
\tablerefs{1=\citet{Cushing2009} 2=\citet{2007AJ....134.1162L} 3=\citet{2007AJ....133.2320S} } 
\tablenotetext{a}{We find the same magnitude of motion as \citet{2007AJ....134.1162L} yet an opposite direction. We conclude there must be a sign error in that work}
\end{deluxetable}

\begin{deluxetable}{lllcccccccccccccccccccccc}
\label{tab:tab1}
\tabletypesize{\tiny}
\rotate
\tablecaption{Photometry of Full Sample\tablenotemark{a}\label{Luminosity}}
\tablewidth{0pt}
\tablehead{
\colhead{Name} &
\colhead{SpT} &
\colhead{SpT} &
\colhead{$\pi$} &
\colhead{J} &
\colhead{H} &
\colhead{K} &
\colhead{Note\tablenotemark{b}} &
\colhead{Ref} \\
 &
\colhead{(OpT)} &
\colhead{(near-IR)} & 
\colhead{(mas)} &
\colhead{(MKO)} &
\colhead{(MKO)} &
\colhead{(MKO)} \\
\colhead{(1)}  &
\colhead{(2)}  &
\colhead{(3)}  &
\colhead{(4)}  &
\colhead{(5)}  &
\colhead{(6)}  &
\colhead{(7)}  &
\colhead{(8)}  &
\colhead{(9)}  \\
}
\startdata
                  2MASSJ00043484-4044058    &           L5.0    &           L4.5    &     76.86   $\pm$      3.97    &     13.06   $\pm$      0.03    &     12.14   $\pm$      0.03    &     11.36   $\pm$      0.03    &         B    &            23,       65,        9,        70\\
                  2MASSJ00250365+4759191    &           L4.0    &            ---    &     23.69   $\pm$      1.08    &     14.72   $\pm$      0.04    &     13.71   $\pm$      0.03    &     12.85   $\pm$      0.06    &         B    &            13,                 94\\
                  2MASSJ00303013-1450333    &           L7.0    &            ---    &     37.42   $\pm$      4.50    &     16.39   $\pm$      0.03    &     15.37   $\pm$      0.03    &     14.49   $\pm$      0.03    &         N    &            15,                            74\\
                  2MASSJ00325584-4405058    &   L0.0$\gamma$    &            ---    &     38.42   $\pm$      4.80    &     14.68   $\pm$      0.04    &     13.89   $\pm$      0.03    &     13.23   $\pm$      0.04    &        LG    &   1,       23,       22\\
                  2MASSJ00325937+1410371    &            ---    &           L8.0    &     30.14   $\pm$      5.16    &     16.58   $\pm$      0.05    &     15.66   $\pm$      0.05    &     14.99   $\pm$      0.05    &         N    &            34,                            74\\
                   ULASJ00340277-0052067    &            ---    &           T8.5    &     79.60   $\pm$      3.80    &     18.15   $\pm$      0.04    &     18.49   $\pm$      0.04    &     18.48   $\pm$      0.04    &         N    &            59\\
                  2MASSJ00345157+0523050    &            ---    &           T6.5    &    105.36   $\pm$      7.48    &     15.11   $\pm$      0.03    &     15.55   $\pm$      0.03    &     15.96   $\pm$      0.03    &         N    &   1,       24,                 19\\
\enddata
\end{deluxetable}

\begin{deluxetable}{lllccccccccccccccccccccc}
\label{tab:tab5}
\tabletypesize{\scriptsize}
\rotate
\tablecaption{Magnitudes for L/T Transition Binaries\label{LTBinaries2}}
\tablewidth{0pt}
\tablehead{
\colhead{Name} &
\colhead{SpT} &
\colhead{SpT} &
\colhead{SpT} &
\colhead{M$_{J}$} &
\colhead{M$_{J}$} &
\colhead{M$_{H}$} &
\colhead{M$_{H}$} &
\colhead{M$_{K}$} &
\colhead{M$_{K}$}&
\colhead{Ref} \\
&
\colhead{Combined} &

\colhead{Primary} &
\colhead{Secondary} &

\colhead{Primary} &
\colhead{Secondary} &

\colhead{Primary} &
\colhead{Secondary} &

\colhead{Primary} &
\colhead{Secondary} \\
\colhead{(1)}  &
\colhead{(2)}  &
\colhead{(3)}  &
\colhead{(4)}  &
\colhead{(5)}  &
\colhead{(6)}  &
\colhead{(7)}  &
\colhead{(8)}  &
\colhead{(9)}  &
\colhead{(10)} &
\colhead{(11)} \\

}
\startdata
           SD0423-0414  &      L7.5  &      L6.5  &      T2.0  &     13.95$\pm$      0.10  &     14.39$\pm$      0.12  &     13.05$\pm$      0.10  &     13.78$\pm$      0.12  &     12.38 $\pm$      0.09  &     13.51$\pm$      0.06&1 \\
         2M0518-2828  &      L7.5  &      L6.0  &      T4.0  &     14.34$\pm$      0.41  &     14.47$\pm$      0.42  &     12.96$\pm$      0.39  &     14.29$\pm$      0.51  &     12.14 $\pm$      0.37  &     14.32$\pm$      0.39 &1\\
         SD1021-0304  &      T3.0  &      T1.0  &      T5.0  &     14.24$\pm$      0.48  &     14.40$\pm$      0.51  &     13.42$\pm$      0.40  &     14.54$\pm$      0.57  &     13.19 $\pm$      0.36  &     14.65$\pm$      0.35 &1\\
         2M1404-3159  &      T2.5  &      T1.0  &      T5.0  &     14.47$\pm$      0.46  &     13.94$\pm$      0.43  &     13.57$\pm$      0.43  &     14.05$\pm$      0.45  &     12.87 $\pm$      0.45  &     14.07$\pm$      0.38 &2\\
         2M1711+2232  &      L6.5  &      L5.0  &      T5.5  &     15.08$\pm$      0.52  &     16.00$\pm$      0.59  &     13.54$\pm$      0.44  &     15.64$\pm$      0.75  &     12.39 $\pm$      0.42  &     15.44$\pm$      0.65 &1\\
         SD1511+0607  &      T0.0  &      L5.5  &      T5.0  &     14.17$\pm$      0.48  &     14.72$\pm$      0.55  &     13.18$\pm$      0.44  &     14.96$\pm$      0.65  &     12.43 $\pm$      0.42  &     15.09$\pm$      0.47 &1\\
             Epsilon Indi  &      T2.5  &      T1.0  &      T6.0  &     14.69$\pm$      0.04  &     15.63$\pm$      0.04  &     13.99$\pm$      0.04  &     15.75$\pm$      0.04  &     13.51 $\pm$      0.04  &     15.69$\pm$      0.01 &3\\
         2M1209-1004  &      T3.0  &      T2.0  &      T7.5  &     14.17$\pm$      0.57  &     15.64$\pm$      0.69  &     13.69$\pm$      0.56  &     16.45$\pm$      0.78  &     13.59 $\pm$      0.56  &     16.85$\pm$      1.04 &4\\
\enddata
\tablerefs{1=\citet{2010ApJ...710.1142B}
2=\citet{2008ApJ...685.1183L}
3=\citet{2004A&A...413.1029M}
4=\citet{2010ApJ...722..311L}
5=\citet{2011A&A...525A.123S}}
\end{deluxetable}

\begin{deluxetable}{lccccccrrrrrrrr}
\label{tab:tab5}
\tabletypesize{\scriptsize}
\tablecaption{Coefficients of Polynomial Fits for L0 -T8 Dwarfs \label{coeffs}}
\tablewidth{0pt}
\tablehead{
\colhead{MKO Filter} &
\colhead{c$_{0}$} &
\colhead{c$_{1}$} &
\colhead{c$_{2}$} &
\colhead{c$_{3}$} &
\colhead{c$_{4}$} &
\colhead{rms} \\
\colhead{(1)}  &
\colhead{(2)}  &
\colhead{(3)}  &
\colhead{(4)}  &
\colhead{(5)}  &
\colhead{(6)}  &
\colhead{(7)}   \\}
\startdata
M$_{J}\tablenotemark{a}$&1.48948e+1    &-1.99967e+0    &2.75734e-1    &-1.29143e-2    &2.03252e-4    &     0.30   \\
M$_{H}\tablenotemark{a}$&1.32372e+1&-1.60036e+0    &2.27363e-1    &-1.08205e-2    &1.75368e-4    &      0.27   \\
M$_{K}\tablenotemark{a}$&9.76100e+0    &-4.85519e-1    &9.40816e-2    &-4.69032e-3    &8.16516e-5    &      0.28   \\
\hline
\enddata
\tablenotetext{a}{Including 8 L/T transition binaries with resolved photometry: SDSS J0423-0414, 2MASS J0518-2828, SDSS J1021-0304, 2MASS J1404-3159, 2MASS J1711+2232, SDSS J1511+0607, Epsilon Indi, SDSS J1534+1615}
\tablecomments{Relations use (MKO) magnitudes.  Polynomial fits to optical L dwarfs and NIR T dwarfs (L dwarfs with no optical spectral type have NIR spectral types) excluding subdwarfs, low gravity dwarfs, and binaries.  Function is defined as M$_{J,H,K}$=$\sum_{i=0}^{n}$ c$_{i}$(SpT)$^{i}$ and is valid for spectral types L0-T8 where 10=L0, 20=T0, etc.}
\end{deluxetable}

\begin{deluxetable}{lccccccrrrrrrrr}
\label{tab:tab1}
\tabletypesize{\scriptsize}
\tablecaption{Absolute Magnitudes With Resolved L/T Binaries\label{absmagswithbin}}
\tablewidth{0pt}
\tablehead{
\colhead{SpT} &
\colhead{BDKP} &
\colhead{Lit} &
\colhead{Total} &
\colhead{Norm} &
\colhead{Bin} &
\colhead{LG,Y} &
\colhead{SD} &
\colhead{M$_{J}$} &
\colhead{M$_{H}$} &
\colhead{M$_{K}$} \\
 &
 &
 &
 &
 &
 &
 &
 &
\colhead{(MKO)} &
\colhead{(MKO)} &
\colhead{(MKO)} \\
\colhead{(1)}  &
\colhead{(2)}  &
\colhead{(3)}  &
\colhead{(4)}  &
\colhead{(5)}  &
\colhead{(6)}  &
\colhead{(7)}  &
\colhead{(8)}  &
\colhead{(9)}  &
\colhead{(10)}  &
\colhead{(11)}  \\
}
\startdata
   M6    &    1    &    8    &    9    &    8    &    0    &    1    &    0    &     10.23 $\pm$      0.43    &      9.67 $\pm$      0.42    &      9.33 $\pm$      0.41    &   \\
   M7    &    0    &    7    &    7    &    4    &    0    &    0    &    3    &     10.93 $\pm$      0.20    &     10.28 $\pm$      0.13    &      9.90 $\pm$      0.09    &   \\
   M8    &    4    &    5    &    9    &    4    &    0    &    4    &    1    &     10.87 $\pm$      0.27    &     10.22 $\pm$      0.24    &      9.84 $\pm$      0.22    &   \\
   M9    &    2    &    5    &    7    &    3    &    0    &    1    &    3    &     11.72 $\pm$      0.48    &     11.06 $\pm$      0.55    &     10.58 $\pm$      0.55    &   \\
\hline
   L0    &    2    &    8    &   10    &    6    &    2    &    2    &    0    &     11.69 $\pm$      0.24    &     11.05 $\pm$      0.21    &     10.51 $\pm$      0.19    &   \\
   L1    &    3    &    5    &    8    &    7    &    0    &    1    &    0    &     12.26 $\pm$      0.34    &     11.52 $\pm$      0.37    &     10.92 $\pm$      0.41    &   \\
   L2    &    3    &    2    &    5    &    4    &    1    &    0    &    0    &     12.82 $\pm$      0.03    &     12.09 $\pm$      0.07    &     11.44 $\pm$      0.16    &   \\
   L3    &    0    &    6    &    6    &    4    &    2    &    0    &    0    &     12.84 $\pm$      0.15    &     11.93 $\pm$      0.10    &     11.33 $\pm$      0.09    &   \\
\enddata
\tablecomments{All objects with parallaxes and absolute mag uncertainty $<$ 0.5.  Note that Bin are Binaries, SD are Subdwarfs, LG are Low-surface gravity dwarfs, Y are young low mass companions}
\end{deluxetable}

\begin{deluxetable}{lccccccrrrrrrrr}
\label{tab:tab5}
\tabletypesize{\scriptsize}
\tablecaption{Coefficients of Piecewise Fits for L0 -T9 Dwarfs \label{coeffs2}}
\tablewidth{0pt}
\tablehead{
\colhead{MKO Filter} &
\colhead{c$_{0}$} &
\colhead{c$_{1}$} &
\colhead{SpT range}  \\
\colhead{(1)}  &
\colhead{(2)}  &
\colhead{(3)}  \\}
\startdata

M$_{J}$&	0.366$\pm$0.002 	& 	8.003$\pm$0.025	&		L0-L9\\
M$_{H}$&0.335  $\pm$    0.002&7.616  $\pm$     0.025&\\
M$_{K}$&0.297  $\pm$     0.002&7.500  $\pm$     0.025&\\
\hline
M$_{J}$&-0.304$\pm$0.020&21.039$\pm$0.438&T0-T4.0\\
M$_{H}$&-0.089  $\pm$     0.020&15.797  $\pm$     0.432&\\
M$_{K}$&0.082  $\pm$     0.019&11.729  $\pm$     0.430&\\
\hline
M$_{J}$&0.892$\pm$0.008&-8.035$\pm$0.212&T5.0-T9.0\\
M$_{H}$&0.977  $\pm$     0.008&-10.119  $\pm$     0.225&\\
M$_{K}$&0.895  $\pm$     0.006&-7.632  $\pm$    0.163&\\
\hline 
\enddata
\tablecomments{See Table ~\ref{coeffs} comments.}
\end{deluxetable}

\begin{deluxetable}{lccccccrrrrrrrr}
\label{tab:tab6}
\tabletypesize{\scriptsize}
\tablecaption{Low Gravity Dwarfs \label{lg}}
\tablewidth{0pt}
\tablehead{
\colhead{Name} &
\colhead{Spt} &
\colhead{M$_{J}$} &
\colhead{M$_{H}$} &
\colhead{M$_{K}$}&
\colhead{$\Delta_{M_{J}}$\tablenotemark{a}}&
\colhead{$\Delta_{M_{H}}$\tablenotemark{a}}&
\colhead{$\Delta_{M_{K}}$\tablenotemark{a}} \\
\colhead{(1)}  &
\colhead{(2)}  &
\colhead{(3)}  &
\colhead{(4)}  &
\colhead{(5)}  &
\colhead{(6)}  &
\colhead{(7)}  &
\colhead{(8)} \\
}
\startdata
2MASS J0422+1530	&	M6	$\gamma$	&	9.6	$\pm$	0.3	&	8.8	$\pm$	0.3	&	8.2	$\pm$	0.3	&	0.7	&	1.0	&	1.1	\\
2MASS J0221-6831		&	M8	$\beta$		&	10.9	$\pm$	0.3	&	10.3	$\pm$	0.3	&	9.8	$\pm$	0.3	&	-0.2	&	-0.2	&	0.0	\\
2MASS J0652-5741		&	M8	$\beta$		&	11.0	$\pm$	0.3	&	10.5	$\pm$	0.2	&	9.9	$\pm$	0.2	&	-0.3	&	-0.3	&	-0.1	\\
2MASS J0608-2753\tablenotemark{b}		&	M8.5	$\gamma$	&	11.1	$\pm$	0.3	&	10.5	$\pm$	0.3	&	9.9	$\pm$	0.3	&	-0.1	&	-0.1	&	0.1	\\
TWA28				&	M8.5	$\gamma$	&	10.2	$\pm$	0.4	&	9.6	$\pm$	0.4	&	9.1	$\pm$	0.4	&	0.8	&	0.7	&	0.9	\\
2MASS J1022+0200	&	M9	$\beta$		&	11.1	$\pm$	1.0	&	10.5	$\pm$	1.0	&	10.0	$\pm$	1.0	&	0.0	&	0.0	&	0.2	\\
TWA 26				&	M9	$\gamma$	&	10.3	$\pm$	0.3	&	9.8	$\pm$	0.3	&	9.2	$\pm$	0.3	&	0.8	&	0.8	&	0.9	\\
																						
\hline																						
2MASS J0518-2756	&	L0	$\gamma$		&	11.8	$\pm$	0.8	&	11.0	$\pm$	0.8	&	10.2	$\pm$	0.8	&	-0.2	&	-0.1	&	0.2	\\
\enddata
\tablenotetext{a}{$\Delta$ values are calculated from the MKO M$_{JHK}$ values from the polynomials presented in Table ~\ref{coeffs}. Negative values indicate under-luminous objects.}
\tablenotetext{b}{2MASS J0608-2753 is a member of the $\beta$ Pictoris moving group as described in \citet{2010ApJ...715L.165R}}
\tablenotetext{c}{Objects that are companions to young ($<<$ ~1 Gyr) main sequence stars}
\end{deluxetable}

\clearpage

\clearpage
\bibliographystyle{apj}
\bibliography{/Users/jfaherty/Dropbox/For_Submit_Revised/Paper2}

\end{document}